\documentclass[11pt, a4paper]{article}

\RequirePackage{amsthm,amsmath,amsfonts,amssymb}
\usepackage[round]{natbib} 
\RequirePackage[colorlinks,citecolor=blue,urlcolor=blue]{hyperref}
\RequirePackage{graphicx}
\RequirePackage{bm, dsfont}
\usepackage{listings}
\usepackage[margin=1in,headheight=13.6pt]{geometry}
\usepackage{authblk}
\usepackage{caption}
\DeclareMathOperator{\E}{\mathbb{E}}
\DeclareMathOperator{\V}{\mathbb{V}}
\DeclareMathOperator{\z}{\mathbf{z}}

\DeclareMathOperator{\betaf}{\bm{\beta}}
\DeclareMathOperator{\sigmaf}{\bm{\sigma}}
\DeclareMathOperator{\tauf}{\bm{\tau}}
\DeclareMathOperator{\lambdaf}{\bm{\lambda}}
\DeclareMathOperator{\deltaf}{\bm{\delta}}
\DeclareMathOperator{\phif}{\bm{\phi}}

\DeclareMathOperator{\xf}{\mathbf{x}}
\DeclareMathOperator{\tf}{\mathbf{t}}
\DeclareMathOperator{\Vvec}{ \{ \mathbf{v} \} }
\DeclareMathOperator{\vf}{\mathbf{v}}
\DeclareMathOperator*{\argmax}{arg\,max}

\usepackage{fancyhdr}
\begin{document}
	
	\title{A Bayesian accelerated failure time model for interval censored three-state screening outcomes}
	\author{Thomas Klausch (\href{t.klausch@amsterdamumc.nl}{t.klausch@amsterdamumc.nl})}
	\author[1] {\\Eddymurphy U Akwiwu}
	\author[1] {Mark A. van de Wiel}
	\author[1] {\mbox{Veerle M. H. Coup\'e}}
	\author[1] {\mbox{Johannes Berkhof}}
	\affil[1]{Amsterdam University Medical Centers, Department of Epidemiology and Data Science, Amsterdam, The Netherlands} 
	
	\maketitle
	
	\begin{abstract}
		Women infected by the Human papilloma virus are at an increased risk to develop cervical intraepithelial neoplasia lesions (CIN). CIN are classified into three grades of increasing severity (CIN-1, CIN-2, and CIN-3) and can eventually develop into cervical cancer. The main purpose of screening is detecting CIN-2 and CIN-3 cases which are usually removed surgically. Screening data from the POBASCAM trial involving 1,454 HPV-positive women are analyzed with two objectives: estimate (a) the transition time from HPV diagnosis to CIN-3; and (b) the transition time from CIN-2 to CIN-3. The screening data have two key characteristics. First, the CIN state is monitored in an interval censored sequence of screening times. Second, a woman's progression to CIN-3 is only observed if the woman progresses to, both, CIN-2 and from CIN-2 to CIN-3 in the same screening interval. We propose a Bayesian accelerated failure time model for the two transition times in this three-state model. To deal with the unusual censoring structure of the screening data, we develop a Metropolis-within-Gibbs algorithm with data augmentation from the truncated transition time distributions.		
	\end{abstract}

	
\section{Introduction} \label{sec::introduction}

Human papilloma virus (HPV) infections are considered the prime cause of cervical cancer, one of the most common cancers in young women world-wide. After the infection has been acquired through sexual contact, a woman is at risk to develop cervical intraepithelial neoplasia (CIN) lesions in the subsequent years, which can eventually develop into cervical cancer. CIN are classified into three grades of increasing severity: CIN-1, CIN-2, and CIN-3. Treatment is recommended for CIN-2 and CIN-3, but CIN-2 is considered an ambiguous diagnosis and many CIN-2 cases are thought to cure spontaneously without treatment. Hence, CIN-3 is the primary target for detection and treatment in cervical cancer screening programs. For designing screening programs it is essential to have information on two time distributions: (a) the total transition time from HPV diagnosis (state 1) to CIN-3 (state 3) which can be used to optimize screening schedules at the time of expected transition; and (b) the transition time from CIN-2 (state 2) to CIN-3 (state 3) which can be used to judge the speed of transition to CIN-3. \\

To estimate these transition times, we re-analyze 1,454 HPV-positive women, recruited between 1999 and 2002 in the \textit{Population Based Screening Study Amsterdam} \citep[POBASCAM;][]{rijkaart_human_2012, dijkstra_safety_2016}. Participants were screened for CIN progression at multiple occasions yielding a series of interval censored observations of the disease status. The key characteristic of screening data, like POBASCAM, is their censoring structure: after a woman is diagnosed with CIN-2, the lesion is radically removed and the screening series is censored, since the patient's progression from CIN-2 to CIN-3 cannot be observed anymore. As a consequence, CIN-3 is found only in those women who progress to, both, CIN-2 and from CIN-2 to CIN-3 in the same time interval. \\ 

\begin{figure}[t]
	\centering
	\includegraphics[scale=.55, trim=0 270 300 50, clip]{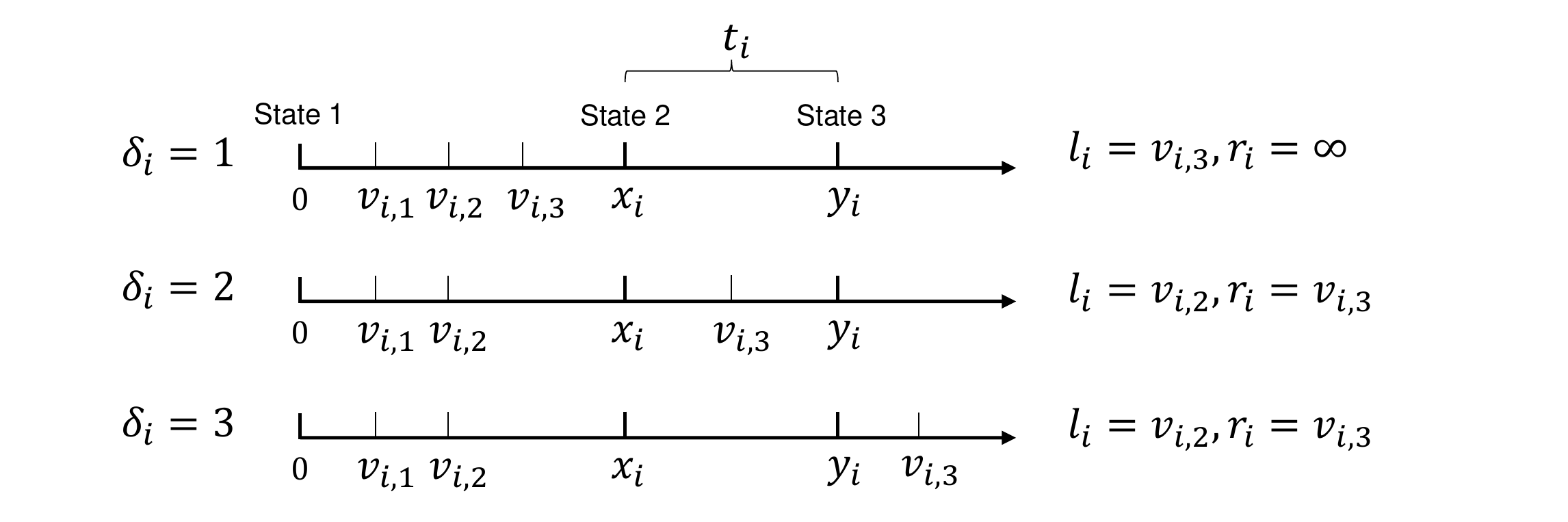}
	\caption{Censoring structure in the present study. The arrows represent three timelines for a person $i$ with transition times $y_i=x_i+t_i$. Superimposed are three hypothetical screening sequences with screening times $(v_{i,1},v_{i,2},v_{i,3})$ leading to the censoring states  $\delta_i \in \{1,2,3\}$, corresponding to HPV+, CIN-2, and CIN-3. The coding of the censoring interval  $\phi_i = (l_i,r_i]$ is shown on the right. }
	\label{fig::setup}
\end{figure}

The type of screening data considered in the present study follow a \textit{progressive three-state model}, where state 2 (CIN-2) follows state 1 (HPV infection), and state 3 (CIN-3) follows state 2 (CIN-2). In a sample of individuals, $i=1,...,n$, the latent onset transition time from state 1 to 2 is $x_i$ and the progression transition time from state 2 to state 3 is $t_i$, where the total transition time is $y_i=x_i+t_i$. In Figure \ref{fig::setup}, we superimpose three hypothetical sequences of screening times $(v_{i,1},v_{i,2},v_{i,3})$ on this natural disease process yielding the censoring states $\delta_i \in \{1,2,3\}$. The screening times may be regular (e.g., planned) or irregular, as shown in Figure \ref{fig::setup}. For $\delta_i = 1$, the individual does not progress to state 2 before the last screening moment ($v_{i,3}$), so that the individual is in state 1 at the end of follow-up.  We use $\phi_i = (l_i,r_i]$ to denote the most recent screening interval and its coding is shown on the right hand of Figure \ref{fig::setup}. We adopt the convention that if $\delta_i = 1$, $l_i$ is equal to the last screening time ($v_{i,3}$) and $r_i = \infty$.  For $\delta_i=2$, a state 2 event is detected at the last screening time before the individual progresses to state 3 ($v_{i,2} < x_i \le v_{i,3} < y_i$). After its detection, CIN-2 is removed and, therefore, we do not observe anymore when the individual would have progressed to CIN-3. For $\delta_i=3$, a state 3 event is observed. This implies that the individual has progressed from state 1 to 2 and the individual has continued progression from state 2 to the detected state 3 ($v_{i,2} < x_i <  y_i \le v_{i,3}$).   \\

In this paper, we develop a Bayesian parametric simultaneous accelerated failure time (AFT) model for the two transition times $x_i$ and $t_i$. Related methodology is reviewed in Section \ref{sec::lit_review}. Our method is novel in that it allows choosing any type of parametric distribution for the transition times. In particular, our model can handle non-constant and non-proportional hazards on both transitions (e.g., Weibull, loglogistic or lognormal distributions). Furthermore, the method is tailored to our censoring setting which is often encountered in screening data. This setting is challenging, since the transition of a woman to CIN-3 is never observed if a CIN-2 lesion is detected during screening, so that estimating $t_i$ is hard. \\   

The Bayesian estimation approach is motivated as follows. First, we develop a Markov Chain Monte Carlo (MCMC) algorithm with a general data augmentation step which can handle any parametric transition time distribution, provided its truncated form exists. Through this, different distributions can be 'plugged in' to the algorithm and standard Bayesian model selection can readily be applied to select the best distributions for the data. We also derive an alternative maximum likelihood (ML) expectation-maximization (EM) estimator to demonstrate that it needs a dedicated implementation for any choice of transition time distributions, rendering it the less flexible approach. Second, we use weakly informative priors to enhance the robustness of parameter estimation through regularization. It is well-known that priors effectively regularize the likelihood and can improve optimization in ill-posed problems (e.g., flat and multi-modal likelihoods) by putting higher weight on a range of realistic values \citep[e.g.,][]{gelman_weakly_2008}. The benefit of regularization is usually greatest in sparse data settings, which indeed emerge due to our censoring structure when estimating the parameters of the distribution of $t_i$. As we show, these parameters are updated only by information from patients experiencing state 2 and 3 events. This problem is aggravated in screening data, where advanced states usually occur infrequently (19\% in POBASCAM). We demonstrate that Bayes can outperform ML-EM in such settings. Third, Gibbs sampling facilitates inference on parameters as well as the cumulative incidence functions (CIFs). In contrast, ML-EM requires re-sampling techniques which can be costly due to numerical integration in the E-step. \\

In the next section, we review related statistical methodology. Subsequently, Sections \ref{sec::model} and \ref{sec:estimation} present the model, the Bayesian estimation procedure, and the ML-EM algorithm.  The proposed method is evaluated in a simulation study (Section \ref{sec::simulation}) and applied to the data from POBASCAM (Section \ref{sec::application}). We discuss the results and the methodology in Section \ref{sec::discussion}. 

\section{Related statistical methodology} \label{sec::lit_review}
A straightforward modeling approach treats the three-state screening data (Figure \ref{fig::setup}) as univariate interval censored. This approach estimates the distribution of $y_i$ by considering an observation right censored at $l_i$ if $\delta_i=1$, right censored at $r_i$ if $\delta_i=2$, and interval censored if $\delta_i=3$. Furthermore, to estimate the distribution of $x_i$ the approach considers $\delta_i=2$ and $3$ as the same event. These data could then readily be analyzed by standard survival methods, such as a parametric interval censored survival regression \citep{jackson_flexsurv_2016} or the non-parametric maximum likelihood estimator \citep[NPMLE;][]{turnbull_empirical_1976}. However, unfortunately, this approach cannot estimate $t_i$ and, in general, fails for $y_i$, as it introduces informative interval censoring. This problem emerges, because the re-coding scheme introduces a dependence of the censoring times and $y_i$; see proof in the Supplemental Material, Section A.  \\ 

More tailored methods for modeling screening data are available, in particular multi-state Markov and semi-Markov models \citep[for reviews see, e.g.,][]{mandel_estimating_2010, asanjarani_estimation_2021}. Multi-state models differ, broadly, concerning their assumptions on (a) the observation process (i.e., the censoring mechanism) and (b) the transition time distributions. These aspects are interlinked. First, if transitions are discretely observed, as in the present study (i.e., interval censoring, also called panel data), information on the time of transition is limited which may necessitate stronger constraints on the transition time distributions for tractable estimation. In Markov models, exponential distributions have been applied which implies that hazards are constrained to be constant. Consequently, estimation of models with arbitrary transitions and censoring structures is tractable \citep[e.g.,][]{jackson_multi-state_2011}. However, we find the constant hazard constraint unrealistic, for instance because HPV infections can clear over time. \\

Second, if the transition times are exactly observed (possibly right or left censored), which they are not in our study, more information is available for modeling distributions with non-constant hazards (semi-Markov models). The \texttt{R} package \texttt{flexsurv} \citep{jackson_flexsurv_2016}, for example, then allows fitting multi-state models with various parametric survival distributions, while the package \texttt{mstate} \citep{wreede_mstate_2011} estimates semi-parametric proportional hazards. \\

Third, the problem of estimating semi-Markov multi-state models with interval censored transitions, addressed in the present study, is challenging, because the involved likelihoods can become intractable \citep{lange_fitting_2013}. Our Bayesian three-state model can be viewed as a semi-Markov model applying to a specific interval censored observation process (Figure \ref{fig::setup}) and allowing arbitrary parametric transition time distributions with potentially non-constant and non-proportional hazards by an AFT specification. To the contrary, previous studies have offered tailored solutions for other censoring structures, admissible transitions, or transition time distributions. Progressive three-state models for 'doubly-interval censored' data, for example, differ from our setting in that each individual has two paired intervals in which $x_i$ and $y_i$ occur, respectively \citep{de_gruttola_analysis_1989}. \cite{komarek_bayesian_2008} proposed a Bayesian AFT model for doubly-interval censored three-state models which allows approximation of the transition time distributions by a Gaussian mixture. Similarly, \cite{boruvka_sieve_2016} addressed doubly-interval censoring in an illness-death model by a sieve estimator. Furthermore, several multi-state semi-Markov models have been developed for Weibull transitions \citep{foucher_semi-markov_2007, foucher_flexible_2010, kang_statistical_2007, wei_semi-markov_2016}. These models either assumed interval censoring of all events, exactly observed absorbing event times (e.g., death), or they required one transition to be exponentially distributed. While the first two requirements are not met by the three-state screening data in our study, we find the Markov assumption too restrictive. \\

Finally, an approach suggested by \cite{titman_semi-markov_2010} and \cite{lange_fitting_2013} can be applied to approximate semi-Markov transitions in multi-state models by specifying multiple latent Continuous Time Markov Chains \citep[see also][]{lange_estimating_2018}. Another approximate method, suggested by \cite{jackson_multi-state_2011} for the \texttt{R} package \texttt{msm}, allows hazards to change at pre-specified time points (knots), so called piece-wise constant hazards. This approach models time inhomogeneous transitions, but cannot, in general, be used to approximate semi-Markov transitions. The two approximate methods require a priori choices on the type of latent transition structures or the number of knots that should be tried during model fitting. In contrast, the proposed Bayesian approach provides inference for semi-Markov transitions with arbitrary parametric time distributions.

\section{Three-state AFT model for screening data} \label{sec::model}

We assume that the random variables $x_i$ and $t_i$ follow two AFT models
\begin{align} 
	&\log x_i = \z_{x,i}' \betaf_x + \sigma_x \epsilon_{i} \label{eq::model_eq1} \\
	&\log t_i = \z_{t,i}' \betaf_t + \sigma_t \xi_{i} \quad \quad \quad \text{for} \ i=1,...,n, \label{eq::model_eq2}
\end{align}
where $n$ denotes the sample size and $\z_i = \z_{x,i} \cup \z_{t,i}$ denotes vectors of covariates of dimensions $(p \times 1)$, $(p_x \times 1)$, and $(p_t \times 1)$. The unknown regression parameters are $\betaf=(\betaf_x', \betaf_t')'$ and the unknown scale parameters are $\sigmaf =(\sigma_x, \sigma_t)$. The covariate vectors are understood to include an intercept on the first entry, respectively. We assume the joint density of the random error terms factors $f_{\epsilon,\xi}(\epsilon,\xi)=f_\epsilon(\epsilon)f_\xi(\xi)$, so that $x_i$ and $t_i$ are conditionally independent given the covariates $\z_i$, but marginally dependent. The pair $(x_i, t_i)$ is never observed so that estimating their association is hard. In the POBASCAM application, we include HPV-sub-types as known important common causes of faster transition in both model equations. In addition, we suggest a sensitivity analysis that relaxes conditional independence; see Section \ref{sec::sensitivity_analysis}. \\

The specific choice of $f_\epsilon$ and $f_\xi$ determines the distributions of $x_i$ and $t_i$, obtained by a change of variables
\begin{align} \label{eq::change_of_var}
	f_x(x_i | \z_{x,i}, \betaf_x, \sigma_x) = f_\epsilon [ (\log x_i - \z_{x,i}' \betaf_x) \sigma_x^{-1} ] (\sigma_x x_i)^{-1} \quad ; x > 0
\end{align}
and the density $f_t$ follows similarly. Common choices for $f_\epsilon$ and $f_\xi$ include the extreme value, logistic, and normal distributions, for which $x_i$ and $t_i$ are distributed Weibull, loglogistic, and lognormal, respectively (Table \ref{tab::change_of_var}). 

\begin{table*}
	\caption{Error distributions and change of variable in AFT models }
	\label{tab::change_of_var}
	\resizebox{\textwidth}{!}{
	\begin{tabular}{@{}lllll@{}}
	Error dist. & $f_\epsilon(\epsilon_i)$ & Dist. $x_i$ & $f_x(x_i)$ & Re-parameterization   \\ \hline
	&&&& \\
	Extreme value & $\exp( \epsilon_i - e^{\epsilon_i})$ & Weibull & $\frac{\eta}{\gamma} ( \frac{x_i}{\gamma} )^{\eta-1} \exp [- (\frac{x_i}{\gamma} )^\eta ]$ & $\eta := \sigma_x^{-1}$ ; $\gamma:= \exp( \z_{x,i}' \betaf_x)$ \\ &&&& \\
	Logistic & $e^{\epsilon_i}(1+e^{\epsilon_i})^{-2} $ & Log-logistic & $\frac{\eta}{\gamma} [\frac{x_i}{\gamma}]^{\eta-1} [1 + (\frac{x_i}{\gamma})^{\eta}]^{-2}$ & $\eta := \sigma_x^{-1}$ ; $\gamma:= \exp( \z_{x,i}' \betaf_x)$ \\  &&&& \\
	Normal & $[\sqrt{2\pi}]^{-1} \exp [ -\frac{1}{2}\epsilon_i^2 ]$ & Log-normal & $[\sigma_x x_i \sqrt{2 \pi}  ]^{-1} \exp [ -(\frac{\log x_i - \mu}{\sqrt{2}\sigma_x})^2 ]$ & $\mu:= \z_{x,i}' \betaf_x$ \\  &&&& \\
	\hline
\end{tabular}
}
\end{table*}

\subsection{Censoring mechanism} \label{sec::censoring}
A fixed or random screening process yields a vector of subsequent potential screening times  $\vf_i = (0, v_{i,1},...,v_{i,m_i}, \infty )$ with first and last entries set to zero and infinity, respectively, and $m_i$ random. We assume non-informative censoring in the sense that $x_i$ and $t_i$ are independent of $\vf_i$, unconditionally or conditionally on the covariates $\z_i$. Then the censoring interval $\phi_i = (l_i,r_i]$ is defined such that
\begin{align} \label{eq::def_phi}
\Pr( l_i = v_{i,j}, r_i = v_{i,j+1} | x_i, \vf_i ) = \mathds{1}_{\{ v_{i,j} < x_i \le v_{i,j+1} \}}
\end{align}	
for $j=1,...,m_i$, where $\mathds{1}$ denotes the indicator function; see \cite{komarek_bayesian_2008} for a similar definition of interval censoring. The corresponding censoring event is then
\begin{align} \label{eq::def_delta}
	\delta_i = \begin{cases}
		1 \ &\text{if} \ \ l_i < x_i < r_i = \infty  \\
		2 \ &\text{if} \ \ l_i < x_i \le r_i < x_i+t_i \\
		3 \ &\text{if} \ \ l_i < x_i < x_i+t_i \le r_i < \infty.
	\end{cases}
\end{align}
Indicator $\delta_i$ relates to states 1 (HPV-positive), 2 (CIN-2), and 3 (CIN-3), respectively. The condition $l_i < x_i $ appears in each level of $\delta_i$ and may also be omitted. Observe that indicator $\delta_i$ is marginally random, but, conditionally on $x_i, t_i$, and $\phi_i$, it is deterministic; for example, $\Pr(\delta_i=3 | x_i, t_i, \phi_i) = \mathds{1}_{\{x_i+t_i \le r_i < \infty \}}$. 

\subsection{Data and prior assumptions} \label{sec:data_and_priors}
The data observed for individual $i$ are $\mathcal{D}_i = \{\z'_i, \delta_i, l_i, r_i \}$ and we write $\bm{\mathcal{D}}$ for the stacked $[n \times (p+3)]$ matrix of $i = 1,...,n$ independent observations. Although the screening process $\vf_i$ is sometimes partly observed, i.e., the screening history $v_{i,1}...v_{i,j-1}$ is known, $\phi_i$ is sufficient to identify the posterior distribution, as discussed below. \\

\begin{figure}[t]
		\centering
	\includegraphics[scale=.40, trim=40 75 55 100, clip, page = 6]{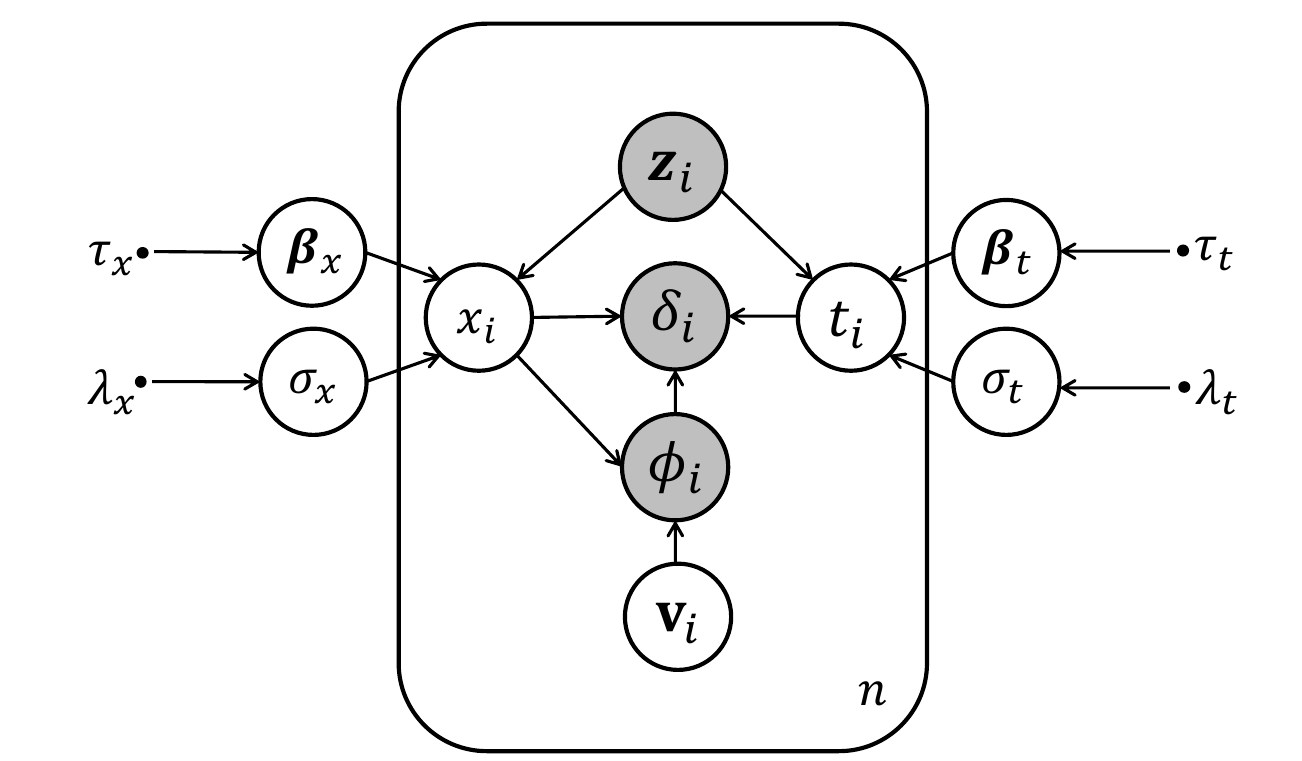}
	\caption{Graphical illustration of the three-state AFT model using plate notation of directed acyclical graph theory. Circles denote random variables, dots denote fixed variables, and filled circles denote observations. Arrows indicate the direction of causation; for example, $x_i$ is generated by $f_x(x_i|\z_i, \betaf_x,\sigma_x)$.}
	\label{fig::GM}
\end{figure}

The model parameters $\betaf$ and $\sigmaf$ are random variables with independent priors
\begin{align} \label{eq::priors}
	\pi( \betaf, \sigmaf | \tauf, \lambdaf) = \pi(\sigma_x|\lambda_x) \pi(\sigma_t|\lambda_t) \bigg[ \prod_{j=1}^{p_x} \pi(\beta_{x,j} | \tau_x) \bigg] \bigg[ \prod_{k=1}^{p_t}  \pi(\beta_{t,k} | \tau_t) \bigg],
\end{align}
where $\tauf=(\tau_x, \tau_t)$ and $\lambdaf=(\lambda_x, \lambda_t)$ are the hyperparameters. Figure \ref{fig::GM} summarizes the hierarchical structure of the model using plate notation. As argued in Section \ref{sec::introduction}, we use weakly informative priors to regularize parameter estimation. In particular, we set Student's t-distribution with four degrees of freedom as prior for $\beta_j$, i.e., $\pi(\beta_{j}|\tau) = student(\tau=4)$, which encodes the prior expectation that $\beta_{j}$ of a standardized $z_j$ lies with 95\% probability in (-2.776, 2.776) or $\exp(\beta_{j})$ in $(0.062,16.062)$, approximately, denoting large effects. Still the prior leaves the possibility for very large effects due to its wide tails. Furthermore, we use a half-normal $\pi(\sigma|\lambda) = N^{+}(0,\lambda=\sqrt{10})$ prior (with $\lambda$ the standard deviation) which encodes 95\% prior probability in the interval $(0.099, 6.198)$. Since the shape of the Weibull and lognormal distributions is governed by $\eta = \sigma^{-1}$ (Table \ref{tab::change_of_var}), the $\sigma$-prior locates $\eta$ with 95\% prior mass in $(0.161, 10.091)$. We believe, this prior generally allows very flexible shape of the time distributions. A prior sensitivity analysis is shown in our case study; see Section \ref{sec::application_sensitivity_prior}. 

\section{Bayesian estimation} \label{sec:estimation}
Let $q$ denote a generic density. Let  $\xf=(x_1,...,x_n)', \tf=(t_1,...,t_n)'$ be $(n \times 1)$ vectors of latent transition times, respectively, and $\Vvec = \{\vf_1,...,\vf_n \}$ the set of screening times vectors. Our objective is sampling from the joint posterior distribution
\begin{align} \label{eq::posterior}
	q( \betaf, \sigmaf, \xf, \tf | \bm{\mathcal{D}}, \Vvec ; \tauf, \lambdaf) =  	q(\betaf, \sigmaf | \xf, \tf, \bm{\mathcal{D}}, \Vvec ; \tauf, \lambdaf) q(\xf,\tf | \bm{\mathcal{D}}, \Vvec ).
\end{align}
The first factor on the right side of (\ref{eq::posterior}) is the complete data posterior which is proportional to the prior, given in (\ref{eq::priors}), times the complete data likelihood, given by
\begin{align} \label{eq:complete_LL}
	L(\betaf, \sigmaf | \xf, \tf, \bm{\mathcal{D}}, \Vvec) &\propto \prod_{i=1}^{n} q(x_i, t_i | \z_i, \betaf, \sigmaf ) q( \phi_i, \delta_i, \vf_i | x_i, t_i,  \z_i, \betaf, \sigmaf) \\ \nonumber
	&= \prod_{i=1}^{n} q(x_i, t_i | \z_i, \betaf, \sigmaf ) q( \phi_i, \delta_i, \vf_i | x_i, t_i, \z_i)  \\ \nonumber
	&\propto \prod_{i=1}^{n}  f_x(x_i | \z_i,\betaf_x, \sigma_x ) f_t(t_i | \z_i,\betaf_t, \sigma_t )  \\ \nonumber
	&= L (\betaf_x,\sigma_x | \xf, \z) \times  L (\betaf_t,\sigma_t | \tf, \z),
\end{align}

where $f_x$ and $f_t$ are defined in (\ref{eq::change_of_var}). We used d-separation based on Figure \ref{fig::GM} to establish independence of $\phif, \deltaf$ of the parameters conditional on $\xf, \tf$ in the second equation. The third equation results from the assumption of conditional independence. We find that the parameters are identified by the vector of latent times, which, however, are unobserved. Furthermore, the second factor on the right side of (\ref{eq::posterior}) is the marginal likelihood of the latent times that is hard to sample directly. Therefore, we employ Gibbs sampling to sample from  (\ref{eq::posterior}), based on the full conditional distributions of the random variables. 

\subsection{Gibbs sampling from the full conditional distributions} \label{sec::gibbs}
After suitable initialization of $\{\betaf, \sigmaf, \xf, \tf \}$, the Gibbs sampler proceeds by updating parameters by the following blocks
\begin{align} 
 \betaf^{(k+1)}, \sigmaf^{(k+1)} &\sim q(\betaf, \sigmaf | \xf^{(k)}, \tf^{(k)}, \bm{\mathcal{D}}, \Vvec; \tauf, \lambdaf) \label{eq::fc_par} \\
 x_i^{(k+1)} &\sim q( x_i |  \betaf^{(k+1)}, \sigmaf^{(k+1)}, t_i^{(k)}, \mathcal{D}_i, \vf_i; \tauf, \lambdaf) & i=1,...,n \label{eq::fc_x} \\
 t_i^{(k+1)} &\sim q( t_i |  \betaf^{(k+1)}, \sigmaf^{(k+1)}, x_i^{(k+1)}, \mathcal{D}_i, \vf_i; \tauf, \lambdaf) & i=1,...,n. \label{eq::fc_t}
\end{align}
We derive the full conditional distributions (\ref{eq::fc_par}-\ref{eq::fc_t}) in the following. From (\ref{eq:complete_LL}) it directly follows for (\ref{eq::fc_par}) that
\begin{align*}
	q(\betaf, \sigmaf | \xf, \tf, \bm{\mathcal{D}}, \Vvec; \tauf, \lambdaf) 
 \propto \ & q(\betaf_x, \sigma_x | \xf, \z; \tau_x, \lambda_x) q(\betaf_t, \sigma_t | \tf, \z; \tau_t, \lambda_t).
\end{align*}

The step in (\ref{eq::fc_par}) thus constitutes sampling from the complete data posteriors of two conditionally independent AFT models. In general, the weakly informative priors considered are not conjugate to the likelihoods. Therefore, we use a Metropolis step for sampling ($\betaf, \sigmaf$) jointly, applying a multivariate normal proposal distribution $\text{N}(\bm{\mu},\Sigma)$ with $\bm{\mu}$ centered at the previous draw $(\betaf^{(k)}, \sigmaf^{(k)})$ and a diagonal proposal variance matrix $\Sigma$.  \\

The steps in (\ref{eq::fc_x}) and (\ref{eq::fc_t}) constitute data augmentation of the latent transition times \citep{albert_bayesian_1993}. Proofs of the following results (\ref{eq::truncX1}-\ref{eq::bounds_cd}) are given in the Supplemental Material, Section A. For (\ref{eq::fc_x}), we have
\begin{align} \label{eq::truncX1}
	q( x_i |  \betaf, \sigmaf, t_i, \mathcal{D}_i,\vf_i; \tauf, \lambdaf) =
	f_{x}(x_i | a(\phi_i,\delta_i,t_i)<x_i<b(\phi_i,\delta_i,t_i), \z_i, \betaf_x, \sigma_x ),
\end{align}
i.e., the full conditional distribution of $x_i$ is a truncated distribution with bounds
\begin{align} \label{eq::bounds_ab}
	(a(\phi_i,\delta_i,t_i), b(\phi_i,\delta_i,t_i))= 
	\begin{cases}
		(l_i , \infty) \quad &\text{if} \ \delta_i=1 \\
		(\max(r_i-t_i,l_i), r_i ] \quad &\text{if} \ \delta_i=2 \\
		(l_i,r_i -t_i ] \quad &\text{if} \ \delta_i=3. \\
	\end{cases} 
\end{align}
Sampling from the truncated distribution is straight forward because the cumulative distribution $F_x$ is known; given $u \sim \text{uniform}(0,1)$, we evaluate $F^{-1}_x$ at $F_x(a) + u (F_x(b)-F_x(a))$. Similarly, for the full conditional (\ref{eq::fc_t}) we have
\begin{align} \label{eq::truncT}
	q( t_i |  \betaf, \sigmaf, x_i, \mathcal{D}_i,\vf_i; \tauf, \lambdaf) &= f_{t}(t_i | c(\phi_i,\delta_i,x_i)<t_i<d(\phi_i,\delta_i,x_i), \z_i, \betaf_t, \sigma_t )
\end{align}
with truncation bounds:
\begin{align} \label{eq::bounds_cd}
	(c(\phi_i,\delta_i,x_i), d(\phi_i,\delta_i,x_i))= 
	\begin{cases}
		(0 , \infty) \quad &\text{if} \ \delta_i=1 \\
		(r_i-x_i, \infty ) \quad &\text{if} \ \delta_i=2 \\
		(0, r_i-x_i ] \quad &\text{if} \ \delta_i=3. \\
	\end{cases} 
\end{align}
Note that these results imply that (\ref{eq::fc_par}-\ref{eq::fc_t}) depend on $\vf_i$ only through interval $\phi_i$. Posterior sampling, therefore, neither requires the full screening history to be part of the data nor the specification of a distribution for the screening process. Furthermore, we observe from (\ref{eq::bounds_cd}) that, for $\delta_i=1$, $t_i$ is not constrained by the data, i.e., $(c,d)=(0,\infty)$; as a result posterior estimation of $(\betaf_t, \sigma_t)$ crucially depends on the frequency of $\delta_i \in \{2,3\}$, i.e., information for updating $t_i$. However, in the POBASCAM data, as in other screening data sets likewise, advanced state events occur infrequently. The Bayesian estimation approach then can benefit from including weak prior information, as described in Sections \ref{sec::introduction} and \ref{sec:data_and_priors}. \\

Gibbs algorithm (\ref{eq::fc_par}-\ref{eq::fc_t}) was implemented in the statistical programming language \texttt{R} as package BayesTSM (\textit{Bayesian three-state model}), available under \url{https://github.com/thomasklausch2/BayesTSM}. \\

\subsection{Maximum likelihood EM-algorithm} \label{sec::EM}
Bayesian estimation is compared to an EM-algorithm used to maximize the observed data likelihood
\begin{align} \label{eq::obs_LL}
	L(\betaf, \sigmaf | \bm{\mathcal{D}}) \propto &\prod_{i: \delta_i=1} \big[ 1- F_x( l_i | \z_i, \betaf_x, \sigma_x) \big] \times \\  \nonumber
	&\prod_{i: \delta_i=2} \big[\int_{l_i}^{r_i} f_x(x_i | \z_i, \betaf_x, \sigma_x) [1 - F_t(r_i - x_i | \z_i, \betaf_t, \sigma_t)] dx_i  \big] \times \\ \nonumber
	&\prod_{i: \delta_i=3} \big[\int_{l_i}^{r_i} f_x(x_i | \z_i, \betaf_x, \sigma_x) F_t(r_i - x_i | \z_i, \betaf_t, \sigma_t) dx_i  \big]. \nonumber
\end{align}
The three product terms in (\ref{eq::obs_LL}) factorize the probabilities of occurrence of $x_i$ and $t_i$ in the observed interval $\phi_i$ given the observed state $\delta_i$. EM finds the maximum of (\ref{eq::obs_LL}) by iterating until convergence
\begin{align} \label{eq::EM.Q}
	(\betaf^{(k+1)}, \sigmaf^{(k+1)}) = \argmax_{\betaf, \sigmaf} \ \big[&\E_{ x_i | \mathcal{D}_i, \betaf^{(k)},\sigmaf^{(k)} } \big[ \log L (\betaf_{x},\sigma_x | \xf, \z) \big ] + \\ & \E_{t_i | \mathcal{D}_i, \betaf^{(k)},\sigmaf^{(k)} } \big[ \log L (\betaf_t,\sigma_t | \tf, \z) \big ] \big], \nonumber
\end{align}
where $L (\betaf_{x},\sigma_x | \xf, \z)$ and $L (\betaf_t,\sigma_t | \tf, \z)$ are the complete data likelihoods in (\ref{eq:complete_LL}). The expectations are taken with respect to the predictive densities $q(x_i | \mathcal{D}_i, \betaf,\sigmaf)$ and $q(t_i | \mathcal{D}_i, \betaf,\sigmaf)$, which are similar to the full conditional densities in (\ref{eq::fc_x}-\ref{eq::fc_t}), but do not condition on $t_i$ and $x_i$, respectively. In the Supplemental Material, Section A, we prove that

\begin{align} \label{eq::preddensx}
	q(x_i|\mathcal{D}_i,\betaf,\sigmaf) \propto g_x(x_i|\z_i,\betaf_{x},\sigma_x) [F_t(d(\phi_i,\delta_i,x_i) | \z_i, \betaf_{t},\sigma_t ) - F_t(c(\phi_i,\delta_i,x_i) | \z_i, \betaf_{t},\sigma_t )],
\end{align}

where $g_x(x_i|\cdot) = f_x(x_i|\cdot)$ if $l_i<x_i\le r_i$ and $g_x(x_i|\cdot) = 0$ else, and

\begin{align} \label{eq::preddenst}
	q(t_i|\mathcal{D}_i,\betaf,\sigmaf) \propto f_t(t_i | \z_i, \betaf_{t},\sigma_t ) [F_x(b(\phi_i,\delta_i,t_i) | \z_i, \betaf_{x},\sigma_x ) - F_x(a(\phi_i,\delta_i,t_i) | \z_i, \betaf_{x},\sigma_x )],
\end{align}
with $a, b, c, d$ the same bounds given in (\ref{eq::bounds_ab}) and (\ref{eq::bounds_cd}) for the data augmentation algorithm. We show in the Supplemental Material, Section B, that the lognormal distribution is a convenient choice for, both, $f_x$ and $f_t$. Under this model, the E-step is fast when $\delta_i =1$ as it only involves the density and percentiles of the normal distribution. However, for  $\delta_i \in \{2,3\}$ the E-step is computationally expensive due to numerical integration. We propose using importance sampling for this purpose. The M-step then has a closed-form, similar to the ordinary least squares estimator.  However, when $f_x$ and $f_t$ are not chosen lognormal, other dedicated implementations need to be derived which can involve numerical optimization in the M-step and other integration problems in the E-step. This result contrasts with the universality of the data augmentation scheme (\ref{eq::fc_x}-\ref{eq::fc_t}) which only requires that sampling from a truncated distribution is feasible. 

Th EM algorithm (was implemented in the statistical programming language \texttt{R} as part of package BayesTSM (\textit{Bayesian three-state model}), available under \url{https://github.com/thomasklausch2/BayesTSM}. \\

\subsection{Model selection} \label{sec::model_selection}
In practice, it is usually a priori unknown which parametric distributions of $x_i$ and $t_i$ are most appropriate. However, the posterior samples generated by Gibbs algorithm (\ref{eq::fc_par}-\ref{eq::fc_t}) generally allow for a variety of model selection techniques. Here we use information criteria: the Deviance Information Criterion (DIC) and the Widely Applicable Information Criterion of which there are two variants (WAIC-1/2). WAIC is superior to DIC under various settings; for example, if the posterior deviates from multivariate normality \citep{gelman_understanding_2014}. For details, see Supplemental Material, Section A. 

\subsection{Posterior predictive cumulative incidence functions}
 The cumulative incidence function (CIF), e.g., $F_x(\tilde{x} | \tilde{\z}, \betaf_x, \sigma_x )$, indicates the cumulative probability of occurrence of an event on or before time $\tilde{x}$ conditional on new covariates $\tilde{\z}$. The conditional posterior predictive CIF given a known value of $\tilde{\z}$ is
\begin{align} \label{eq::ppd_conditional}
		F_x( \tilde{x} | \tilde{\z}, \bm{\mathcal{D}})  = \iint_{\Omega_x} F_x(\tilde{x} | \tilde{\z}, \betaf_x, \sigma_x ) q(\betaf_x,\sigma_x | \bm{\mathcal{D}}) d\betaf_x d\sigma_x,
\end{align}

where $\Omega_x = \{\betaf_x \in \mathbb{R}^{p_x}, \sigma_x \in \mathbb{R}^+ \}$. To sample from (\ref{eq::ppd_conditional}), we make use of the draws $ (\betaf_{x}^{(k)}, \sigma_{x}^{(k)})$, $k=1,...,K$, provided by Gibbs sampler (\ref{eq::fc_par}-\ref{eq::fc_t}) by the push-forward transform $F_x(\tilde{x} | \tilde{\z}, \betaf_{x}^{(k)}, \sigma_{x}^{(k)} )$ which can readily be used, because $F_x$ has closed form. For $F_t$ we proceed similarly. Since the convolution $f_y$ does not have closed form in general, we obtain $F_y$ by Monte Carlo integration. For each $ (\betaf_{x}^{(k)}, \sigma_{x}^{(k)})$, we take a large sample $j=1,...,J$ of $\tilde{x}^{(k)}_j \sim f_x(\tilde{x} | \tilde{\z}, \betaf_{x}^{(k)}, \sigma_{x}^{(k)} )$ and $\tilde{t}^{(k)}_j \sim f_t(\tilde{t} | \tilde{\z}, \betaf_{t}^{(k)}, \sigma_{t}^{(k)} )$, so that $\tilde{y}^{(k)}_j = \tilde{x}^{(k)}_j + \tilde{t}^{(k)}_j$. The percentile $F_y(\tilde{y}|\tilde{\z}, \betaf^{(k)}, \sigmaf^{(k)} )$ is then obtained by the proportion of  $\{\tilde{y}_{1}^{(k)},...,\tilde{y}_{J}^{(k)}\}$ smaller than $\tilde{y}$. \\

Furthermore, the marginal posterior predictive CIF does not condition on $\tilde{\z}$ by integrating $F_x(\tilde{x} | \betaf_x, \sigma_x ) = \int F_x(\tilde{x} | \z, \betaf_x, \sigma_x ) q_{\z}(\tilde{\z})d\z$, so that
 \begin{align}\label{eq::ppd_unconditional}
	F_x( \tilde{x} |  \bm{\mathcal{D}})  = \iint_{\Omega_x} F_x(\tilde{x} | \betaf_x, \sigma_x ) q(\betaf_x,\sigma_x | \bm{\mathcal{D}}) d\betaf_x d\sigma_x.
\end{align}
We use Monte Carlo integration across $\tilde{\z}_j \sim q_{\z}(\tilde{\z})$, $j=1,...,J$, where $q_{\z}$ is the empirical distribution of the covariates. Given $(\betaf_{x}^{(k)},\sigma_{x}^{(k)})$, we sample $\tilde{x}^{(k)}_j \sim f_x(\tilde{x} |\tilde{\z}_j,\betaf_{x}^{(k)},\sigma_{x}^{(k)} )$. The percentile $F_x(\tilde{x} | \betaf_{x}^{(k)}, \sigma_{x}^{(k)} )$ is obtained by the proportion of values $\{\tilde{x}_{1}^{(k)},...,\tilde{x}_{J}^{(k)}\}$ smaller than $\tilde{x}$. Inference for $F_t( \tilde{t} |  \bm{\mathcal{D}})$ and $F_y( \tilde{y} |  \bm{\mathcal{D}})$ follows similarly. 

\subsection{Sensitivity analysis} \label{sec::sensitivity_analysis}
In Section \ref{sec::model}, we assume that $(x_i, t_i)$ in model (\ref{eq::model_eq1}-\ref{eq::model_eq2}) are conditionally independent given $\z_i$ because the random errors $(\epsilon_{i}, \xi_{i})$ are assumed independent. In this section, we propose a sensitivity analysis that relaxes this assumption. We assume that an unobserved common cause $w_i \sim p_w(w)$ is present such that 
\begin{align} 
	\log x_i = \z_{x,i}' \betaf_x + \beta_{x,w} w_i + \sigma_x \epsilon_{i} \label{eq::model_eq1_sens1} \\
	\log t_i = \z_{t,i}' \betaf_t + \beta_{t,w} w_i + \sigma_t \xi_{i}. \label{eq::model_eq2_sens1}
\end{align}
Latent variable $w_i$ is equivalent to an omitted covariate, also called an unobserved confounder. Coefficients $(\beta_{x,w},\beta_{t,w})$ are considered known and subjected to a sensitivity analysis by varying the coefficients across a range of plausible values in independent runs of Gibbs sampler (\ref{eq::fc_par}-\ref{eq::fc_t}); alternatively a prior may be imposed. We implemented (\ref{eq::model_eq1_sens1}-\ref{eq::model_eq2_sens1}) by sampling $w_i \sim N(0,1)$ at the start of the Gibbs sampler and holding $(\beta_{x,w},\beta_{t,w})$ fixed at their pre-specified values. Note that we still assume independence of $(\epsilon_{i}, \xi_{i})$, but now $x_i, t_i$ are independent conditionally on $\z_i$ and $w_i$. Results of the sensitivity analysis for the POBASCAM data are shown in Section \ref{sec::application_sensitivity}. 

\section{Simulation} \label{sec::simulation}
We conducted a series of experiments to assess the performance of Bayesian estimation compared to ML-EM and two alternative models: a progressive three-state Markov model and the NPMLE; see Section \ref{sec::lit_review}. Both methods are selected for comparison, as they represent obvious choices for practitioners due to their availability and ease of use.  

\subsection{Experimental set-up}
We generated data from the models
\begin{align*}
	\log(x_i) &= 3 + \beta_{1,x}  z_{i,1}+ \beta_{2,x} z_{i,2} + 0.2 \epsilon_{i}  \\
	\log(t_i) &= 1.2 + \beta_{1,t}  z_{i,1}+ \beta_{2,t} z_{i,2} + 0.3 \xi_{i},
\end{align*}
where errors followed a normal distribution, so that $x_i,t_i$ were lognormal (Table \ref{tab::change_of_var}). We sampled $\z_{i,1} \sim N(0,1)$ and $\z_{i,2} \sim Bernoulli(0.5)$. Parameters $\beta_{0,x}=3$ and $\beta_{0,t}=1.2$ are the intercepts and $\sigma_x=0.2$ and $\sigma_t=0.3$ the scale parameters. We then varied the following factors systematically:
\begin{enumerate}
	\item Sample size: $n=\{1000,2000$\}
	\item Number of covariates: $\beta_x = \beta_t = 0$ (i.e., $p=0$) or $\beta_x=\beta_t=0.5$ (i.e., $p=2$)
	\item Strength of right censoring (medium vs. strong, see below)
\end{enumerate}
This gave $2 \times 2 \times 2 = 8$ separate simulation conditions. Time $x_i$ had greater median (20.1)  than $t_i$ (3.3) under $p=0$; similarly, for $p=2$ the medians were 25.8 and 4.3. Furthermore, the times were marginally independent under $p=0$ but strongly correlated under $p=2$ (Pearson $r=.799$), because the same covariates $\z_i$ were included in both models. See the Supplementary Material, Section C, for further details. \\ 

The observed data were generated by the following screening process. First, we simulated $v_{i,1} \sim \text{uniform}(c_{min},c_{max})$, where $c_{min}<c_{max}$, and then recursively
\begin{align*}
	v_{i,j+1} \sim \text{uniform}(v_{i,j}+c_{min},v_{i,j}+c_{max})
\end{align*}
until $v_{i,j+1} > v_{i,rc}$ where $v_{i,rc} \sim \text{exp}(\theta^{-1})$ was the time of right censoring of $x_i$. The generated screening sequence thus was $\vf_i=(0,v_{i,1},...,v_{i,m_i},\infty)$ with $m_i = j$. Interval bounds $\phi=(l_i,r_i]$ and censoring event $\delta_i$ then followed from $(\ref{eq::def_phi}-\ref{eq::def_delta})$.  Parameters $c_{min}, c_{max}$ can be regarded as the minimum and maximum time between screening moments, whereas $\theta$ determines the strength of right censoring of $x_i$. We defined a 'medium' and a 'strong' censoring setting, where we set $\theta$ to smaller values in order to obtain smaller mean times to right censoring in the 'strong' condition. To do so, note that, given the transition times $x_i$ and $t_i$, the distribution of $\delta_i \in \{1,2,3\}$ depends on censoring parameters $c_{min},c_{max},\theta$. We manually tuned these parameters in a large ($n=10^6$) data setting to obtain approximately equal proportions of $\delta_i$ across the conditions with $p=0$ and $p=2$ covariates (Table \ref{tab::simcens}). We did so in order to study the impact of censoring independently of $p$.

\begin{table*}[h]
	\centering
	\caption{Parameters of the screening times process and resulting distributions of $\delta_i$ \\($^*$estimated by Monte Carlo with $10^6$ replications). Parameters $c_{min}$ and $c_{max}$ indicate the minimum and maximum time between screening moments and $\theta$ indicates the mean time to right censoring of $x_i$}
	\label{tab::simcens}
	\begin{tabular}{@{}l c c l c c @{}}
	& \multicolumn{2}{c}{Medium censoring} && \multicolumn{2}{c}{Strong censoring} \\
	\cline{2-3}
	\cline{5-6}
	& $p=0$ & $p=2$ &&  $p=0$ & $p=2$  \\ 
	\hline
	$c_{min}$ & 1 & 1 && 1 & 1 \\  
	$c_{max}$ & 8 & 8.7 && 7 & 6.5 \\  
	$\theta$ & 40 & 56 && 20 & 26.1 \\
	$\Pr(\delta_i = 1 | c_{min},c_{max},\theta )^*$ & 0.368 & 0.370 && 0.600 & 0.601 \\
	$\Pr(\delta_i = 2 | c_{min},c_{max},\theta)^*$ & 0.426 & 0.426 && 0.298 & 0.296 \\
	$\Pr(\delta_i = 3 | c_{min},c_{max},\theta)^*$ & 0.206 & 0.204 && 0.103 & 0.103 \\	
	\hline
\end{tabular}
\end{table*}

\subsection{Implementation} \label{sec::sim_implementation}
We generated 500 data sets for each experimental condition and estimated the three-state model using MCMC sampler (\ref{eq::fc_par}-\ref{eq::fc_t}) with weakly informative priors (Section \ref{sec:data_and_priors}). Three randomly initialized chains were ran for $10^5$ iterations, initially. We continued sampling, if necessary, until the Gelman-Rubin convergence statistic $R$ decreased under the threshold of $R<1.1$ and the effective sample size (of posterior draws) was at least 30 for \textit{each} model parameter. We discarded half of the iterations as warm-up (burn-in) period, respectively, before calculating these statistics \citep[choices recommended by][p. 285-287]{gelman_bayesian_2013}. For computational efficiency, we limited the maximum number of draws to $2 \times 10^6$, after which the Gibbs sampler stopped. Non-convergence until $2 \times 10^6$ draws occurred in 119 out of the $8 \times 500$ data sets; these cases were replaced by additional runs. The proposal variance of the Metropolis step was tuned using a heuristic search to achieve the optimal acceptance rate of approximately 23\% \citep[][p. 296]{gelman_bayesian_2013}. This was done for \textit{each} data set separately to ensure comparability of convergence rates across simulation conditions; see Supplementary Material, Section C. Implementation notes for the three-state Markov model and the NPMLE are also given in the Supplement. The ML-EM algorithm was considered converged when the change in observed data likelihood between successive iterations came below a value of $10^{-5}$ while using $5 \times 10^3$ importance samples for the numerical integration in the E-step. 

\subsection{Results}
\begin{figure}[h]
		\centering
	\includegraphics[scale=.50, trim=170 160 170 190, clip, page = 1]{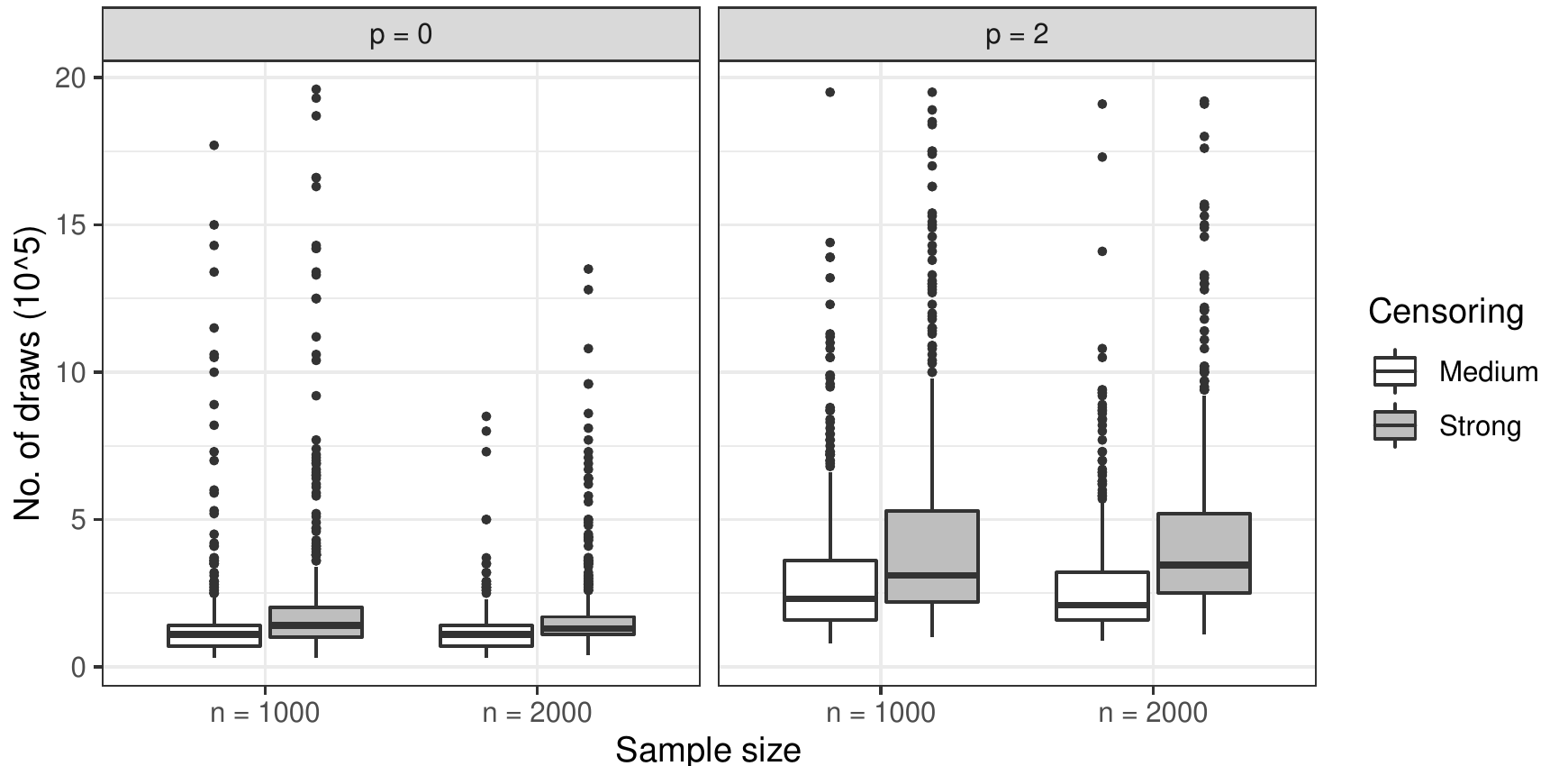}
	\caption{Number of MCMC draws required to obtain Gelman-Rubin convergence statistic $R<1.1$ and an effective sample size of at least 30 for all parameters. 500 replicated data sets. Convergence evaluated every $10^4$ draws.}
		\label{fig::sim_convergence}
\end{figure}

We compared MCMC convergence across the simulation conditions (Figure \ref{fig::sim_convergence}). Without covariates ($p=0$), convergence was obtained in less than $2.5 \times 10^5$ draws in most replications. With covariates ($p=2$), more draws were needed, where the data sets exhibiting strong censoring required most draws. Still, at most $10^6$ draws were sufficient to achieve convergence in most cases in the $p=2$ setting. We also studied the rate of convergence by model parameters and found that the parameters of $f_t$ had slower convergence than those of $f_x$; see Supplementary Material, Section C.  \\
 
\begin{figure}[h]
	\centering
	\includegraphics[scale=.60, trim=130 120 100 150, clip, page =1]{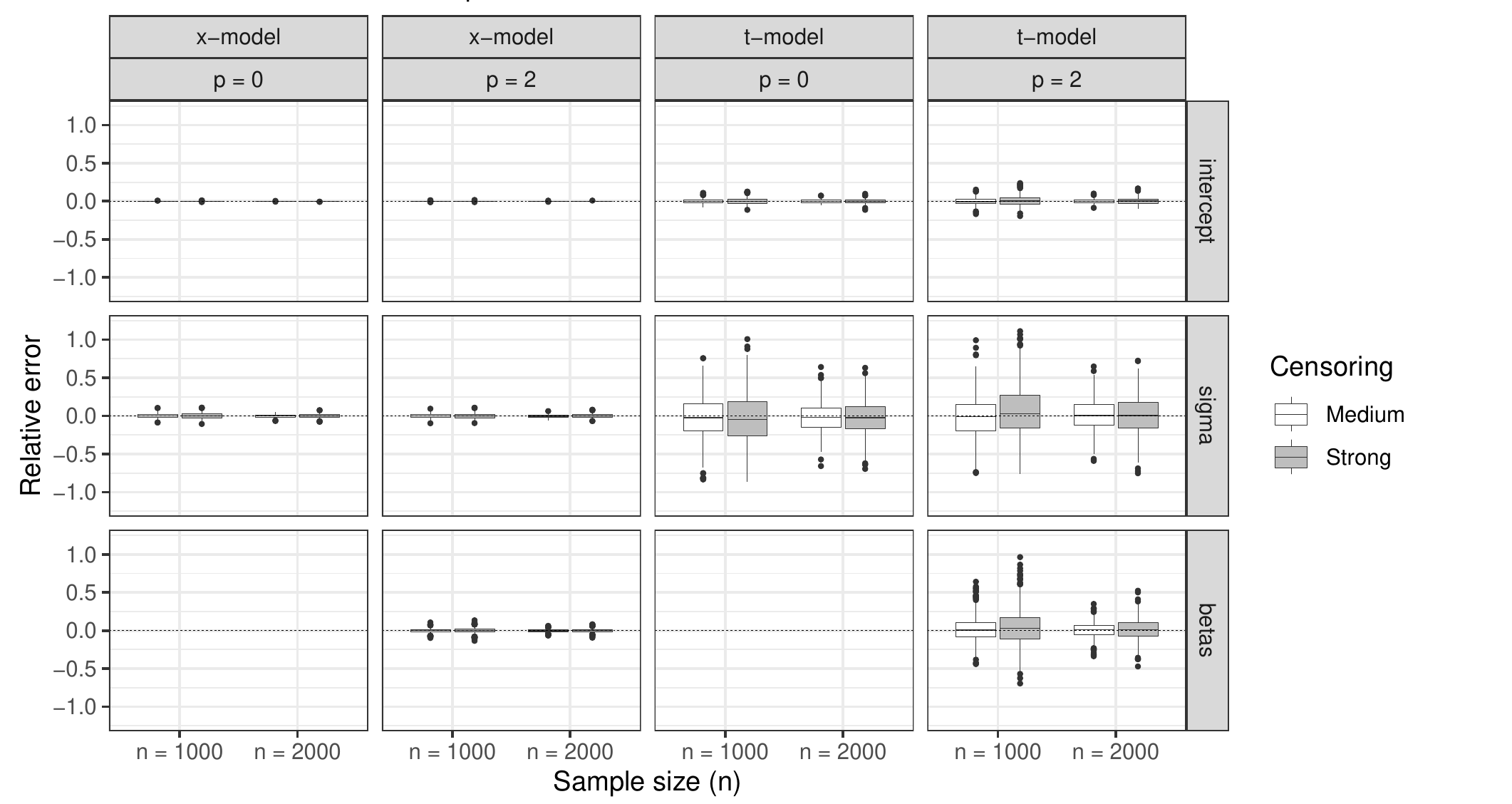}
	\caption{ Relative error of Bayesian estimation compared to ML-EM for the parameters of the model for $t_i$ across 500 replications. Results for the parameters $\beta_1$ and $\beta_2$ are reported jointly as 'betas'.}
	\label{fig::mcmcvsem_t}
\end{figure}

We compared the relative error of the parameter estimates of the Bayesian Gibbs sampler to those of ML-EM. Relative error was defined as $[\hat{\kappa}-\kappa]/\kappa$  where $\kappa$ is the true value of the parameter and $\hat{\kappa}$ a posterior median estimate or a ML estimate. Importantly, the performance characteristics of the Bayesian and the ML-EM estimators were equivalent. Figure \ref{fig::mcmcvsem_t} shows the relative errors of Bayes across the simulated replications; comparisons with ML-EM are given in the Supplemental Material, Section C. All parameters of $f_x$ and $f_t$ had approximately zero error in central tendency (i.e., the estimators were approximately unbiased). The estimates of the parameters of $f_x$ were more precise than the estimates of the parameters of $f_t$, where strong right censoring and introducing covariates ($p=2$) increased the variance of all parameter estimates. Estimates of $\sigma_t$ had higher variance than those of $\beta_{0,t}$ and $\beta_{1,t},\beta_{2,t}$. Precision increased with sample size, as can be seen by slightly narrower inter-quartile ranges in the $n=2000$ conditions (in addition, mean squared error estimates are given in the Supplemental Material, Section C). We also studied the estimated frequentist coverage probability of the 95\% credible intervals which was at approximately nominal level for all parameters; see the Supplemental Material. \\

\begin{figure} [h]
		\centering
	\includegraphics[scale=.60, trim=130 120 100 135, clip, page = 6]{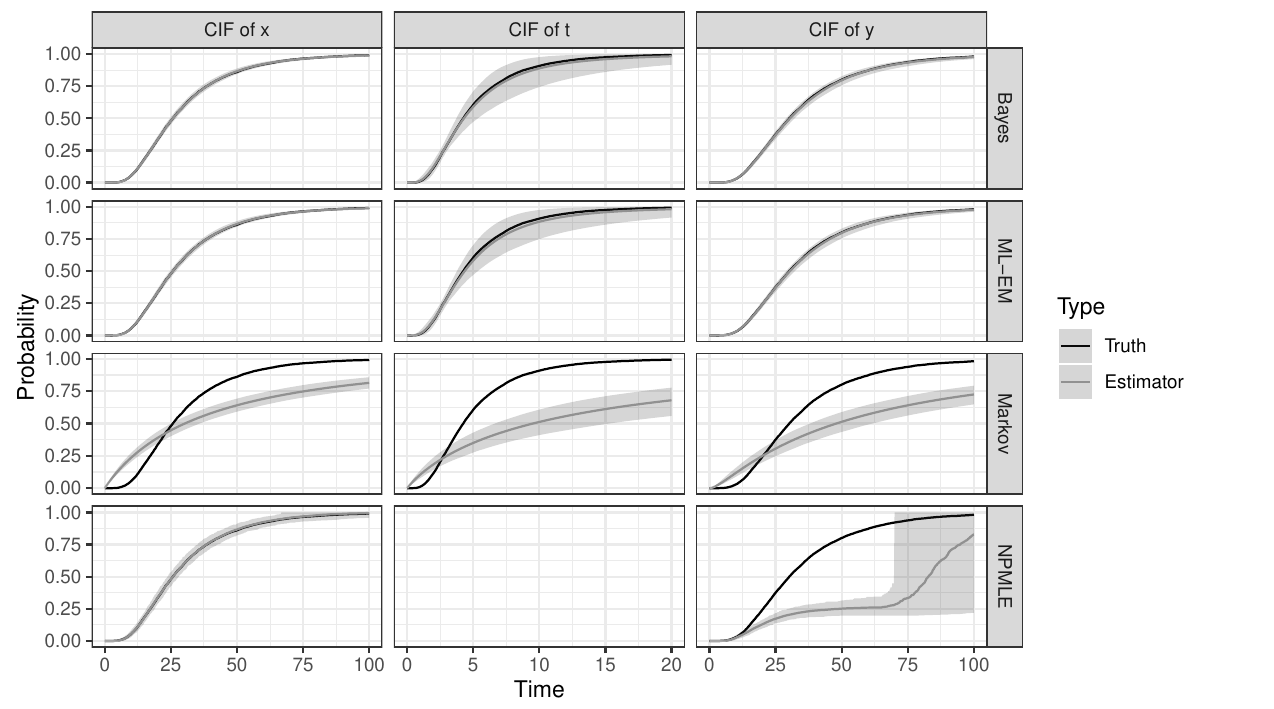}
	\caption{Marginal cumulative incidence functions (CIFs) compared for Bayesian estimation, ML-EM, the three-state Markov model, and the NPMLE. CIFs are point-wise averaged over 500 replications in the setting with $p=2$, $n = 1000$, and strong censoring; for other conditions see Supplemental Material, Section C. 95\% simulation intervals shown. }
	\label{fig::CIFcomparison}
\end{figure}

Bayesian estimation and ML-EM were compared with two existing methods: a progressive three-state constant hazard Markov model, implemented by \texttt{R} package \texttt{msm} \citep{jackson_multi-state_2011}, and the NPMLE, implemented by package \texttt{interval} \citep{fay_exact_2010}. Figure \ref{fig::CIFcomparison} shows estimated marginal cumulative incidence functions (\ref{eq::ppd_unconditional}) averaged point-wise over the replications for the setting with $p=2$, $n = 1000$, and strong censoring (see Supplental Material, Section C, for all other conditions). The Markov model had a strong bias for all transition times, while the NPMLE approximated the CIF of $x_i$ correctly and failed for $y_i$. These results are plausible, as the Markov model assumes exponentially distributed transition times while the true distributions were lognormal. Furthermore, the NPMLE introduces informed censoring when estimating the CIF of $y_i$ (see Section \ref{sec::lit_review}). To the contrary, our Bayesian and ML-EM estimation methods yielded approximately unbiased CIF estimates. These results were similar under the other simulation conditions. \\

Finally, we evaluated whether the information criteria WAIC-1, WAIC-2, and DIC could select the correct model among various misspecified models; see the Supplemental Matierial, Section C. In particular, we misspecified the true model with lognormal $x_i$ and lognormal $t_i$ as either lognormal-Weibull, lognormal-exponential, Weibull-lognormal or Weibull-Weibull. The true distribution of $x_i$ was reliably selected, while the true distribution of $t_i$ was selected in 66\% to 69\% of runs correctly (with Weibull selected in the remainder), unless the sample size was $n=1000$ and censoring was strong (50\%). However, even in the models using the Weibull for $t_i$ (or $x_i$), the regression coefficients and the CIFs were still estimated with negligible bias. 

\section{Application to POBASCAM data} \label{sec::application}

POBASCAM was a randomized trial based on a random sample of 44,102 women from the Dutch population recruited in the time between 1999 and 2002. The trial compared a new screening protocol for CIN lesions to standard of care. The new protocol required patients to be screened earlier in time for CIN progression after an initial HPV diagnosis. As an important result, CIN-3 lesions were found earlier in women screened by the new protocol than by standard of care. In this study, we re-analyzed a subset of $n=1,454$ women from POBASCAM who were diagnosed as HPV-positive without cytological abnormalities at the start of trial (baseline). These women are considered a high-risk group for progression to CIN-2 and CIN-3. Our first objective was to estimate the distribution of the onset time from baseline until development of a CIN-2 lesion ($x_i$), the progression time from CIN-2 to CIN-3 ($t_i$), as well as the total transition time ($y_i=x_i+t_i$). Our second objective was identifying population heterogeneity by regressing the transition times on patient characteristics ($\z_i$). \\

Women were followed for a maximum of 14.3 years (Table \ref{tab::data}). The mean time to right censoring was 8.6 years. A total of 96 women (6.6\%) were diagnosed with CIN-2; 176 women (12.1\%) were diagnosed with CIN-3. Of the 176 women, 17 women had also developed a cancer. For this reason, we refer to the CIN-3 category as CIN-3+ in the following. CIN-2 and CIN-3+ lesions were found at screening times averaging 4.9 and 5.3 years, respectively. Patients had a mean age of 38.1 years at baseline. About half (52.7\%) of women were diagnosed with a virus sub-type HPV-16, HPV-18 or HPV-31 which are associated with faster progression. 

\begin{table*}[h]
		\centering
	\caption{Description of the analysis set of HPV-positive women from POBASCAM. Categories indicate the state at the time of censoring (HPV-positive, CIN-2 or CIN-3+)}
	\label{tab::data}
	\resizebox{\textwidth}{!}{
	\begin{tabular}{@{} l r r r r @{}}
		& Total ($n=1,454$) & HPV-pos. ($n=1,182$) & CIN-2 ($n=96$) & CIN-3+ ($n=176$)  \\ 
		\hline
		\% of total & 100 & 81.3 & 6.6 & 12.1 \\
		Max. screening time (years) & 14.3 & 14.3 & 13.9 & 12.4 \\
		Time to censoring (mean/sd) & 8.0 (3.0) & 8.6 (2.6) & 4.9 (3.5) & 5.3 (2.9) \\ 
		Age (mean/sd) & 38.1 (8.5) & 38.7 (8.7) & 35.2 (6.8) & 35.8 (6.9) \\
		HPV-16 (\%) & 28.0 & 23.9 & 32.3 & 53.4 \\
		HPV-18 (\%) & 9.1 & 8.5 & 11.5 & 11.4 \\
		HPV-31 (\%) & 15.6 & 14.9 & 19.8 & 18.2 \\		
		\hline
	\end{tabular}
}
\end{table*}

\subsection{Implementation and analysis}
We modeled (\ref{eq::model_eq1}-\ref{eq::model_eq2}), i.e., the log of the transition times from baseline to CIN-2 ($x_i$) and from CIN-2 to CIN-3+ ($t_i$), as linear combinations of patient age (z-standardized) and the HPV sub-type indicators (Table \ref{tab::data}). The HPV sub-types are known causes of faster progression to CIN-3 and we assume that after including them as covariates, the conditional independence assumption discussed in Section \ref{sec::model}, holds. In addition, a sensitivity analysis was conducted; see Section \ref{sec::application_sensitivity}. We specified the error distributions of the AFT model (\ref{eq::model_eq1}-\ref{eq::model_eq2}) as either extreme value with $\sigma$ constrained to 1, extreme value (with $\sigma$ free), logistic, or normal, so that $x_i$ and $t_i$ were exponentially, Weibull, loglogistically or lognormally distributed (Table \ref{tab::change_of_var}). We fitted all possible combinations, leading to $4 \times 4 = 16$ models and subsequently performed model selection; see Section \ref{sec::model_selection}. We ran the Gibbs sampler (\ref{eq::fc_par}-\ref{eq::fc_t}) with fifteen randomly initialized MCMC chains and continued sampling until $R<1.1$ and the effective sample size was at least 1,000 for all parameters, dropping the first half of the draws. The MCMC chains were inspected for convergence. 

\subsection{Results}
The lognormal-loglogistic model fitted the data best (i.e., the model using the lognormal for $x_i$ and the loglogistic for $t_i$), but models that used either a loglogistic or lognormal distribution for one of the transition times had almost identical fit (Table \ref{tab::IC}). Models involving the Weibull distribution for any transition time fitted slightly worse and models involving the exponential distribution had substantially worse fit. Model convergence was reached, depending on the model, between 4 and $11 \times 10^5$ MCMC draws.   \\

\begin{table*}[h]
		\centering
	\caption{Information criteria ordered by WAIC-1 and the total number of MCMC draws until convergence (first half dropped for warm-up). Reading example: lognormal-loglog is the model with lognormal $x_i$ and loglogistic $t_i$. All statistics were computed using $10^5$ posterior draws }
	\label{tab::IC}
	\begin{tabular}{@{} l r r r c @{}}
		Model & WAIC-1 & WAIC-2 & DIC  & Draws ($\times 10^5$) \\ 
		\hline
lognormal-loglogistic      &1979.1& 1979.4& 1979.3 & 8 \\
lognormal-lognormal        &1979.6& 1979.9& 1979.7 & 7 \\
loglogistic-loglogistic    &1980.8& 1981.1& 1980.9 & 6 \\
loglogistic-lognormal      &1980.9& 1981.2& 1981.0 & 6 \\
lognormal-Weibull          &1981.2& 1981.6& 1981.2 & 7 \\
loglogistic-Weibull        &1982.6& 1983.0& 1982.6 & 6 \\
Weibull-loglogistic        &1983.2& 1983.4& 1983.3 & 8 \\
Weibull-lognormal          &1983.2& 1983.5& 1983.3 & 7 \\
Weibull-Weibull            &1985.2& 1985.5& 1985.2 & 7 \\
loglogistic-exponential    &2016.8& 2017.3& 2015.3 & 5 \\
lognormal-exponential      &2016.9& 2017.4& 2015.3 & 5 \\
Weibull-exponential        &2018.4& 2018.9& 2017.0 & 8 \\
exponential-loglogistic    &2154.8& 2155.1& 2154.3 & 11 \\
exponential-lognormal      &2154.9& 2155.2& 2154.4 & 8 \\
exponential-Weibull        &2156.7& 2157.0& 2156.0 & 8 \\
exponential-exponential    &2175.1& 2175.5& 2173.5 & 4 \\
		\hline
	\end{tabular}
\end{table*}

\begin{figure} [h]
		\centering
	\includegraphics[scale=.60, trim=100 160 100 150, clip, page = 1]{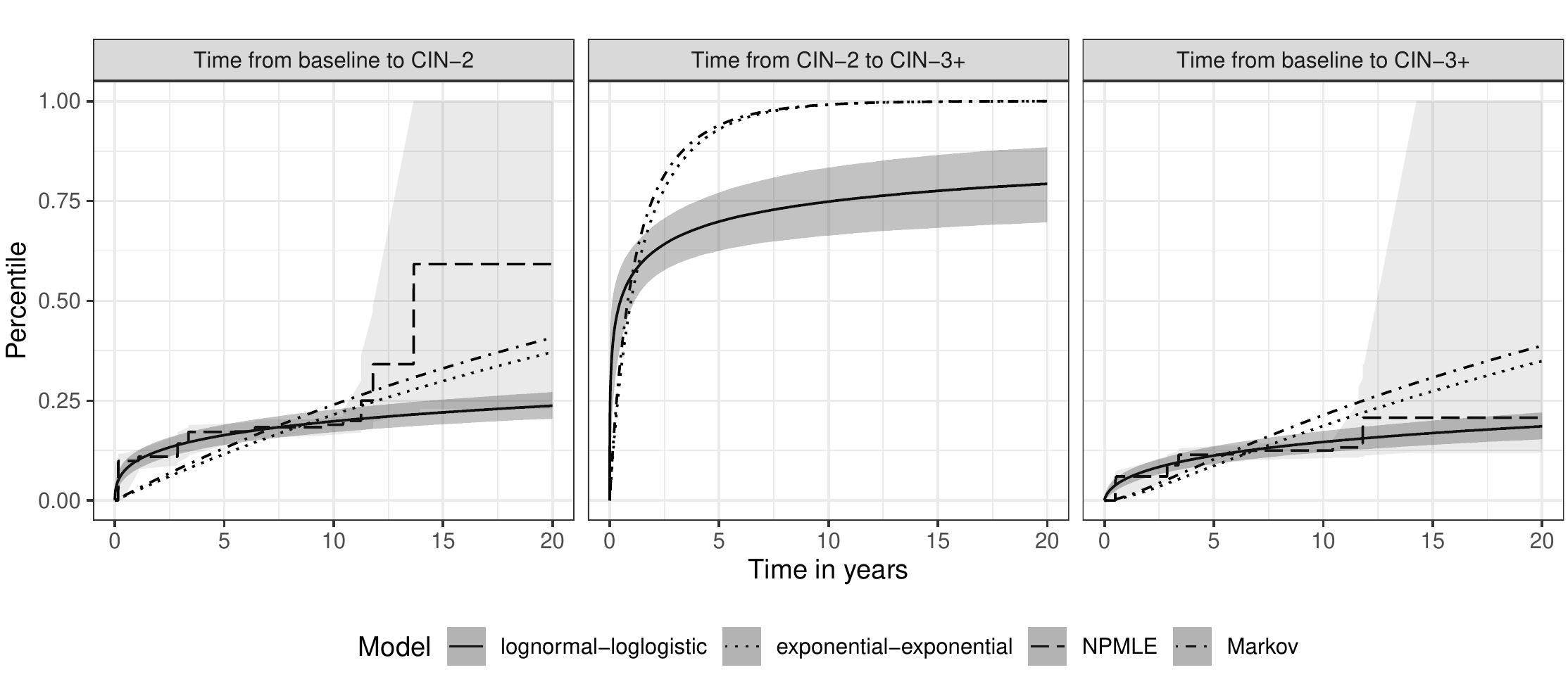}
	\caption{Posterior median of predictive CIFs (\ref{eq::ppd_unconditional}) for the Bayesian lognormal-loglogistic model (best performance acc. to all three information criteria) and the exponential-exponential model (worst performance) giving the cumulative probability to progress until a point in time. CIFs from a progressive three-state Markov model and the NPMLE are superimposed.  95\% credible intervals of the lognormal-loglogistic model shown as dark shaded region. Bootstrapped 95\% confidence intervals of the NPMLE shown as gray shaded region. }
	\label{fig::res_ppdplots_modcomp}
\end{figure}

We compared the marginal predictive CIFs (\ref{eq::ppd_unconditional}) of the lognormal-loglogistic model to the NPMLE, the progressive three-state Markov model, and the exponential-exponential model (Figure \ref{fig::res_ppdplots_modcomp}). The exponential-exponential model had worst fit (Table \ref{tab::IC}) and had similar CIFs as the Markov model. This result was expected, because the Markov model also assumes exponentially distributed transition times. However, the CIFs of both models deviated strongly from those of the lognormal-loglogistic model. In particular, the lognormal-loglogistic model CIFs had long tails suggesting that a part of the population never progresses from baseline to CIN-2 or from CIN-2 to CIN-3+ in their lifetime. The CIFs obtained by the NPMLE were similar to those of the lognormal-loglogistic model on the time from baseline to CIN-2 ($x_i$) up until the last observed screening time (14.3 years). Beyond this time point the estimator became imprecise. This result can be viewed as a validation of model fit, as the NPMLE is unbiased for this transition  (Sections \ref{sec::lit_review} and \ref{sec::simulation}). Surprisingly, the CIF of $y_i$ was also close to the Bayesian fit, which was unexpected, as the NPMLE is biased for this transition. We suspect that the bias of the NPMLE was mitigated as the proportion of CIN-2 events was small (6.6\%). \\

Based on the lognormal-loglogistic model, we estimated the probability for HPV-positive women to develop a CIN-2 lesion in the first twenty years after diagnosis of their HPV infection at 23.7\% (95\% credible interval (CI): [20.5, 27.2]; left plot,  Figure \ref{fig::res_ppdplots_modcomp}). Progression from CIN-2 to CIN-3+ was much more rapid (middle plot); probability of progression within the first ten years after development of CIN-2 was 74.8\% (95\% CI: [66.2, 83.3]). Also, already at the end of the first year women had over 50\% chance to have progressed. The time from baseline to CIN-3+ is the convolution of these two processes (right plot). CIN-3+ progression probability from baseline within the first twenty years was 18.6\% (95\% CI: [15.3, 22.0]). \\

\begin{table*}[b]
		\centering
	\caption{Percentiles of the posterior distribution of the lognormal-loglogistic model\\ ($x_i$: time to CIN-2; $t_i$: time from CIN-2 to CIN-3+)}
	\label{tab::parameters}
	\begin{tabular}{@{} l r r r r r r r @{}}
		& \multicolumn{3}{c}{Model for $\log x_i$} && \multicolumn{3}{c}{Model for $\log t_i$} \\
		\cline{2-4}
		\cline{6-8}
		Parameter & $p_{.025}$ & $p_{.50}$ & $p_{.975}$ && $p_{.025}$ & $p_{.50}$ & $p_{.975}$ \\
		\hline		
		Intercept & 6.82 &  7.94 & 9.09 && -1.31 & -0.02 & 0.78 \\
		Age (standardized) &   0.41 &  0.83 & 1.23 && -1.23 & -0.34 & 0.34 \\
		HPV-16* &      -4.09&  -3.04& -2.29&& -4.45 & -2.12 & -1.01 \\
		HPV-18* &      -2.68&  -1.40& -0.44&& -2.19 &  -0.47&  0.77 \\
		HPV-31* &      -2.48&  -1.37& -0.56&& -1.83 & -0.33&  0.80 \\
		sigma   &   4.09&   4.96&  5.85&&  1.55 &  2.71&  4.58 \\
		\hline
		\multicolumn{8}{l}{ * Reference: other HPV sub-type} \\
	\end{tabular}
\end{table*}

\begin{figure} [h]
		\centering
	\includegraphics[scale=.60, trim=100 160 100 150, clip, page = 1]{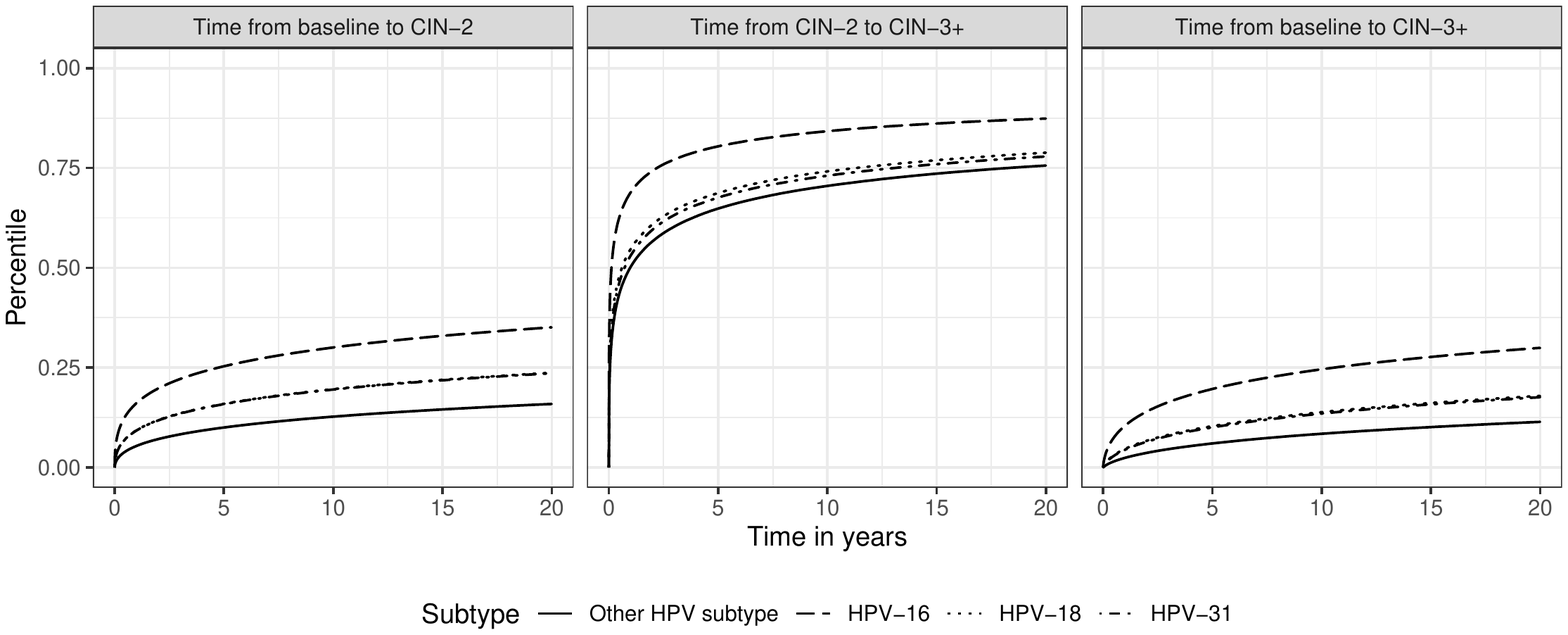}
	\caption{Posterior median of conditional predictive cumulative incidence functions (\ref{eq::ppd_conditional}) indicating the cumulative probability to progress until a point in time, given HPV sub-type and a mean age of 38.1 years (lognormal-loglogistic model).}
	\label{fig::res_condppdplots}
\end{figure}

Table \ref{tab::parameters} displays posterior median parameter estimates for the parameters in (\ref{eq::model_eq1}-\ref{eq::model_eq2}) as well as 95\% credible intervals (CI). Age had a positive effect and the three HPV sub-types had negative effects on the log-time from baseline to CIN-2 ($\log x_i$), where HPV-16 had the strongest effect ($\beta=-3.04$). Among the factors contributing to progression from CIN-2 to CIN-3+ ($\log t_i$) we, again, identified HPV-16 as the dominating factor ($\beta = -2.12$). There was no evidence for an influence of the other covariates. The effect of the HPV sub-types at a mean age (38.1 years) is illustrated by separate conditional predictive CIFs (\ref{eq::ppd_conditional}) in Figure \ref{fig::res_condppdplots}. For example, women with HPV-16 sub-type had 30.0\% risk of progression to CIN-3+ within 20 years (95\% CI: [24.8, 35.5]), whereas women with types 18, 31, or a different subtype ('other') had 18.0\%, 17.6\% and 11.5\% risk, respectively. Similarly, progression from CIN-2 to CIN-3+ was faster for HPV-16 than for all other HPV sub-types; for example, 10-year progression risk was 84.2\% for HPV-16 (95\% CI: [74.6, 91.8]) and 70.5\% for 'other' types (95\% CI: [61.2, 80.1]).

\clearpage
\subsection{Model fit by ML-EM}
Estimation of the lognormal-lognormal model by ML-EM (Section \ref{sec::EM}) is discussed in detail in the Supplemental Material, Section D. We used $2 \times 10^4$ importance samples in the numerical integration of the E-step and a convergence limit of $10^{-5}$. The maximum of the likelihood was determined by starting one hundred EM runs with random initialization. These runs converged at similar likelihood values but yielded quite different estimates of $\sigma_t$ and $\betaf_t$ which was indicative of a flat and/or multi-modal likelihood function. The maximum likelihood estimates (i.e., the estimates from the run maximizing the likelihood across all runs) were similar to the posterior median estimates from the Bayesian lognormal-lognormal model. The CIFs matched those of the Bayesian model closely. 

\subsection{Conditional independence sensitivity analysis} \label{sec::application_sensitivity}

We assessed the sensitivity of our findings to the conditional independence assumption of model (\ref{eq::model_eq1}-\ref{eq::model_eq2}), as described in Section \ref{sec::sensitivity_analysis}, for the lognormal-loglogistic model. In particular, we specified a grid of parameters 
\begin{align*}
	(\beta_{x,w},\beta_{t,w}) = \{ (-3,3),(-2,2),(-1,1),(0,0),(1,1),(2,2),(3,3) \}
\end{align*}
in model (\ref{eq::model_eq1_sens1}-\ref{eq::model_eq2_sens1}), where $(0,0)$ denotes the conditional independence model (\ref{eq::model_eq1}-\ref{eq::model_eq2}). The results of this analysis are shown in the Supplemental Material, Section D. The deviations from conditional independence caused a stronger curvature in the predictive CIFs, especially those of $t_i$. However, the conclusions of the main analysis did not change. Also the sign and significance of the regression coefficients did not change as compared to conditional independence.

\subsection{Prior sensitivity analysis} \label{sec::application_sensitivity_prior}
\begin{figure}[h]
		\centering
	\includegraphics[scale=.60, trim=100 170 100 150, clip, page = 1]{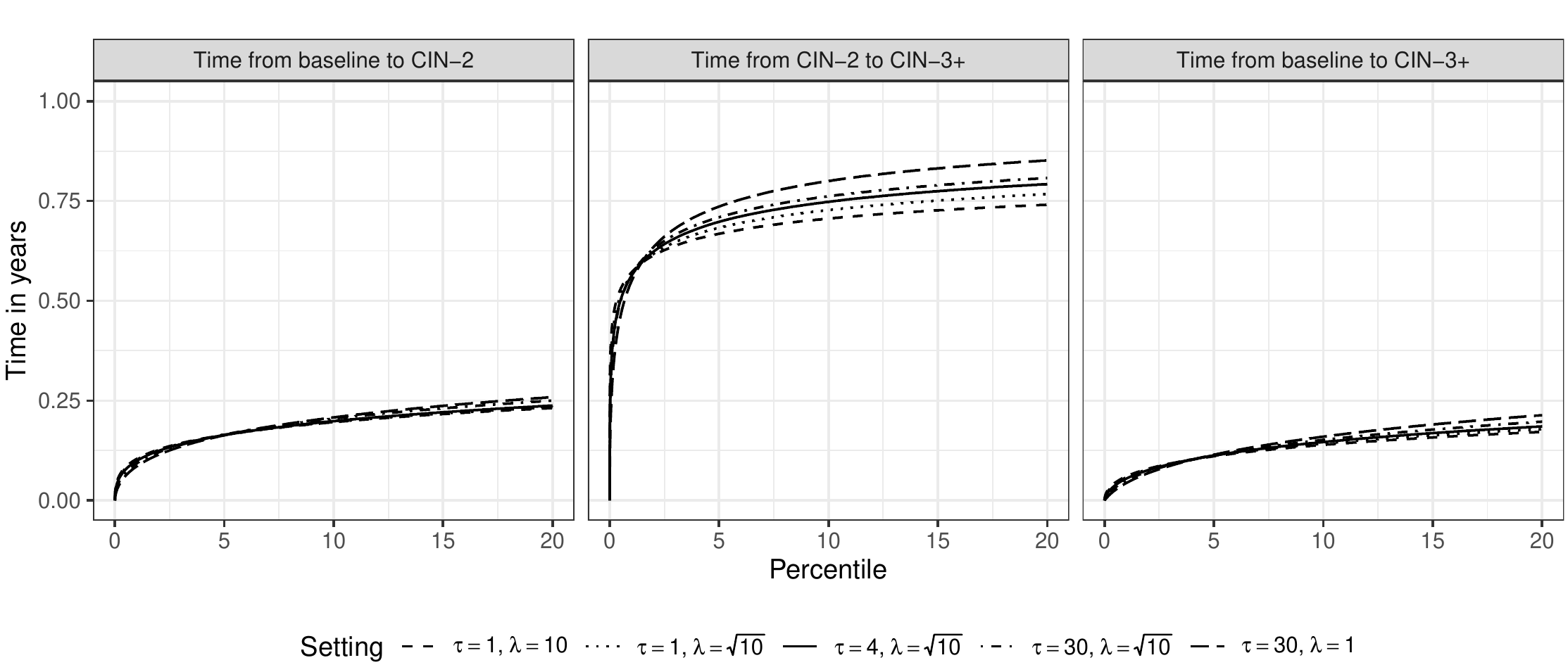}
	\caption{Posterior median of predictive cumulative incidence functions (\ref{eq::ppd_unconditional}) for different weakly informative hyperparameter choices.  The main analysis model uses $\tau=4$, $\lambda = \sqrt{10}$ (lognormal-loglogistic)).}
	\label{fig::sens_ppd_prior}
\end{figure}

\begin{figure}[h]
	\centering
	\includegraphics[scale = .57, trim=110 165 100 195, clip, page = 3]{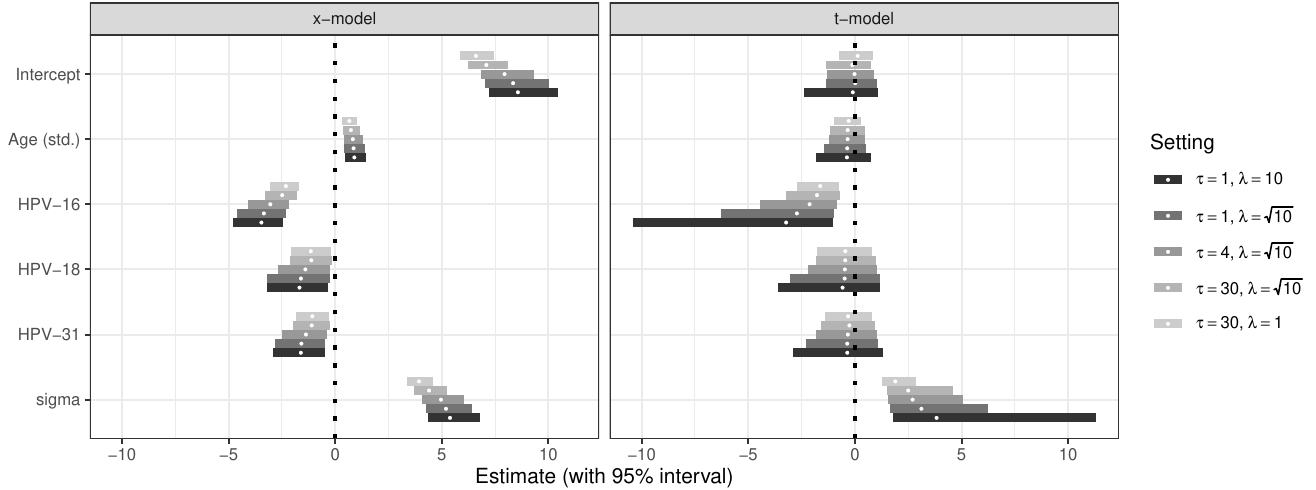}	
	\caption{Posterior parameter sensitivity to different weakly informative hyperparameter choices. The main analysis model uses $\tau=4$, $\lambda = \sqrt{10}$ (lognormal-loglogistic).}
	\label{fig::sens_prior_par}
\end{figure}

We assessed the sensitivity of the findings to alternative specifications of the weakly informative priors in the lognormal-loglogistic model. In particular, we considered two different t-distributions for $\beta$: $student(\tau=1)$ and $student(\tau=30)$, as opposed to $student(\tau=4)$ in our main analysis. These choices denote the central Cauchy and an approximate standard-normal distribution, respectively. For $student(\tau=30)$ we expect stronger regularization due to flat tails, whereas $student(\tau=1)$ has substantially wider tails than $student(\tau=4)$. Furthermore, the main analysis assumed a half-normal $N(0,\lambda=\sqrt{10})$ prior for $\sigma_x$ and $\sigma_t$. As alternatives, we considered $\lambda = 1$ and $\lambda = 10$ which are more and less restrictive, respectively. The predictive CIFs of $x_i$ and $y_i$ were insensitive to the prior alternatives (Figure \ref{fig::sens_ppd_prior}). Furthermore, also the predictive CIF of $t_i$ was not sensitive to changes in $\tau$, but exhibited moderate sensitivity to $\lambda$. The influence of the choice of $\tau$ on the posterior distributions of the parameters was negligible, although $\tau=30$ caused slightly stronger shrinkage, as expected (Figure \ref{fig::sens_prior_par}). However, $\lambda$ had a more profound impact: the posterior variance of $\sigma_t$ and $\beta_t$ of HPV-16 increased strongly under $\lambda=10$, illustrating the benefits of regularizing the parameters more strongly by setting $\lambda = \sqrt{10}$.

\section{Discussion} \label{sec::discussion}

The screening data considered in the present study have a very specific censoring structure (Figure \ref{fig::setup}): if state 2 of a disease or a condition is found at a screening time, the condition is treated (e.g., a CIN-2 lesion is surgically removed). As a result, the progression time ($t_i$) to state 3 remains unobserved. It is bounded only by the most recent screening interval if the patient progresses to, both, states 2 and 3, in the same interval. This data structure was present in the POBASCAM trial from the screening of HPV-positive women. However, there are many more health conditions where screening programs yield similar data; examples are colon cancer (e.g., with states adenoma and cancer) and breast cancer (e.g., with states ductal carcinoma in situ and cancer). In both cases, screening and prevention programs are in place to timely detect and treat the onset condition (state 2). The proposed three-state model, therefore, is anticipated to have wider applicability. \\

To estimate the model, we suggested a Bayesian Gibbs sampler (Section \ref{sec::gibbs}) and we also considered using a frequentist estimator (Section \ref{sec::EM}) that finds the maximum of the likelihood by an expectation-maximization algorithm (ML-EM). In simulations, both estimators had similarly good performances. However, Bayesian estimation, in our view, is the preferable approach for three reasons. First, the Bayesian data augmentation scheme (\ref{eq::fc_x}-\ref{eq::fc_t}) allows fitting flexible AFT models for arbitrary transition time distributions including models for non-constant and non-proportional hazards. ML-EM requires specific implementations for any combination of distributions, where we focused on the lognormal-lognormal model. Second, Bayesian estimation enables posterior inference for parameters and CIFs, while ML-EM has to rely on re-sampling techniques. The Bayesian credible intervals were shown to have good frequentist performance. Third, while the approaches had similar performance in the simulation study, ML-EM suffered from convergence problems in the application to the POBASCAM data. Across repeated randomly initialized runs, the algorithm converged at similar likelihood values that, however, yielded different estimates for the parameters of $f_t$ indicating that the likelihood function was almost flat and/or multi-modal. When the likelihood is almost flat, the convergence of the EM algorithm at maximum likelihood estimates generally hinges on setting the convergence limit to a small enough value while making sure that the numerical integration in the E-step is sufficiently precise. Hence, the procedure can become computationally very expensive. The Bayesian answer to these problems is to apply regularized parameter estimation by including weak prior information. This measure strongly improved estimation performance, in that a single random initialization of multiple MCMC chains converged to the posterior distribution. However, we emphasize that the suggested weakly informative priors are important for the performance of the Gibbs sampler and that the priors should not, in general, be chosen non-informative (i.e., flat). This is indicated by the prior sensitivity analysis (Figure \ref{fig::sens_prior_par}) demonstrating that the less informative prior choice ($\lambda_t=10$) inflated the posterior variance. A non-informative prior (large $\lambda$) would likely further increase this variance implying slower convergence and less robust estimates.  \\ 

We compared Bayesian estimation also to a three-state Markov model and a NPMLE, which are convenient choices for practitioners but can have bias. The NPMLE can be used to estimate the CIF of $x_i$ but cannot be used for $y_i$ (Sections \ref{sec::lit_review} and \ref{sec::simulation}). Furthermore, Markov models can be easily fit to a wide range of transition and censoring structures under a constant hazards assumption \citep{jackson_multi-state_2011}. However, in many applications hazards change over time. HPV infections can clear, for example, so that hazards decrease. This leads to CIFs approaching a ceiling value below one, where non-progressed cases subsumed in the tail are not expected to progress within relevant time frames (e.g., 20 or 30 years). Markov models lack the flexibility to fit such transitions (Figures \ref{fig::CIFcomparison} and \ref{fig::res_ppdplots_modcomp}). The estimated transition probabilities by our model, still, depend on a correct model specification and also have to be judged in the context of the maximum follow-up (14.3 years). Within these constraints, we predicted that about 24\% and 19\% of women progress to, respectively, CIN-2 and CIN-3+ within 20 years after their HPV diagnosis. Progression from CIN-2 to CIN-3+ is rapid: more than half of women progress within a year and about 75\% of women progress within 10 years. The remaining women have low probability to progress in their lifetime. Our model also allows  including covariates to uncover population heterogeneity. We found that women with HPV-16, -18, and -31 sub-types had a strongly decreased transition time, where the effect of HPV-16 was strongest. Also, younger women showed faster progression to CIN-2. Determining an optimal screening schedule for different screening sub-groups is an important topic for further research. For example, the posterior samples of transition times from our model may be used in simulations that compare different screening schedules. \\

The latent Continuous-Time Markov Chains model (CTMC) implemented in \texttt{R} package  \texttt{cthmm} by \cite{lange_fitting_2013} and the piece-wise constant hazard model in package \texttt{msm} \citep{jackson_multi-state_2011} can be viewed as alternative modeling approaches. The methods make approximations of non-constant hazards under proportional hazard assumptions. Their accuracy hinges on the appropriateness of this assumption and good choices of latent transition structures or knots, respectively, which can require expert knowledge for successful implementations. Users of the proposed Bayesian method do not need to make similar decisions. However, our method makes several assumptions whose appropriateness has to be judged in each application. First, the transition and censoring structure (\ref{eq::def_phi}-\ref{eq::def_delta}) has to be appropriate. Second, the AFT specification (\ref{eq::model_eq1}-\ref{eq::model_eq2}) assumes independent errors conditionally on the covariates. In our study, the most important cause of faster transition, HPV sub-type, was included and a sensitivity analysis was conducted. Third, the error distributions need to be correctly specified and the information criteria (e.g., WAIC) allow model selection. However, our simulation indicated that even misspecified semi-Markov transitions (e.g., Weibull for a true lognormal transition) only cause minor bias on the parameter estimates and CIFs. In addition, the NPMLE for $x_i$ can be used as an indication of goodness of fit. Fourth, we assumed non-informative interval censoring in the sense that the transition times $x_i$ and $t_i$ are independent of the screening times conditionally on covariates. This means that within strata defined by age groups and HPV sub-types the screening visits were not patient initiated due to symptomatic disease states. In POBASCAM, this assumption was plausible as CIN-2/3 lesions are asymptomatic and screening schedules were planned in advance. Fifth, we assumed that the disease state of a patient can be ascertained without error. In POBASCAM, women were screened by cytology combined with PCR-DNA test which has high accuracy or by cytology alone having moderate accuracy. However, the effect of ignoring the imperfect sensitivity on the estimated incidence functions is expected to be limited \citep{Witte2017}. Still, extending the Bayesian model to allow for patient initiated screening times and misclassified states are viable paths for further research. \\

\clearpage
\newpage
\bibliographystyle{plainnat} 
\bibliography{bibliography_bayes3S}

\end{document}


\title{Supplemental Material to: \\ A Bayesian accelerated failure time model for interval censored three-state screening outcomes}
	\author{Thomas Klausch (\href{t.klausch@amsterdamumc.nl}{t.klausch@amsterdamumc.nl})}
	\author[1] {\\Eddymurphy U Akwiwu}
	\author[1] {Mark A. van de Wiel}
	\author[1] {\mbox{Veerle M. H. Coup\'e}}
	\author[1] {\mbox{Johannes Berkhof}}
	\affil[1]{Amsterdam University Medical Centers, \mbox{Department of Epidemiology and Data Science}, Amsterdam, The Netherlands} 

\maketitle
\newpage
\tableofcontents
\newpage

\section{Supplement A: Proofs and Information Criteria} 

\subsection{Proof for Section 2 (Informative Interval Censoring)} \label{sec::infcens}
We first define informative interval censoring in the standard interval censored survival analysis setting. For this consider the random survival time variable $Y$ with density $f_Y(y)$ which is observed for $n$ units $i=1,...,n$, subject to interval censored observations at times $\vf_i = (0, v_{i,1},...,v_{i,m_i}, \infty )$ with first and last entries set to zero and infinity, respectively, and $m_i$ random. In particular, the censoring interval $\phi_i = (l_i,r_i]$ is defined such that
\begin{align*}
	\Pr( l_i = v_{i,j}, r_i = v_{i,j+1} | y_i, \vf_i ) = \mathds{1}_{\{ v_{i,j} < y_i \le v_{i,j+1} \}}.
\end{align*}

Non-informative interval censoring is defined as \citep[][p. 79]{kalbfleisch_statistical_2002}
\begin{align} \label{eq::noninf}
	\Pr( y_i \in (l_i, r_i] | l_i, r_i) = \Pr( y_i \in (l_i, r_i]).
\end{align}

This situation occurs when the screening times $\vf$ are generated independently of $y_i$ or are fixed in advance of data collection. Informative censoring occurs when (\ref{eq::noninf}) does not hold. Standard survival analysis methods for interval censored data usually assume non-informative censoring and yield biased estimates and wrong inference when (\ref{eq::noninf}) does not hold. \\

Now consider the data generating mechanism of the three-state model for screening outcomes defined in Section 3 of the main manuscript. We have that $y_i = x_i + t_i$ and 
\begin{align*}  
	\Pr( l_i = v_{i,j}, r_i = v_{i,j+1} | x_i, \vf_i ) = \mathds{1}_{\{ v_{i,j} < x_i \le v_{i,j+1} \}}
\end{align*}	
with censoring states 
\begin{align*}  
	\delta_i = \begin{cases}
		1 \ &\text{if} \ \ l_i < x_i < r_i = \infty  \\
		2 \ &\text{if} \ \ l_i < x_i \le r_i < y_i \\
		3 \ &\text{if} \ \ l_i < x_i < y_i \le r_i < \infty,
	\end{cases}
\end{align*}
where we assume \textit{non-informative interval censoring in the three state model}, i.e.,
\begin{align} \label{eq::noninf3s}
	q(\vf_i | x_i, t_i) = q(\vf_i).
\end{align}

In the main manuscript and Section \ref{sec::proof4.1} of the Supplemental Material we additionally show that the estimation methods discussed in the main manuscript only require (\ref{eq::noninf3s}) to hold conditionally on a vector of covariates $\z_i$, which are suppressed here for brevity. \\

A naive approach to using the three-state data for modeling the time distribution of $y_i$ first transforms the censoring times $\phi_i$ to standard interval censoring of $y_i$ by the rules
\begin{align*}
	\tilde{l}_i = \tilde{l}(l_i,r_i,\delta_i, y_i) &= \begin{cases}
		l_i\ &\text{if} \ \ \delta_i = 1 \\
		r_i\ &\text{if} \ \ \delta_i = 2 \\
		l_i\ &\text{if} \ \ \delta_i = 3 
	\end{cases} \\
   &= \begin{cases}
		l_i\ &\text{if} \ \ r_i = \infty \\
		r_i\ &\text{if} \ \ r_i < y_i \\
		l_i\ &\text{if} \ \ y_i \le r_i 
	\end{cases}
\end{align*}
and 
\begin{align*}
	\tilde{r}_i = \tilde{r}(r_i,\delta_i,y_i) &= \begin{cases}
		\infty\ &\text{if} \ \ \delta_i = 1 \\
		\infty\ &\text{if} \ \ \delta_i = 2 \\
		r_i\ &\text{if} \ \ \delta_i = 3 
	\end{cases} \\ 
    &= \begin{cases}
	\infty\ &\text{if} \ \ r_i = \infty \\
	\infty\ &\text{if} \ \ r_i < y_i \\
	r_i\ &\text{if} \ \ y_i \le r_i.
\end{cases}
\end{align*}

However, we then have that

\begin{align*}
	\Pr( y_i \in (\tilde{l}_i, \tilde{r}_i] | \tilde{l}_i, \tilde{r}_i) \neq \Pr( y_i \in ( \tilde{l}_i,  \tilde{r}_i])
\end{align*}
in violation of condition (\ref{eq::noninf}) due to the dependence of $\tilde{l}(l_i,r_i,\delta_i, y_i) $ and $\tilde{r}(r_i,\delta_i,y_i)$ on $y_i$. Thus, we expect standard survival analysis methods using the interval $(\tilde{l}(l_i,r_i,\delta_i, y_i), \tilde{r}(r_i,\delta_i,y_i)$] to yield biased estimates of parameters of models of $y_i$. \\

Consider now the problem to model $x_i$ instead of $y_i$ by standard survival analysis methods for interval censored data. Then we recognize that $x_i$ is bounded by the interval $\phi_i$ so that no pre-processing is needed and the interval can directly be used for standard survival analysis methods. Now we notice that $l_i, r_i$ are independent of $x_i$ by the non-informative censoring assumption of the three-state model in (\ref{eq::noninf3s}) and we have
\begin{align*} 
	\Pr( x_i \in (l_i, r_i] | l_i, r_i) = \Pr( x_i \in (l_i, r_i]).
\end{align*}

Thus, we expect standard survival analysis methods using the interval $(l_i, r_i]$ to yield unbiased estimates of parameters of standard survival analysis models of $x_i$.

\subsection{Proofs for Section 4.1 (Truncation Bounds)} \label{sec::proof4.1}  
We first proof the proposition
\begin{align*} 
	q( x_i |  \betaf, \sigmaf, t_i, \mathcal{D}_i,\vf_i; \tau, \lambdaf) =
	f_{x}(x_i | a(\phi_i,\delta_i,t_i)<x_i<b(\phi_i,\delta_i,t_i), \z_i, \betaf_x, \sigma_x )
\end{align*}
with truncation bounds
\begin{align} \label{eq::bounds_ab}
	(a(\phi_i,\delta_i,t_i), b(\phi_i,\delta_i,t_i))= 
	\begin{cases}
		(l_i , \infty) \quad &\text{if} \ \delta_i=1 \\
		(\max(r_i-t_i,l_i), r_i ] \quad &\text{if} \ \delta_i=2 \\
		(l_i,r_i -t_i ] \quad &\text{if} \ \delta_i=3. \\
	\end{cases}
\end{align}

We have
\begin{align}
	&q( x_i |  \betaf, \sigmaf, t_i, \mathcal{D}_i,\vf_i; \tauf, \lambdaf)  = \frac{ q(x_i, \betaf, \sigmaf, t_i, \z_i, \phi_i, \delta_i, \vf_i; \tauf, \lambdaf)
	}{ \int q(x_i, \betaf, \sigmaf, t_i, \z_i, \phi_i, \delta_i, \vf_i; \tauf, \lambdaf) d x_i} \nonumber \\
	\label{eq:app1} = &\frac{f_x(x_i | \z_i, \betaf_x, \sigma_x )
		f_t(t_i | \z_i, \betaf_t, \sigma_t )\
		\Pr( \delta_i | x_i, t_i, \phi_i)
		q(\phi_i | x_i, \vf_i )
		q(\vf_i)
	}{ \int f_x(x_i | \z_i, \betaf_x, \sigma_x )
		f_t(t_i | \z_i, \betaf_t, \sigma_t )
		\Pr( \delta_i | x_i, t_i, \phi_i)
		q(\phi_i | x_i, \vf_i )
		q(\vf_i) d x_i} \\
	\label{eq:app2} = &\frac{f_x(x_i | \z_i, \betaf_x, \sigma_x )
		\Pr( \delta_i | x_i, t_i, \phi_i)
		q(\phi_i | x_i, \vf_i )
	}{ \int f_x(x | \z_i, \betaf_x, \sigma_x )
		\Pr( \delta_i | x_i, t_i, \phi_i)
		q(\phi_i | x_i, \vf_i )
		d x_i} \\
	\label{eq:app3} = &\frac{f_x(x_i | \z_i, \betaf_x, \sigma_x )
		\mathds{1}_{\{  a(\phi_i,\delta_i,t_i) < x_i < b(\phi_i,\delta_i,t_i) \}}
	}{ \int f_x(x | \z_i, \betaf_x, \sigma_x )
		\mathds{1}_{\{a(\phi_i,\delta_i,t_i) < x_i < b(\phi_i,\delta_i,t_i) \}}
		d x_i} \\
	\label{eq:app4} = &f_{x}(x_i | a(\phi_i,\delta_i,t_i)<x_i<b(\phi_i,\delta_i,t_i), \z_i, \betaf_x, \sigma_x ).
\end{align}

Equation (\ref{eq:app1}) follows from the factorization implied by the data generating process that is visualized by the DAG in Figure 2 of the main manuscript. In equation (\ref{eq:app2}), terms independent of $x_i$ cancel, among which the density of the screening history $q(\vf_i)$. This result still holds if $q(\vf_i)$ instead depends on covariates $\z_i$, i.e., $q(\vf_i|\z_i)$, as noted in Section "Censoring mechanism" in the main manuscript. Equation (\ref{eq:app4}) follows from the definition of a truncated distribution. Therefore, it remains to be shown that $\Pr( \delta_i | x_i, t_i, \phi_i)q(\phi_i | x_i, \vf_i ) = \mathds{1}_{\{  a(\phi_i,\delta_i,t_i) < x_i < b(\phi_i,\delta_i,t_i) \}}$ holds in (\ref{eq:app3}), where the bounds $a$ and $b$ are defined in (\ref{eq::bounds_ab}). For this we note that $\Pr( \delta_i | x_i, t_i, \phi_i) q(\phi_i | x_i, \vf_i )$ (densities defined in Section "Censoring mechanism" in the main manuscript) is a product of the degenerate distributions
\begin{align} \label{eq:app5} 
	\Pr( \delta_i | x_i, t_i, \phi_i) q(\phi_i | x_i, \vf_i ) = \Pr( \delta_i | x_i, t_i, \phi_i) =
	\begin{cases}
		\mathds{1}_{\{ l_i < x_i <\infty \}} \quad &\text{if} \ \delta_i=1 \\
		\mathds{1}_{\{ l_i < x_i \le r_i <  x_i+t_i  \}} \quad &\text{if} \ \delta_i=2 \\
		\mathds{1}_{\{ l_i < x_i < x_i+t_i \le r_i < \infty \}} \quad &\text{if} \ \delta_i=3. \\
	\end{cases} 
\end{align}

The first equation holds, because $q(\phi_i | x_i, \vf_i ) $  evaluates to one whenever $\Pr( \delta_i | x_i, t_i, \phi_i)$ does. To find the bounds $(a(\phi_i,\delta_i,t_i),b(\phi_i,\delta_i,t_i))$, the inequalities in (\ref{eq:app5}) are solved for $x_i$, yielding (\ref{eq::bounds_ab}). \\

Next, we proof the proposition that
\begin{align*} 
	q( t_i |  \betaf, \sigmaf, x_i, \mathcal{D}_i,\vf_i; \tauf, \lambdaf) &= f_{t}(t_i | c(\phi_i,\delta_i,x_i)<t_i<d(\phi_i,\delta_i,x_i), \z_i, \betaf_t, \sigma_t )
\end{align*}
with truncation bounds 
\begin{align} \label{eq::bounds_cd}
	(c(\phi_i,\delta_i,x_i), d(\phi_i,\delta_i,x_i))= 
	\begin{cases}
		(0 , \infty) \quad &\text{if} \ \delta_i=1 \\
		(r_i-x_i, \infty ) \quad &\text{if} \ \delta_i=2 \\
		(0, r_i-x_i ] \quad &\text{if} \ \delta_i=3. \\
	\end{cases}
\end{align}

Here we have
\begin{align*} 
	q( t_i |  \betaf, \sigmaf, x_i, \mathcal{D}_i,\vf_i; \tauf, \lambdaf) =  &\frac{f_t(t_i | \z_i, \betaf_t, \sigma_t )
		\Pr( \delta_i | x_i, t_i, \phi_i) 
	}{ \int f_t(t | \z_i, \betaf_t, \sigma_t )
		\Pr( \delta_i | x_i, t_i, \phi_i) 
		d t_i} \\
	 = &\frac{f_t(t_i | \z_i, \betaf_t, \sigma_t )
		\mathds{1}_{\{  c(\phi_i,\delta_i,x_i) < t_i < d(\phi_i,\delta_i,x_i) \}}
	}{ \int f_t(t | \z_i, \betaf_t, \sigma_t )
		\mathds{1}_{\{c(\phi_i,\delta_i,x_i) < t_i < d(\phi_i,\delta_i,x_i) \}}
		d t_i},
\end{align*}
where we again solve the inequalities in (\ref{eq:app5}), but now for $t_i$, to find the bounds $c(\phi_i,\delta_i,x_i) $ and $d(\phi_i,\delta_i,x_i)$ given in (\ref{eq::bounds_cd}).

\subsection{Proofs for Section 4.2 (Predictive Densities)} 
We first proof the proposition that

\begin{align*} 
	q(x_i|\mathcal{D}_i,\betaf,\sigmaf) \propto g_x(x_i|\z_i,\betaf_{x},\sigma_x) [F_t(d(\phi_i,\delta_i,x_i) | \z_i, \betaf_{t},\sigma_t ) - F_t(c(\phi_i,\delta_i,x_i) | \z_i, \betaf_{t},\sigma_t )],
\end{align*}
where $g_x(x_i|\cdot) = f_x(x_i|\cdot)$ if $l_i<x_i\le r_i$ and $g_x(x_i|\cdot) = 0$ else. We have by the data generating mechanism implied by the DAG in Figure 2 of the main manuscript
\begin{align}
	q(x_i|\phi_i,\delta_i,\z_i,\betaf, \sigmaf) &= \frac{ \int q(x_i, t_i, \z_i, \phi_i, \delta_i,  \betaf, \sigmaf) dt_i
	}{ \iint q(x_i, t_i, \z_i, \phi_i, \delta_i, \betaf, \sigmaf) d x_i dt_i} \nonumber \\
    &= \frac{\int f_x(x_i | \z_i, \betaf_x, \sigma_x )
	f_t(t_i | \z_i, \betaf_t, \sigma_t )\
	\Pr( \delta_i | x_i, t_i, \phi_i)
	q(\phi_i | x_i, \vf_i )
	q(\vf_i) dt_i
}{  \iint  f_x(x_i | \z_i, \betaf_x, \sigma_x )
	f_t(t_i | \z_i, \betaf_t, \sigma_t )
	\Pr( \delta_i | x_i, t_i, \phi_i)
	q(\phi_i | x_i, \vf_i )
	q(\vf_i) d x_i dt_i} \nonumber \\
	& = \frac{ f_x(x_i | \z_i, \betaf_x, \sigma_x ) \int 
		f_t(t_i | \z_i, \betaf_t, \sigma_t )\
		\Pr( \delta_i | x_i, t_i, \phi_i) dt_i
	}{ \int  f_x(x_i | \z_i, \betaf_x, \sigma_x)  \int
		f_t(t_i | \z_i, \betaf_t, \sigma_t )
		\Pr( \delta_i | x_i, t_i, \phi_i)  d x_i dt_i}. \label{eq::condpredx}
\end{align}

Now, the numerator of (\ref{eq::condpredx}) depends on $\delta_i$; in particular
\begin{align*}
	&f_x(x_i | \z_i, \betaf_x, \sigma_x ) \int 
	f_t(t_i | \z_i, \betaf_t, \sigma_t )\
	\Pr( \delta_i | x_i, t_i, \phi_i) d t_i \\ 
	&= f_x(x_i|\z_i,\betaf_x, \sigma_x) \mathds{1}_{\{ l_i < x_i <r_i \}} \times 
	\begin{cases}
		1 \quad \ &\text{if} \ \delta_i = 1 \\
		\int \mathds{1}_{\{r_i-x_i \le t_i\}} f_t(t_i | \z_i, \betaf_t, \sigma_t ) dt_i \quad \ &\text{if} \ \delta_i = 2 \\
		\int \mathds{1}_{\{r_i-x_i>t_i\}} f_t(t_i | \z_i, \betaf_t, \sigma_t ) dt_i \quad \ &\text{if} \ \delta_i = 3 \\
	\end{cases} \\
	&= g_x(x_i|\z_i,\betaf_x, \sigma_x) \times \begin{cases}
		1 \quad \ &\text{if} \ \delta_i = 1 \\
		[ 1 - F_t(r_i-x_i| \z_i, \betaf_t, \sigma_t )] \quad \ &\text{if} \ \delta_i = 2 \\
		F_t(r_i-x_i| \z_i, \betaf_t, \sigma_t ) \quad \ &\text{if} \ \delta_i = 3. \\ 
	\end{cases}  \\
	&= g_x(x_i|\z_i,\betaf_x, \sigma_x) \times \begin{cases}
	F_t(\infty) - F_t(0) \quad \ &\text{if} \ \delta_i = 1 \\
	[ F_t(\infty) - F_t(r_i-x_i| \z_i, \betaf_t, \sigma_t )] \quad \ &\text{if} \ \delta_i = 2 \\
	F_t(r_i-x_i| \z_i, \betaf_t, \sigma_t ) - F_t(0) \quad \ &\text{if} \ \delta_i = 3. \\ 
	\end{cases} 
\end{align*}

We proceed similarly to derive the predictive density of $t_i$, i.e.
\begin{align*} 
	q(t_i|\mathcal{D}_i,\betaf,\sigmaf) \propto f_t(t_i | \z_i, \betaf_{t},\sigma_t ) [F_x(b(\phi_i,\delta_i,t_i) | \z_i, \betaf_{x},\sigma_x ) - F_x(a(\phi_i,\delta_i,t_i) | \z_i, \betaf_{x},\sigma_x )].
\end{align*}

We have
\begin{align}
	q(t_i|\phi_i,\delta_i,\z_i,\betaf, \sigmaf) = & \frac{ \int q(x_i, t_i, \z_i, \phi_i, \delta_i,  \betaf, \sigmaf) dx_i
	}{ \iint q(x_i, t_i, \z_i, \phi_i, \delta_i, \betaf, \sigmaf) d x_i dt_i} \nonumber \\
	= & \frac{ f_t(t_i | \z_i, \betaf_t, \sigma_t ) \int 
		f_x(x_i | \z_i, \betaf_x, \sigma_x )\
		\Pr( \delta_i | x_i, t_i, \phi_i) dx_i
	}{ \int  f_t(t_i | \z_i, \betaf_t, \sigma_t)  \int
		f_x(x_i | \z_i, \betaf_x, \sigma_x )
		\Pr( \delta_i | x_i, t_i, \phi_i)  d x_i dt_i}. \label{eq::condpredt}
\end{align}

The numerator in (\ref{eq::condpredt}) then simplifies as follows
\begin{align*}
	&f_t(t_i | \z_i, \betaf_t, \sigma_t ) \int 
	f_x(x_i | \z_i, \betaf_x, \sigma_x )\
	\Pr( \delta_i | x_i, t_i, \phi_i) dx_i \\ 
	&= f_t(t_i | \z_i, \betaf_t, \sigma_t ) \times 
	\begin{cases}
		\int \mathds{1}_{\{l_i < x_i \le r_i\}} f_x(x_i|\z_i,\betaf_x, \sigma_x) dx_i \quad \ &\text{if} \ \delta_i = 1 \\
		\int \mathds{1}_{\{l_i < x_i \le r_i\}} \mathds{1}_{\{ x_i > r_i-t_i\}} f_x(x_i|\z_i,\betaf_x, \sigma_x) dx_i \quad \ &\text{if} \ \delta_i = 2 \\
		\int \mathds{1}_{\{l_i < x_i \le r_i\}} \mathds{1}_{\{ x_i \le r_i-t_i\}} f_x(x_i|\z_i,\betaf_x, \sigma_x) dx_i \quad \ &\text{if} \ \delta_i = 3 \\
	\end{cases} \\
	&= f_t(t_i | \z_i, \betaf_t, \sigma_t ) \times
	\begin{cases}
		F_x(r_i|\z_i,\betaf_x, \sigma_x) - F_x(l_i|\z_i,\betaf_x, \sigma_x)  \quad \ &\text{if} \ \delta_i = 1 \\
		F_x(r_i|\z_i,\betaf_x, \sigma_x) - F_x(\max(l_i,r_i-t_i)|\z_i,\betaf_x, \sigma_x) \quad \ &\text{if} \ \delta_i = 2 \\
		F_x(r_i-t_i|\z_i,\betaf_x, \sigma_x) - F_x(l_i|\z_i,\betaf_x, \sigma_x)  \quad \ &\text{if} \ \delta_i = 3. \\
	\end{cases} 
\end{align*}

\subsection{Details on the information criteria and their implementation}
The following exposition of the information criteria WAIC and DIC follows closely the discussion in \cite{gelman_understanding_2014}. WAIC and DIC require evaluating the posterior expectation or the posterior variance of the observed data likelihood that is given by
\begin{align} \label{eq::obs_LL}
	L(\betaf, \sigmaf | \bm{\mathcal{D}}) \propto &\prod_{i: \delta_i=1} \big[ 1- F_x( l_i | \z_i, \betaf_x, \sigma_x) \big] \times \\  \nonumber
	&\prod_{i: \delta_i=2} \big[\int_{l_i}^{r_i} f_x(x_i | \z_i, \betaf_x, \sigma_x) [1 - F_t(r_i - x_i | \z_i, \betaf_t, \sigma_t)] dx_i  \big] \times \\ \nonumber
	&\prod_{i: \delta_i=3} \big[\int_{l_i}^{r_i} f_x(x_i | \z_i, \betaf_x, \sigma_x) F_t(r_i - x_i | \z_i, \betaf_t, \sigma_t) dx_i  \big]. \nonumber
\end{align}

The integration in (\ref{eq::obs_LL}) was implemented by adaptive quadrature using \texttt{R} function \texttt{integrate}. Then, 
\begin{align*}
	\text{DIC} = 2 [\log L( \hat{\betaf}, \hat{\sigmaf} | \bm{\mathcal{D}}) - 2 \E_{\betaf, \sigmaf| \bm{\mathcal{D}} }(\log L(\betaf, \sigmaf | \bm{\mathcal{D}}))],
\end{align*}
where $\hat{\betaf}$, $\hat{\sigmaf}$ are the posterior means and $\E_{\betaf, \sigmaf| \bm{\mathcal{D}} }(\log L(\betaf, \sigmaf | \bm{\mathcal{D}}))]$ denotes the posterior expectation of the likelihood, computed as
\begin{align*}
	\frac{1}{K} \sum_{k=1}^{K} \sum_{i=1}^{n} L(\betaf^{(k)}, \sigmaf^{(k)} | \mathcal{D}_i).
\end{align*}

Furthermore,
\begin{align*}
	\text{WAIC}_1 = -2 \sum_{i=1}^n \bigg[ - \log \E_{\betaf, \sigmaf| \bm{\mathcal{D}} }(L(\betaf, \sigmaf | \mathcal{D}_i)) + 2 \E_{\betaf, \sigmaf| \bm{\mathcal{D}} }(\log L(\betaf, \sigmaf | \mathcal{D}_i)) \bigg],
\end{align*}
where $\log \E_{\betaf, \sigmaf| \bm{\mathcal{D}} }(L(\betaf, \sigmaf | \mathcal{D}_i))$ is computed as
\begin{align*}
	\log \bigg[ \frac{1}{K} \sum_{k=1}^{K} L(\betaf^{(k)}, \sigmaf^{(k)} | \mathcal{D}_i)\bigg]
\end{align*}
and $\E_{\betaf, \sigmaf| \bm{\mathcal{D}} }(\log L(\betaf, \sigmaf | \mathcal{D}_i))$ is computed as
\begin{align*}
	\frac{1}{K} \sum_{k=1}^{K} \log L(\betaf^{(k)}, \sigmaf^{(k)} | \mathcal{D}_i).
\end{align*}

Finally, 
\begin{align*}
	\text{WAIC}_2 = -2 \sum_{i=1}^n \bigg[  \log \E_{\betaf, \sigmaf| \bm{\mathcal{D}} }(L(\betaf, \sigmaf | \mathcal{D}_i))  - \V_{\betaf, \sigmaf| \bm{\mathcal{D}} }( \log L(\betaf, \sigmaf |\mathcal{D}_i)) \bigg],
\end{align*}
where $\V_{\betaf, \sigmaf| \bm{\mathcal{D}} }$ is computed by the sample variance across the posterior draws
\begin{align*}
	\frac{1}{K-1} \sum_{k=1}^{K} (\log L(\betaf^{(k)}, \sigmaf^{(k)} | \mathcal{D}_i) - \bar{L})^2,
\end{align*}
where 
\begin{align*}
	\bar{L} = \frac{1}{K} \sum_{k=1}^{K} \log L(\betaf^{(k)}, \sigmaf^{(k)} | \mathcal{D}_i).
\end{align*}

\newpage

\section{Supplement B: EM algorithm for the lognormal-lognormal model}\label{appn}

We now present the EM algorithm for one specific choice of error distribution, the lognormal-lognormal model, which sets $f_{\epsilon}$ and $f_{\xi}$ as standard normal. In doing so, it is useful to introduce notation for the log of transition and censoring times as follows:

\begin{align*}
\tilde{x}_i := \log x_i \quad  \quad \tilde{t}_i := \log t_i \\
\tilde{l}_i := \log l_i \quad  \quad \tilde{r}_i := \log r_i
\end{align*}

\subsection{EM algorithm}
\textbf{M-step.}
The M-step of the lognormal-lognormal model updates:

\begin{align*}
&\betaf_x^{(k+1)} = (Z_x'Z_x)^{-1}Z_x \E_{x | \mathcal{D}, \betaf_x^{(k)}, \sigma_x^{(k)} } [ \tilde{\xf} ]\\
&\betaf_t^{(k+1)} = (Z_t'Z_t)^{-1}Z_t \E_{t | \mathcal{D}, \betaf_t^{(k)}, \sigma_t^{(k)} } [ \tilde{\tf} ]
\end{align*}

and

\begin{align*}
&\sigma_x^{(k+1)} = \bigg[ \frac{1}{n} \sum_{i=1}^{n} \big( \E_{x | \mathcal{D}, \betaf_x^{(k)}, \sigma_x^{(k)} } [\tilde{x}_i^2] - 2 \E_{x | \mathcal{D}, \betaf_x^{(k)}, \sigma_x^{(k)} } [\tilde{x}_i] \z_{x,i}' \betaf_x^{(k)} + (\z_{x,i}' \betaf_x^{(k)})^2 \big) \bigg]^{\frac{1}{2}} \\
&\sigma_t^{(k+1)} = \bigg[ \frac{1}{n} \sum_{i=1}^{n} \big( \E_{t | \mathcal{D}, \betaf_t^{(k)}, \sigma_t^{(k)} } [\tilde{t}_i^2] - 2 \E_{t | \mathcal{D}, \betaf_t^{(k)}, \sigma_t^{(k)} } [\tilde{t}_i] \z_{t,i}' \betaf_t^{(k)} + (\z_{t,i}' \betaf_t^{(k)})^2 \big) \bigg]^ {\frac{1}{2}}.
\end{align*}

The E-step only involves evaluation of the normal density and the normal distribution function and is, therefore, fast.  \\

\textbf{E-step.}
If $\delta_i = 1$, the E-step consists of computing the following quantities:

\begin{align*}
&\E_{x | \mathcal{D}, \betaf_x^{(k)}, \sigma_x^{(k)} } [\tilde{x}_i] = \mu_{x,i}^{(k)}  + \sigma_x^{(k)}  \Psi \big(\frac{\tilde{l}_i-\mu_{x,i}^{(k)} }{\sigma_x^{(k)} }\big)  \\
&\E_{x | \mathcal{D}, \betaf_x^{(k)}, \sigma_x^{(k)} } [\tilde{x}_i^2] = (\mu_{x,i}^{(k)})^2  + (\sigma_x^{(k)})^2 +  \sigma_x^{(k)} (\tilde{l}_i + \mu_{x,i}^{(k)}) \Psi \big(\frac{\tilde{l}_i-\mu_{x,i}^{(k)}}{\sigma_x^{(k)}} \big)  \\
&\E_{t | \mathcal{D}, \betaf_t^{(k)}, \sigma_t^{(k)}} [\tilde{t}_i] = \mu_{t,i}^{(k)} \\
&\E_{t | \mathcal{D}, \betaf_t^{(k)}, \sigma_t^{(k)} } [\tilde{t}_i^2] = (\mu_{t,i}^{(k)})^2  + (\sigma_t^{(k)})^2, 
\end{align*}
where $\mu_{x,i}^{(k)} = \z_{x,i}' \beta_x^{(k)}$, $\mu_{t,i}^{(k)} = \z_{t,i}' \beta_t^{(k)}$, $\Psi(x) = \phi(x|0,1)/\Phi(x|0,1)$, with  $\phi(x|0,1)$ and $\Phi(x|0,1)$ the standard normal density and standard  normal cumulative distributions functions (i.e., $N(0,1)$). It can, therefore, be seen that the E-step has partly closed form when $\delta_i=1$ and is generally fast. \\

If $\delta_i = 2$, we compute:
\begin{align}
&\E_{x | \mathcal{D}, \theta^{(t)} } [\tilde{x}_i] = h_{x,i}^{-1} \int_{\tilde{l}_i}^{\tilde{r}_i} \tilde{x}_i \phi(\tilde{x}_i|\mu_{x,i}^{(k)},\sigma_x^{(k)}) [1-\Phi(\log( r_i- e^{\tilde{x}_i})|\mu_{t,i}^{(k)},\sigma_t^{(k)})] d\tilde{x}_i   \label{eq::int_xd2} \\
&\E_{x | \mathcal{D}, \theta^{(t)} } [\tilde{x}_i^2] = h_{x,i}^{-1} \int_{\tilde{l}_i}^{\tilde{r}_i} \tilde{x}_i^2 \phi(\tilde{x}_i|\mu_{x,i}^{(k)},\sigma_x^{(k)}) [1-\Phi(\log( r_i-e^{\tilde{x}_i})|\mu_{t,i}^{(k)},\sigma_t^{(k)})] d\tilde{x}_i  \label{eq::int_x2d2} \\
&\E_{t | \mathcal{D}, \theta^{(t)} } [\tilde{t}_i] =  h_{t,i}^{-1} \int_{-\infty}^{\infty} \tilde{t}_i  \phi(\tilde{t}_i|\mu_{t,i}^{(k)},\sigma_t^{(k)}) [\Phi(\tilde{r}_i|\mu_{x,i}^{(k)},\sigma_x^{(k)})-\Phi(\log(\max(l_i,r_i-e^{\tilde{t}_i}))|\mu_{x,i}^{(k)},\sigma_x^{(k)})] d\tilde{t}_i  \label{eq::int_td2} \\ 
&\E_{t | \mathcal{D}, \theta^{(t)} } [\tilde{t}^2_i] = h_{t,i}^{-1} \int_{-\infty}^{\infty} \tilde{t}_i^2  \phi(\tilde{t}_i|\mu_{t,i}^{(k)},\sigma_t^{(k)}) [\Phi(\tilde{r}_i|\mu_{x,i}^{(k)},\sigma_x^{(k)})-\Phi(\log(\max(l_i,r_i-e^{\tilde{t}_i}))|\mu_{x,i}^{(k)},\sigma_x^{(k)})] d\tilde{t}_i. \label{eq::int_t2d2} 
\end{align}

Quantities $h_{x,i}$ and $h_{t,i}$ are the normalization constants defined as

\begin{align*}
	&h_{x,i} = \int_{\tilde{l}_i}^{\tilde{r}_i} \phi(\tilde{x}_i|\mu_{x,i}^{(k)},\sigma_x^{(k)}) [1-\Phi(\log( r_i- e^{\tilde{x}_i})|\mu_{t,i}^{(k)},\sigma_t^{(k)})] d\tilde{x}_i   \\
	&h_{t,i} = \int_{-\infty}^{\infty} \tilde{t}_i  \phi(\tilde{t}_i|\mu_{t,i}^{(k)},\sigma_t^{(k)}) [\Phi(\tilde{r}_i|\mu_{x,i}^{(k)},\sigma_x^{(k)})-\Phi(\log(\max(l_i,r_i-e^{\tilde{t}_i}))|\mu_{x,i}^{(k)},\sigma_x^{(k)})] d\tilde{t}_i.
\end{align*}

The integration of all quantities above is done numerically, as described below. \\

Furthermore, if $\delta_i = 3$, we compute
\begin{align}
	&\E_{x | \mathcal{D}, \theta^{(t)} } [\tilde{x}_i] = h_{x,i}^{-1} \int_{\tilde{l}_i}^{\tilde{r}_i} \tilde{x}_i \phi(\tilde{x}_i|\mu_{x,i}^{(k)},\sigma_x^{(k)}) \Phi(\log( r_i-e^{\tilde{x}_i})|\mu_t,\sigma_t^{(k)}) d\tilde{x}_i  \label{eq::int_xd3} \\
	&\E_{x | \mathcal{D}, \theta^{(t)} } [\tilde{x}_i^2] = h_{x,i}^{-1} \int_{\tilde{l}_i}^{\tilde{r}_i} \tilde{x}_i^2 \phi(\tilde{x}_i|\mu_{x,i}^{(k)},\sigma_x^{(k)}) \Phi(\log( r_i-e^{\tilde{x}_i})|\mu_t,\sigma_t^{(k)}) d\tilde{x}_i \label{eq::int_x2d3} \\
	&\E_{t | \mathcal{D}, \theta^{(t)} } [\tilde{t}_i] =  h_{t,i}^{-1} \int_{-\infty}^{\log(r_i-l_i)} \tilde{t}_i \phi(\tilde{t}_i|\mu_t,\sigma_t^{(k)}) [\Phi( \log(r_i-e^{\tilde{t}_i})|\mu_{x,i}^{(k)},\sigma_x^{(k)})-\Phi(\tilde{l}_i|\mu_{x,i}^{(k)},\sigma_x^{(k)})] d\tilde{t}_i  \label{eq::int_td3} \\ 
	&\E_{t | \mathcal{D}, \theta^{(t)} } [\tilde{t}^2_i] = h_{t,i}^{-1} \int_{-\infty}^{\log(r_i-l_i)} \tilde{t}_i^2  \phi(\tilde{t}_i|\mu_t,\sigma_t^{(k)}) [\Phi(\log(r_i-e^{\tilde{t}_i})|\mu_{x,i}^{(k)},\sigma_x^{(k)})-\Phi(\tilde{l}_i|\mu_{x,i}^{(k)},\sigma_x^{(k)})] d\tilde{t}_i. \label{eq::int_t2d3}
\end{align}

Quantities $h_x$ and $h_t$ are again the normalization constants defined as

\begin{align*}
	&h_{x,i} = \int_{\tilde{l}_i}^{\tilde{r}_i}\phi(\tilde{x}_i|\mu_{x,i}^{(k)},\sigma_x^{(k)}) \Phi(\log( r_i-e^{\tilde{x}_i})|\mu_t,\sigma_t^{(k)}) d\tilde{x}_i   \\
	&h_{t,i} =  \int_{-\infty}^{\log(r_i-l_i)} \phi(\tilde{t}_i|\mu_t,\sigma_t^{(k)}) [\Phi(\log(r_i-e^{\tilde{t}_i})|\mu_{x,i}^{(k)},\sigma_x^{(k)})-\Phi(\tilde{l}_i|\mu_{x,i}^{(k)},\sigma_x^{(k)})] d\tilde{t}_i .
\end{align*}

\subsection{Implementation and numerical integration}
We implemented the lognormal-lognormal EM algorithm in \texttt{R} as function \texttt{em.lognorm.r} which is available as part of the \texttt{Bayes3S} package. An important technical issue is the integration in the E-step when $\delta_i \in (2, 3)$. General-purpose integration software in \texttt{R}, such as function \texttt{integrate}, is often prohibitive for integrating functions which are zero or close to zero across ranges of the integration space, such as densities (see, e.g., \texttt{?integrate}). We therefore use importance sampling for integration. For this we first note that the expectations in  (\ref{eq::int_xd2} - \ref{eq::int_t2d3}) are all of the form

\begin{align*}
	h^{-1} \int l(x) u(x) dx,
\end{align*}

where $h$ is the normalization constant, $l(x) = x$ or $x^2$ and $u(x) = \phi(x) \Phi(x)$ or $\phi(x) (1-\Phi(x))$ the remainder terms (i.e., the products of a normal density and a cumulative normal distribution function), where we suppress the dependence on the parameters in the notation for brevity. The same holds for $t$. The expectations are approximated as follows:

\begin{enumerate}
	\item Generate samples $s_i$, $i=1,...,m$, from a truncated normal proposal distribution with density $\phi(s|a \le s \le b) = v(s) / [\Phi(b)-\Phi(a)] $ where $v(s) = \phi(s)$ if $a \le s \le b$ and 0 else,  $\phi(s)$ has equal mean and standard deviation as $\phi(x)$ in $u$, and the truncation bounds $(a,b)$ are
	\begin{itemize}
		\item $(\tilde{l}, \tilde{r})$ in (\ref{eq::int_xd2} - \ref{eq::int_x2d2}) and (\ref{eq::int_xd3} - \ref{eq::int_x2d3}), 
		\item $(-\infty,\infty)$ in (\ref{eq::int_td2} - \ref{eq::int_t2d2}), and
		\item $(-\infty,\log(r_i-l_i))$ in (\ref{eq::int_td3} - \ref{eq::int_t2d3}). \\
	\end{itemize}
\item Compute importance weights:
\begin{align*}
	w(s_i) = u(s_i)/\phi(s_i|a \le s \le b).
\end{align*} \\
\item Then approximate:
\begin{align*}
	\int l(x) u(x) dx \ \hat{=} \ m^{-1} \sum_{i=1}^{m} w(s_i) l(s_i)
\end{align*}
and 
\begin{align*}
	h^{(-1)} \ \hat{=} \ m^{-1} \sum_{i=1}^{m} w(s_i).
\end{align*}
\\

\end{enumerate}

\subsection{Accuracy of integration and convergence} \label{sec::mlem_remarks}

The E- and M-steps are iterated until the change in observed data likelihood is below a threshold value (the convergence limit). However, since we have to approximate the integrals in  (\ref{eq::int_xd2} - \ref{eq::int_t2d3}), the success of the EM algorithm in finding a maximum of the observed data likelihood at a given convergence limit strongly depends on the accuracy of the integration, which in turn depends on the number of importance samples $m$. The choice of $m$ strongly impacts computation time, since it has to be carried out for all $\delta_i \in (2,3)$. Therefore, $m$ and the convergence limit have to be chosen appropriately in any application. In general, choosing smaller limits and larger $m$ seems preferable, but can be computationally expensive.

\newpage

\section{Supplement C: Additional details on the simulation set-up and simulation results}

\subsection{Details on the simulation set-up}
\begin{figure}[h!]
	\centering
	\includegraphics[scale=.40, trim=40 75 55 120, clip, page = 1]{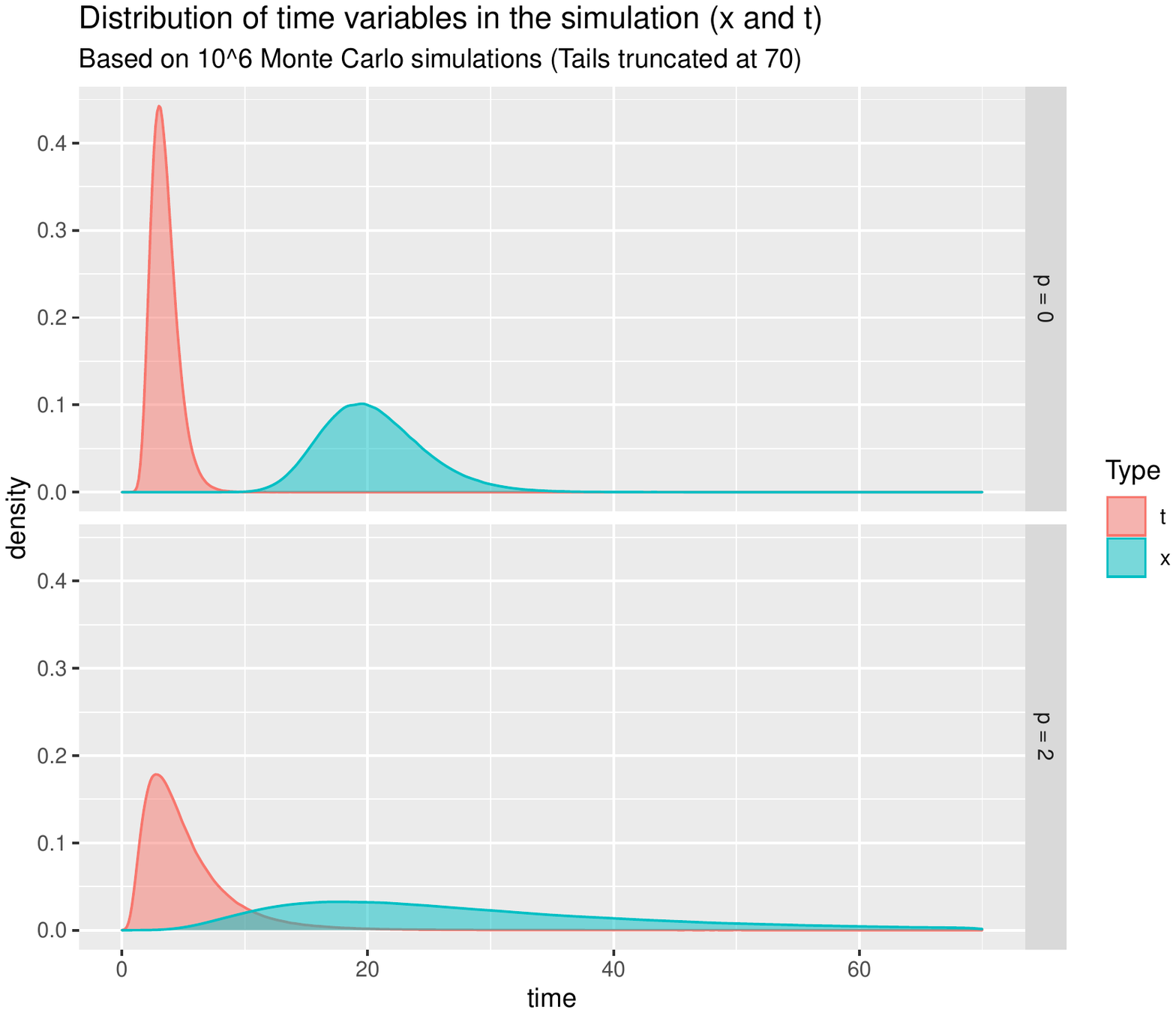}
	\caption{Transition time distributions of $x_i$ and $t_i$ based on a sample of size $n=10^6$ .}
	\label{fig::times}
\end{figure}

\begin{figure}[h!]
	\centering
	\includegraphics[scale=.40, trim=40 75 55 120, clip, page = 2]{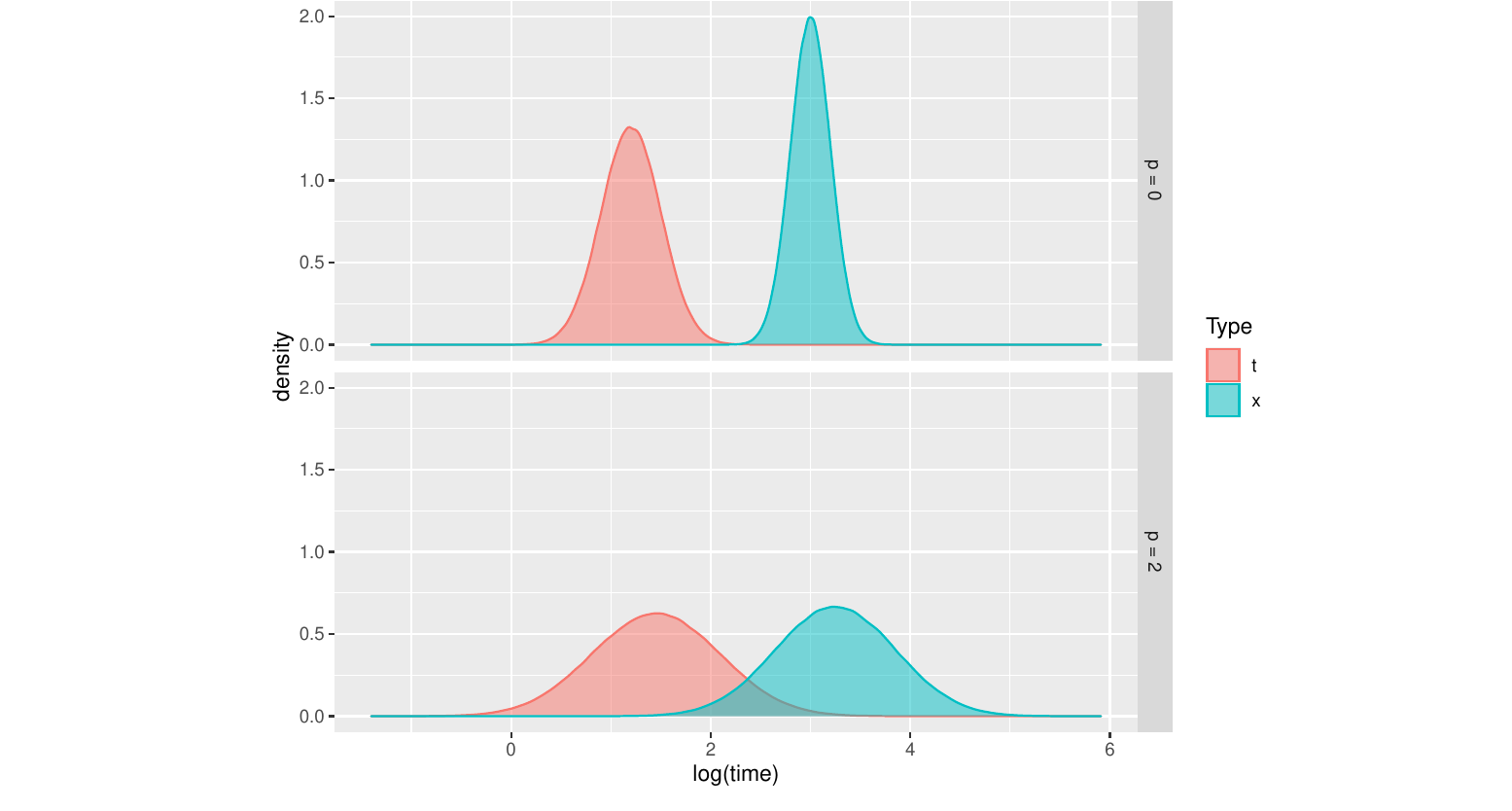}
	\caption{Log transition time distributions of $x_i$ and $t_i$ based on a sample of size $n=10^6$.}
	\label{fig::logtimes}
\end{figure}

In this Section, we provide further details on the transition time and censoring distributions of the simulation study as described in Section 5.1 of the main paper.  We generated data according to the following models: 

\begin{align*}
	\log(x_i) &= 3 + \beta_{1,x}  z_{i,1}+ \beta_{2,x} z_{i,2} + 0.2  \epsilon_{i}   \\
	\log(t_i) &= 1.2 + \beta_{1,t}  z_{i,1}+ \beta_{2,t} z_{i,2} + 0.3 \xi_{i}
\end{align*}

with $p=0$, we have $\beta_x = \beta_t = 0$, or with $p=2$, we set $\beta_x=\beta_t=0.5$. A plot of the resulting marginal transition time densities $f_x(x|\betaf_x, \sigma_x)$ and $f_t(t|\betaf_t, \sigma_t)$ is given in Figure \ref{fig::times} and the log-times are given in Figure \ref{fig::logtimes}.


Time $x_i$ had greater median (20.1)  than $t_i$ (3.3) under $p=0$; similarly, for $p=2$ the medians were: 25.8 and 4.3, respectively. However, the introduction of covariates translates into stronger heterogeneity resulting in distributions exhibiting stronger left skewed forms and longer right tails. Note that with $p=0$ the two distributions are lognormal, whereas with $p=2$ the distributions are a mixture. The times were marginally independent under $p=0$ but strongly correlated under $p=2$ (Pearson $r=.799$). This (marginal) correlation emerged, because the same covariates $\z_i$ were included in both models. Conditionally on the covariates, the times were independent. \\

The simulation was designed to mimic the practical setting where the disease has long onset time $x_i$ but progresses more rapidly once state 2 (e.g., CIN-2) has occurred. A similar finding is reported for the POBASCAM data in the main manuscript, Section 6. We did not intend to replicate the same distributions from the application, but we simulate a similar disease process in terms of a strong difference in progression speeds.  \\

We introduced the censoring times $(l_i, r_i)$ by a visiting process that first simulated $v_{i,1} \sim \text{uniform}(c_{min},c_{max})$, where $c_{min}<c_{max}$, and then recursively:
\begin{align}
	v_{i,t+1} \sim \text{uniform}(v_t+c_{min},v_t+c_{max})
\end{align}
until $v_{i,t+1} > v_{i,rc}$ where $v_{i,rc} \sim \text{exp}(\theta^{-1})$ was the transition time of right censoring of $x_i$ (end of follow-up). In particular, the bounds $(c_{min},c_{max})$ denote, respectively, the minimum and maximum time that elapsed between recurring visits. For example, subjects may be required not to be screened more frequently than every year for a particular disease, $c_{min} = 1$, but are expected or known to attend screening at least every seven years, $c_{max} = 7$. However, in the simulation study we needed to obtain approximately equal probabilities of censoring states $\delta_i \in \{1,2,3\}$ while the distributions of the underlying transition times of $x_i$ and $t_i$ differed across the number of covariates $p$. We tuned the visiting process parameters $c_{min},c_{max},\theta$ to achieve this goal. \\

\begin{table*} [h]
	\centering
	\caption{Parameters of the screening times process and resulting distributions of $\delta_i$ \\($^*$estimated by Monte Carlo with $10^6$ replications). Parameters $c_{min}$ and $c_{max}$ indicate the minimum and maximum time between screening moments and $\theta$ indicates the mean time to right censoring of $x_i$}
	\label{tab::simcens}
	\begin{tabular}{@{}l c c l c c @{}}
		& \multicolumn{2}{c}{Medium censoring} && \multicolumn{2}{c}{Strong censoring} \\
		\cline{2-3}
		\cline{5-6}
		& $p=0$ & $p=2$ &&  $p=0$ & $p=2$  \\ 
		\hline
		$c_{min}$ & 1 & 1 && 1 & 1 \\  
		$c_{max}$ & 8 & 8.7 && 7 & 6.5 \\  
		$\theta$ & 40 & 56 && 20 & 26.1 \\
		$\Pr(\delta_i = 1 | c_{min},c_{max},\theta )^*$ & 0.368 & 0.370 && 0.600 & 0.601 \\
		$\Pr(\delta_i = 2 | c_{min},c_{max},\theta)^*$ & 0.426 & 0.426 && 0.298 & 0.296 \\
		$\Pr(\delta_i = 3 | c_{min},c_{max},\theta)^*$ & 0.206 & 0.204 && 0.103 & 0.103 \\	
		\hline
	\end{tabular}
\end{table*}

The final choice of parameters is reproduced \ref{tab::simcens} from the main manuscript. As can be seen, the parameters differ between $p=0$ and $p=2$ for the medium and strong censoring conditions, respectively. This was done to achieve approximate equivalence in the probability mass of $\delta_i$ between the levels of $p$. The main difference introduced between the medium and strong censoring conditions is due to parameter $\theta$ which determines the mean time to right censoring. This time is substantially shorter for the strong censoring condition (i.e., 20/26.1 as opposed to 40/56) so that the probability of $\delta_i = 1$ is strongly elevated. \\ 

\begin{figure}[h!]
	\includegraphics[scale=.60, trim=40 75 55 120, clip, page = 3]{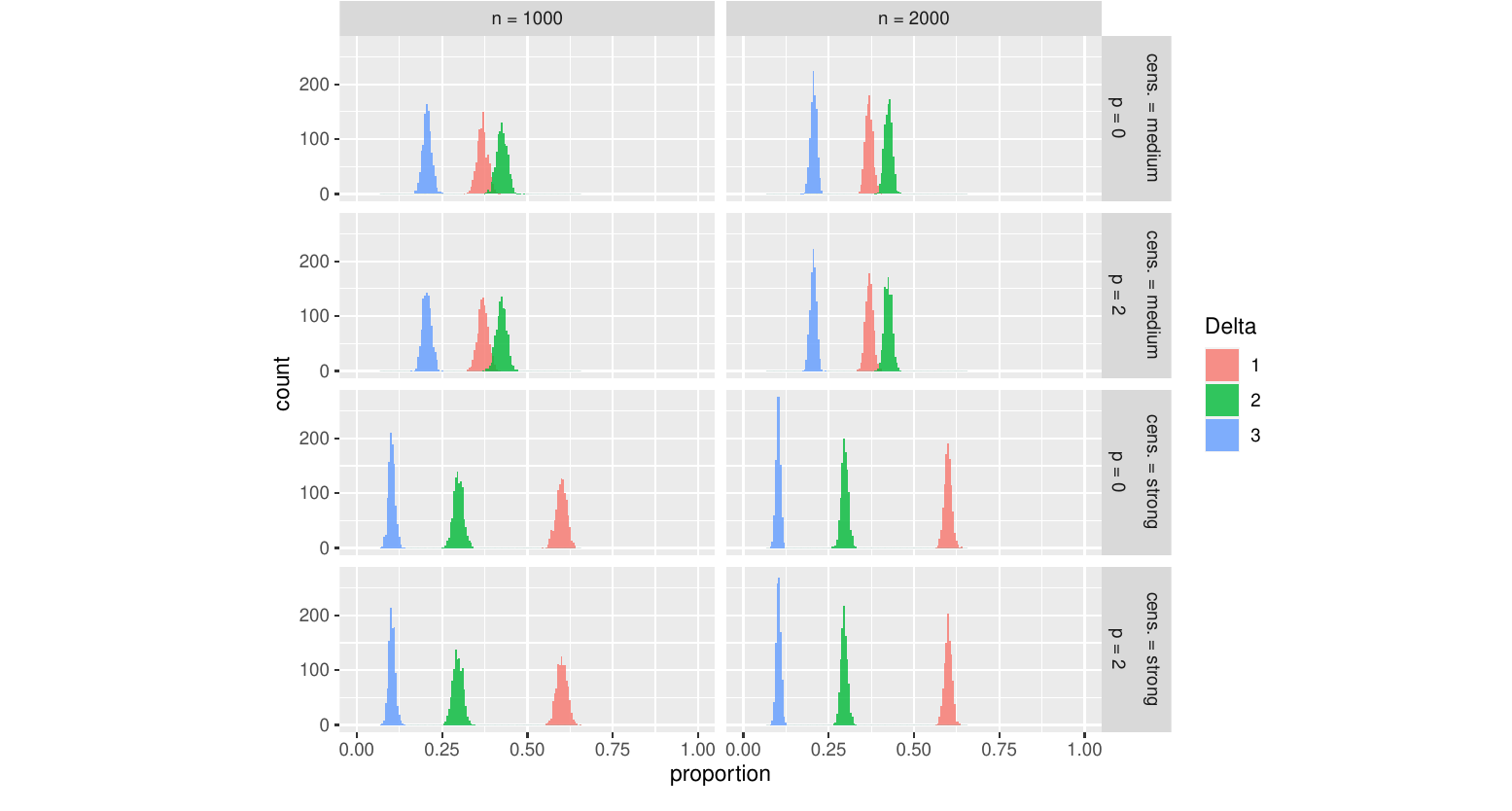}
	\caption{Histograms of the proportion of cases in categories of $\delta_i$ over 1,000 replicated data sets illustrating sampling variance of the sample proportions of the censoring events.}
	\label{fig::delta_variance}
\end{figure}

It is useful to note that the probabilities given in Table \ref{tab::simcens} are large-sample statistics that were determined in a sample of size $10^6$. In samples of size $n=1000$ or $n=2000$, as used in the simulation study, the observed proportion of cases with $\delta_i=j$ had a sampling variance. This is illustrated by Figure \ref{fig::delta_variance}. 

\subsection{Setting the jumping distribution proposal variance}
In the main manuscript, we study convergence of the MCMC algorithm, where the Gibbs algorithm employed a Metropolis step with an independent multivariate normal jumping distribution. It is well-known that the proposal variance of the jumping distribution impacts the acceptance rate of the Metropolis sampler and thus its speed of convergence. It was, therefore, essential to have approximately equal acceptance rates across experimental conditions to have comparable convergence rates. In general, both, too large and too small acceptance rates are problematic, where for multivariate posterior distributions the 'optimal' proposal variance invokes an acceptance rate of approximately 23\% \citep[][p. 296]{gelman_bayesian_2013}. \\

We tuned the proposal variance to every data set separately. For this we employed an automated heuristic search as part of the simulation code, described as follows:

\begin{enumerate}
	\item Set target interval $J = [J_1,J_2]$ for the acceptance rate $w$ to $J=[0.22,0.24]$.
	\item Run the Gibbs sampler with $m=3$ chains for an initial $D=5,000$ draws with proposal standard deviation $s=0.1$ in the Metropolis step.
	\item Determine $w$ as the proportion of accepted draws from the proposal distribution across $m$.
	\item If $w \in J$ go to step 5. Else set: $s_{new} : = s ( 1-(J_1 - w)/J_1 )$ if $s \ge J_1$ and  $s_{new} : = s ( 1+(J_2 - w)/J_2 )$ if $s < J_2$.
	\item Run the Gibbs sampler for another $D = 3,000$ draws and calculate $w$ as the proportion of accepted draws from the proposal distribution across $m$ and the last $D$ draws.
	\item If $w \in J$, go to step 7. Else set $s_{new}$ as described in step 4 and go to step 5.
	\item Repeat steps 5 and 6 for $D=6,000$ and $D=12,000$. Stop if $D=12,000$ and $w \in J$. 
\end{enumerate}

We emphasize that this is a heuristic search algorithm that mimics manual tuning of the proposal variance with the goal to achieve an optimal sampling rate. The reason this step was automated is that it is run as part of the simulation study, which required tuning the acceptance rate to $8 \times 100$ data sets under different experimental conditions.  Alternatives to manual tuning of the proposal variance include adaptive Metropolis Hastings algorithms which could be considered in future work. However, in initial tests with adaptive algorithms we found degenerate MCMC chains when used within the Gibbs sampler of the 3-state model. We therefore gave preference to manual tuning of the proposal variance which proved to reliably achieve convergence. \\

\begin{figure}[h]
	\centering
	\includegraphics[scale=.60, trim=50 280 55 295, clip]{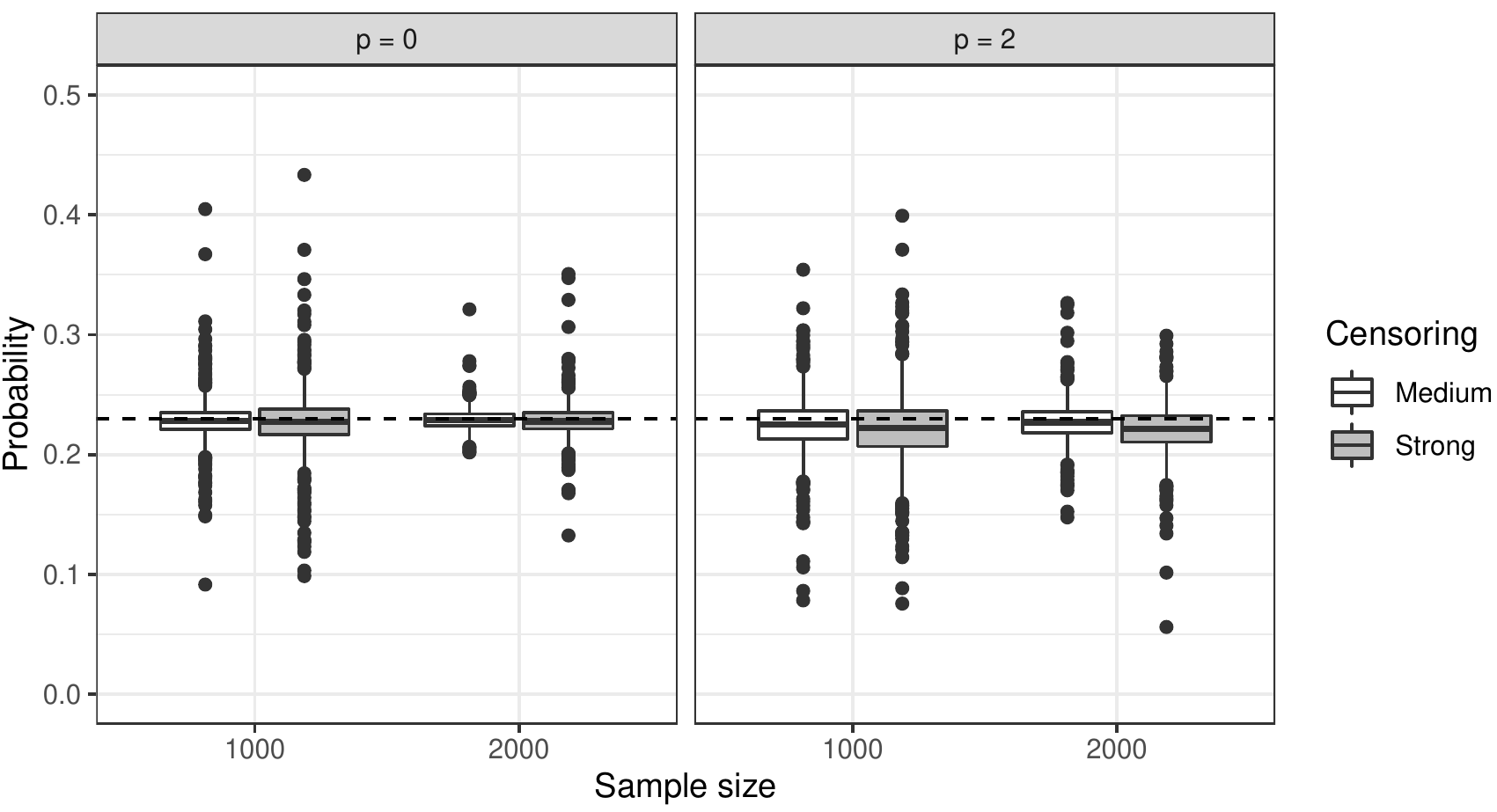}
	\caption{ Metropolis sampling acceptance probability over the simulation replications (proportion of accepted draws from the proposal distribution after discarding half of samples for burn-in). Dashed line indicates 'optimal' acceptance rate of 23\%.}
	\label{fig::acc_rate}
\end{figure}

After the search is completed the final proposal variance $s^2$ is used in a new run of the MCMC sampler (that is, discarding any draws from the search algorithm). The proposal variance thus is kept constant at the determined value across the whole run of the MCMC sampler. We compare the acceptance rates obtained from the various experimental conditions in the simulations in Figure \ref{fig::acc_rate}. As can be seen, the procedure was successful in obtaining acceptance rates close to 23\% in all conditions. 

\subsection{Implementation details for EM, MSM, and NPMLE} \label{sec::simimpl}

Besides the Bayesian MCMC algorithm, we modeled the simulated data also using the lognormal-lognormal maximum likelihood EM algorithm (Supplemental Material, Section B), the  \texttt{msm} package \citep{jackson_multi-state_2011} and a non-parametric maximum likelihood estimator \citep[NPMLE;][]{turnbull_empirical_1976}. \\

We implemented the EM algorithm for the lognormal-lognormal model in \texttt{R} as function \texttt{em.lognorm.r}; see Supplemental Material, Section B. For an initial, approximate run, we first set $m=100$ (the number of samples used in importance sampling) and considered the EM algorithm as converged when the change in observed data likelihood came below a value of $10^{-4}$. The approximate solution can typically be obtained fast but may be inaccurate. Therefore, we refined the approximate solution by more precise ML-EM runs with higher integration accuracy. For this, we set $m=5,000$ and started a new EM run at the values estimated by the initial approximate run. The convergence limit for this run was set to $10^{-5}$. \\

For \texttt{msm}, data were first transformed into the format required by the package. We then defined the following transition matrix (\texttt{qmatrix}):
\begin{lstlisting}[language=R]
     [,1] [,2]  [,3]
[1,]    0 0.01 0.000
[2,]    0 0.00 0.001
[3,]    0 0.00 0.000
\end{lstlisting}
which lets \texttt{msm} estimate a 3-state model with allowed transitions 1 to 2 and 2 to 3; all other transitions (zero entries) are impossible. The specific values in the transition matrix are starting values. We passed the \texttt{qmatrix} and data to \texttt{msm} and left all other values at their defaults. To estimate the marginal cumulative incidence functions (CIF), we proceeded as follows. First we estimated the conditional transition probabilities using function \texttt{pmatrix.msm} for all observed covariate data ($\z$). Let $P_i$ denote the estimated transition probability matrix for individual $i$ with $\z_i$ at time \texttt{t1} with entry $p_{ijk}$, then the CIF percentile of $x_i$ is obtained by the sum of matrix entries $p_{i\cdot}=p_{i12} + p_{i13}$, the percentile of $t_i$ is obtained by $p_{i13}$ and the percentile of $y_i$ is obtained by $p_{i23}$. The corresponding percentile of the marginal CIF is finally obtained by the average of all $p_{i\cdot}$ and $p_{i23}$ across $i=1,...,n$. \\

Furthermore, we used function \texttt{icfit} from package \texttt{interval} \citep{fay_exact_2010} to obtain a NPMLE of the marginal CIFs of $x$ and $y$, see Supplemental Material Section \ref{sec::infcens}. In particular, NPMLE was implemented for $x$ by treating $\delta=2$ and $3$ as the same event (since $\delta=3$ occurring in an interval suggests that state 2 occurred in the same interval). This was achieved by simply passing $l$ to argument \texttt{L} and $r$ to \texttt{R} using the same coding applied in the main manuscript. For $y$, we applied the recoding scheme described in  Section \ref{sec::infcens}, before passing $\tilde{l}$ and $\tilde{r}$ to the arguments \texttt{L} and \texttt{R}.

\subsection{MCMC convergence rates by parameters}
\begin{figure}[h]
		\centering
	\includegraphics[scale=.70, trim=210 140 210 170, clip, page = 2]{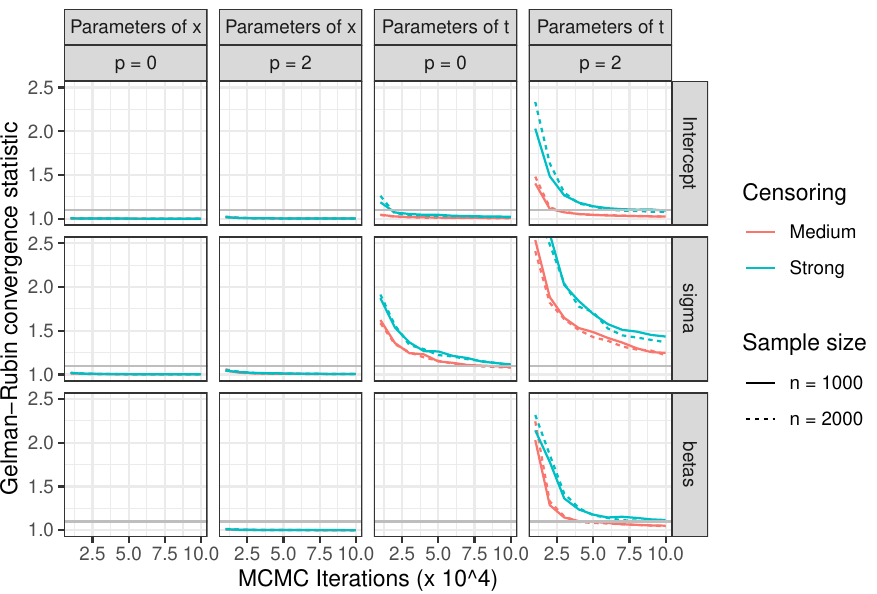}
	\caption{Gelman-Rubin convergence diagnostic $R$ across the first $10^5$ MCMC draws (medians of $R$ over 500 replications per condition). The speed of descent indicates the rate of convergence for the parameters of $f_x$ and $f_t$ ($\beta_1$ and $\beta_2$ reported jointly as 'betas').}
	\label{fig::sim_convergence}
\end{figure}

We evaluated the rate of convergence of the MCMC chains for each of the parameters of $f_x$ and $f_t$ in the 3-state model, i.e., the intercepts $\beta_0$, the slope coefficients $\beta_1$ and $\beta_2$, and the scale coefficients $\sigmaf$. We did so for the first $10^5$ MCMC draws using the Gelman Rubin convergence statistic (Figure \ref{fig::sim_convergence}). We found that the chains of the parameters of $f_x$ quickly converged whereas the chains of the parameters of $f_t$ needed substantially more draws, in particular in the $p=2$ condition. The MCMC chain of parameter $\sigma_t$ had the slowest rate of convergence. We also found that stronger censoring decreased the rate of convergence for all parameters. We note again that eventually all MCMC chains converged, as described above and shown in the main manuscript.

\subsection{Frequentist coverage probability of the Bayesian credible intervals}

\begin{figure}[h]
		\centering
	\includegraphics[scale=.60, trim=130 120 100 150, clip]{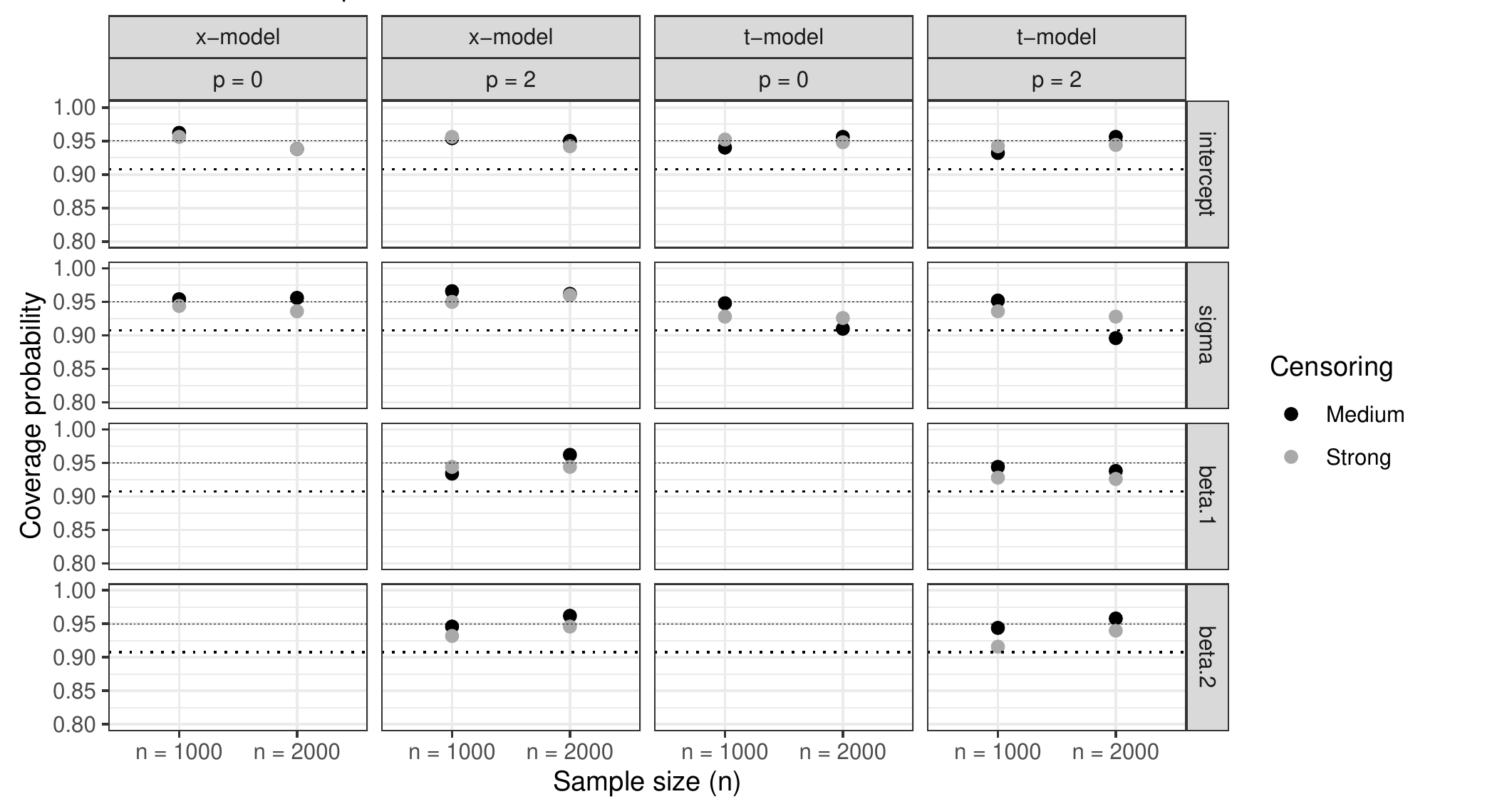}
	\caption{ True parameter coverage rate of 95\% credible intervals over 500 replicated data sets. Dashed line indicates nominal coverage rate (0.95). Dotted line at 0.908 indicates the minimum value at which a normal approximation 95\% confidence interval for the estimated coverage rate still includes the nominal (true) coverage rate of 0.95 after adjusting for multiple testing using Bonferroni correction. }
	\label{fig::coverage}
\end{figure}

We studied the frequentist coverage properties of Bayesian credible intervals as follows. In the simulation, we computed the $95\%$ credible intervals for each parameter by determining the 0.025 and 0.975 quantiles of the posterior distribution (i.e., equal tailed credible intervals). We then determined per parameter and simulation condition the proportion out of all 500 credible intervals that included the true parameter (Figure \ref{fig::coverage}). \\

Figure \ref{fig::coverage} displays a horizontal dotted line per sub-plot which indicates the minimum value at which a normal approximation 95\% confidence interval for the estimated coverage rate still includes the nominal (true) coverage rate of 0.95. This cut-off value was determined by finding the zero point of
\begin{align*}
	\bigg(s + \Phi^{-1} \bigg (1- \frac{0.05}{48 \times 2}\bigg ) \sqrt{\frac{s(1-s)}{K}} - 0.95 \bigg)^2 \quad ; \quad s \in (0,1)
\end{align*}
across $s$, with $s$ the lower bound of interest, $\Phi^{-1}$ the quantile function of cumulative normal distribution and $K=500$ the number of replications in the simulation. The ratio $\frac{0.05}{48 \times 2}$ carries out a Bonferroni correction for multiple testing (48 credible intervals are evaluated). We determined $s$ at approximately $0.908$ by \texttt{optimize} in \texttt{R}. It can be seen that in almost all of the conditions the estimated proportion was greater than this lower bound, indicating that the coverage probability of the credible intervals was at nominal level for most parameters and under all conditions. However, for $\sigma_t$ we found slight under-coverage in the condition with $n=2000$ and $p=2$. In addition, coverage of this parameter seemed to be slightly lower across all conditions in the larger sample size setting ($n=2000$).

\clearpage
\subsection{Simulation results EM compared to Bayes}
\begin{figure}[h]
		\centering
	\includegraphics[scale=.60, trim=130 120 100 150, clip, page =1]{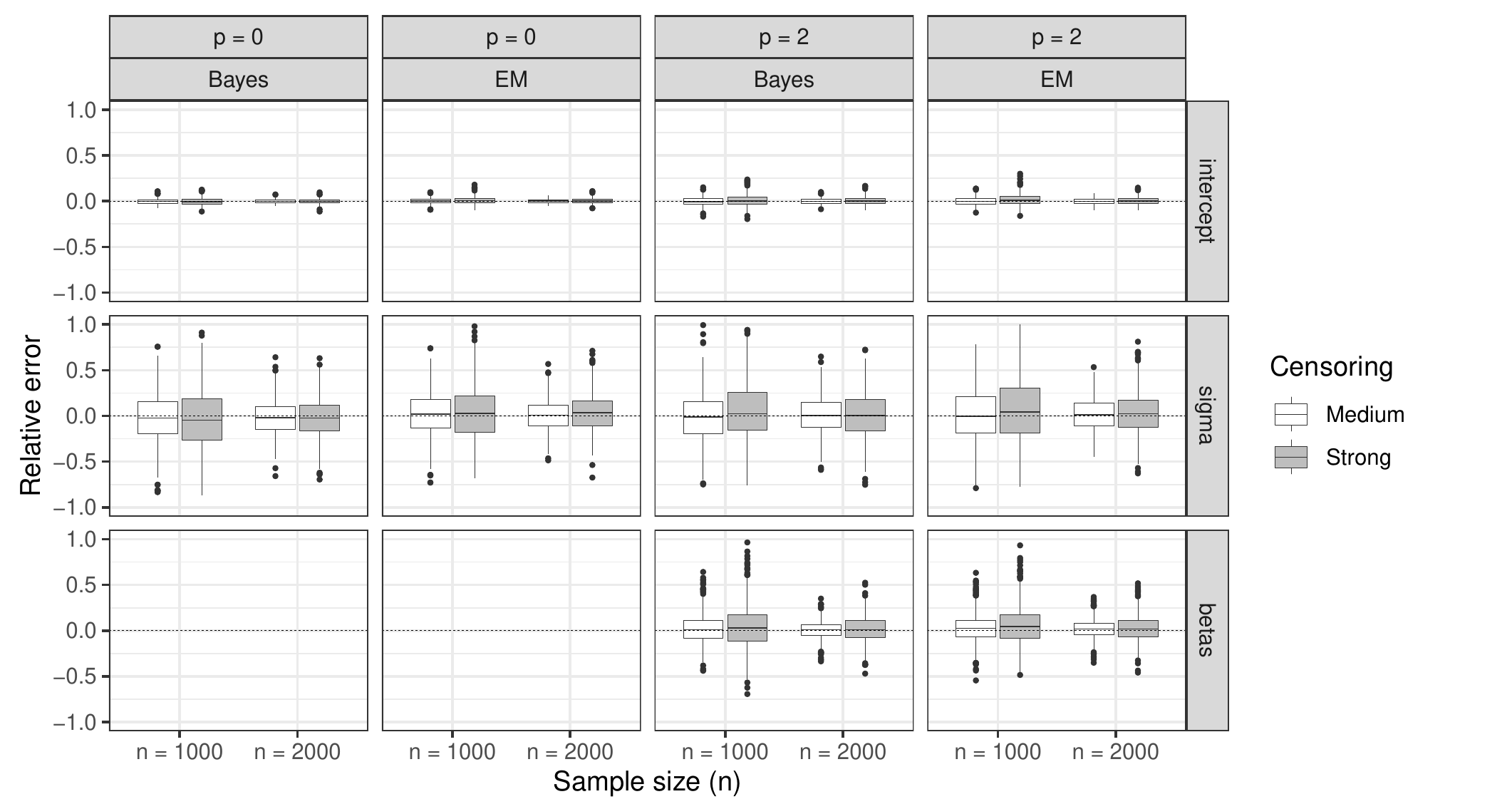}
	\caption{ Relative error of maximum likelihood EM compared to Bayesian estimation over 500 replications for the parameters of the transition time $x_i$.  }
	\label{fig::mcmcvsem_x}
\end{figure}

\begin{figure}[h]
		\centering
	\includegraphics[scale=.60, trim=130 120 100 150, clip, page =3]{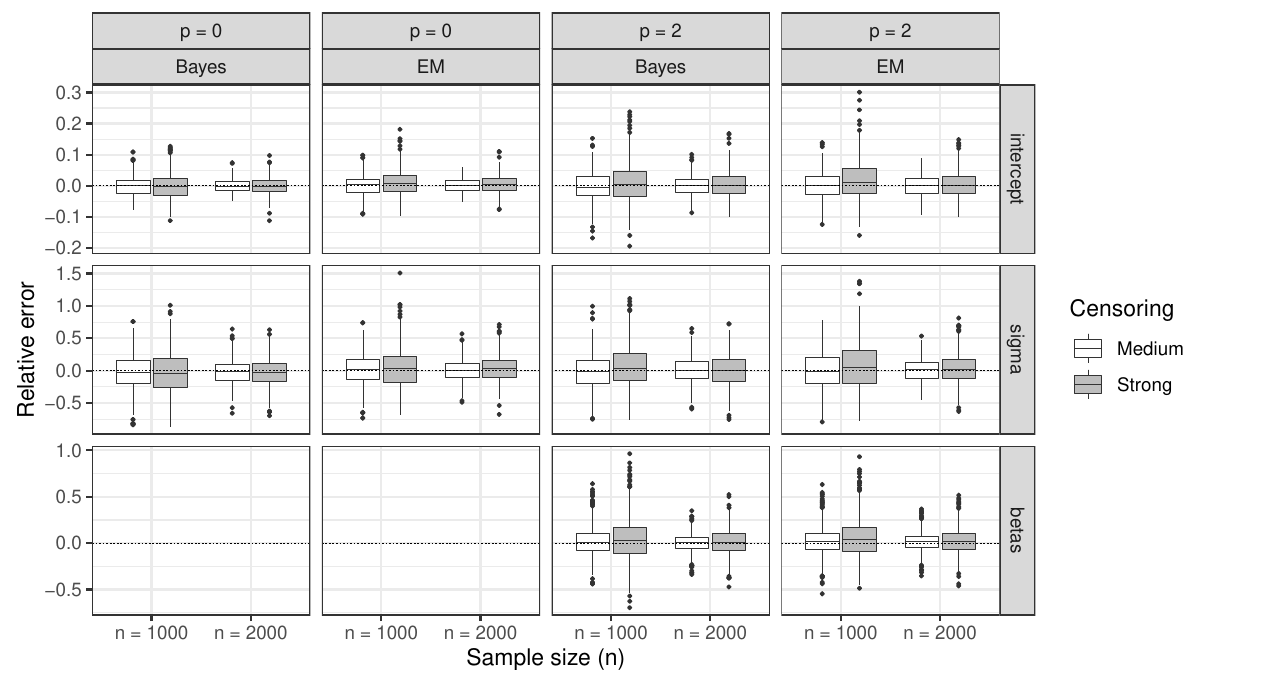}
	\caption{ Relative error of maximum likelihood EM compared to Bayesian estimation over 500 replications for the parameters of the transition time $t_i$.}
	\label{fig::mcmcvsem_t_out}
\end{figure}

\begin{figure}[h]
	\centering
	\includegraphics[scale=.60, trim=130 120 100 152, clip, page =4]{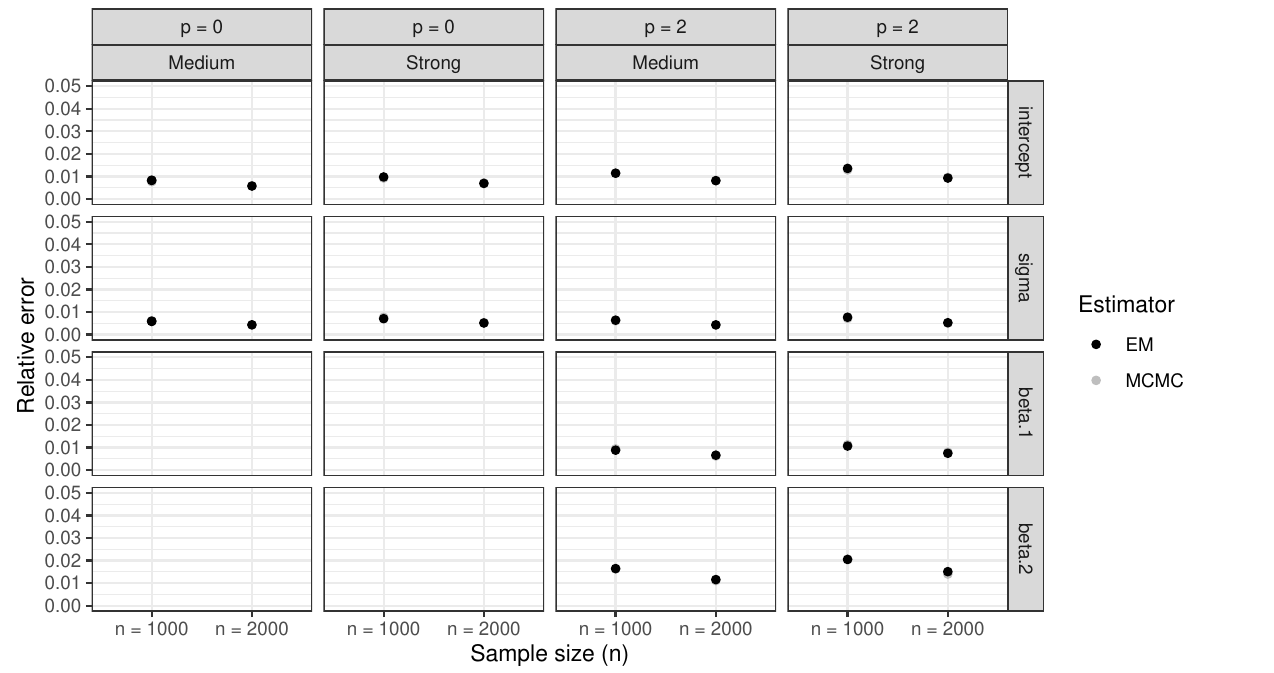}
	\caption{ Root mean squared error of maximum likelihood EM compared to Bayesian estimation over 500 replications for the parameters of the transition time $x_i$.}
	\label{fig::mcmcvsem_x_rmse}
\end{figure}

\begin{figure}[h]
	\centering
	\includegraphics[scale=.60, trim=130 120 100 152, clip, page =5]{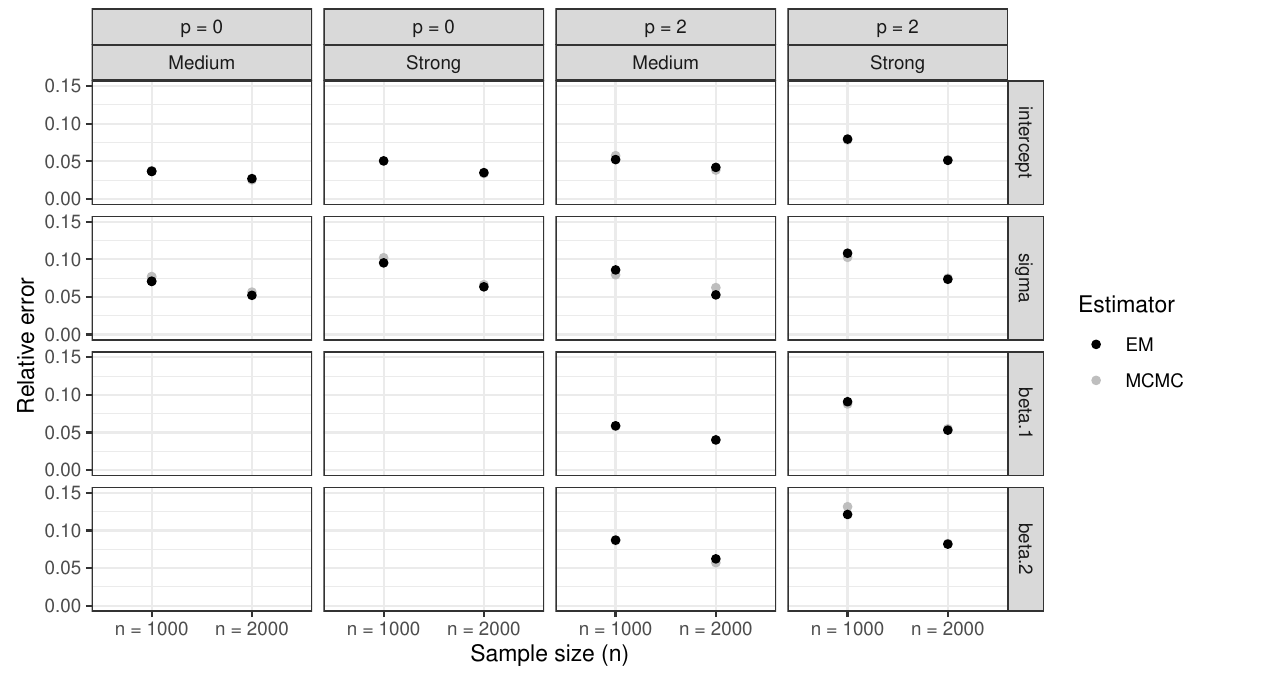}
	\caption{ Root mean squared error of maximum likelihood EM compared to Bayesian estimation over 500 replications for the parameters of the transition time $t_i$.}
	\label{fig::mcmcvsem_t_rmse}
\end{figure}

We compared the relative error of the parameter estimates of the Bayesian Gibbs sampler to those of ML-EM (Figures \ref{fig::mcmcvsem_x}-\ref{fig::mcmcvsem_t_out}). Relative error was defined as $[\hat{\kappa}-\kappa]/\kappa$  where $\kappa$ is the true value of the parameter and $\hat{\kappa}$ a posterior median estimate or a ML estimate. Figure \ref{fig::mcmcvsem_x} shows the relative errors for the parameters of the model of $x_i$. As can be seen, EM and Bayes had largely the same error sizes. In particular, the error for all parameters was zero in central tendency (approximately unbiased estimators) under all conditions. The same result was found for the parameters of the model of $t_i$ (Figure \ref{fig::mcmcvsem_t_out}). In addition, we estimated the root of the mean squared error (root MSE)  for all parameters, defined as $\sqrt{\frac{1}{K}\sum_{k=1}^K(\hat{\kappa}_k-\kappa)^2}$. As can be seen from Figures \ref{fig::mcmcvsem_x_rmse} and \ref{fig::mcmcvsem_t_rmse}, the estimators had very similar root MSE sizes.

\clearpage
\subsection{Comparison of cumulative incidence functions  models}

In this section, we give the cumulative incidence functions compared across estimation methods for all simulation scenarios. For the simulation setting, see the labeling on top of the figures.

\begin{figure}[h]
	\centering
		\includegraphics[scale=.60, trim=130 120 100 100, clip, page =1]{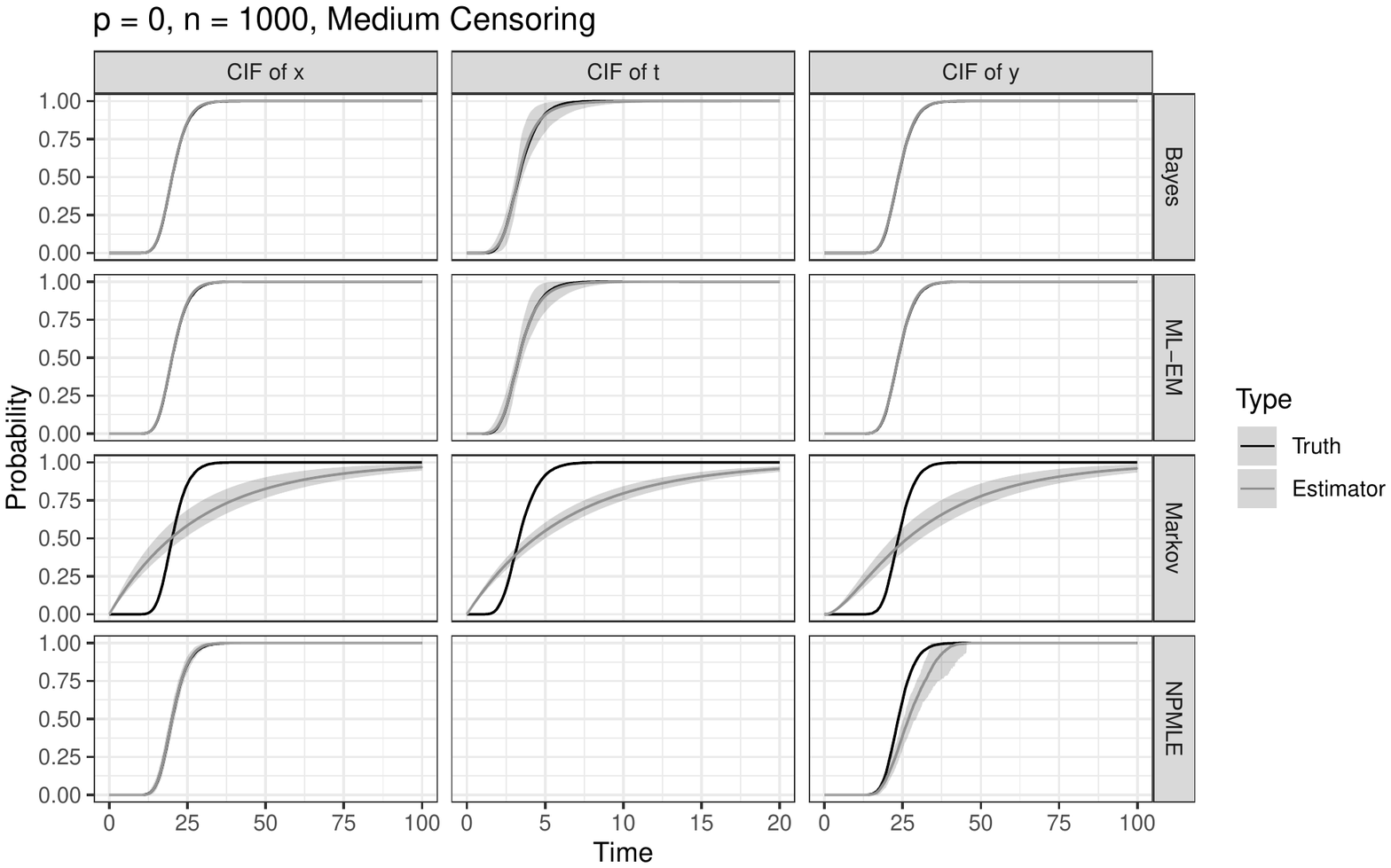}
	\caption{Marginal cumulative incidence functions (CIFs) compared for Bayesian estimation, EM, Markov model, and NPMLE. CIFs are point-wise averaged over 500 replications. For simulation setting, see information on top of the plot. }
\end{figure}
\begin{figure}
		\centering
	\includegraphics[scale=.60, trim=130 120 100 100, clip, page =2]{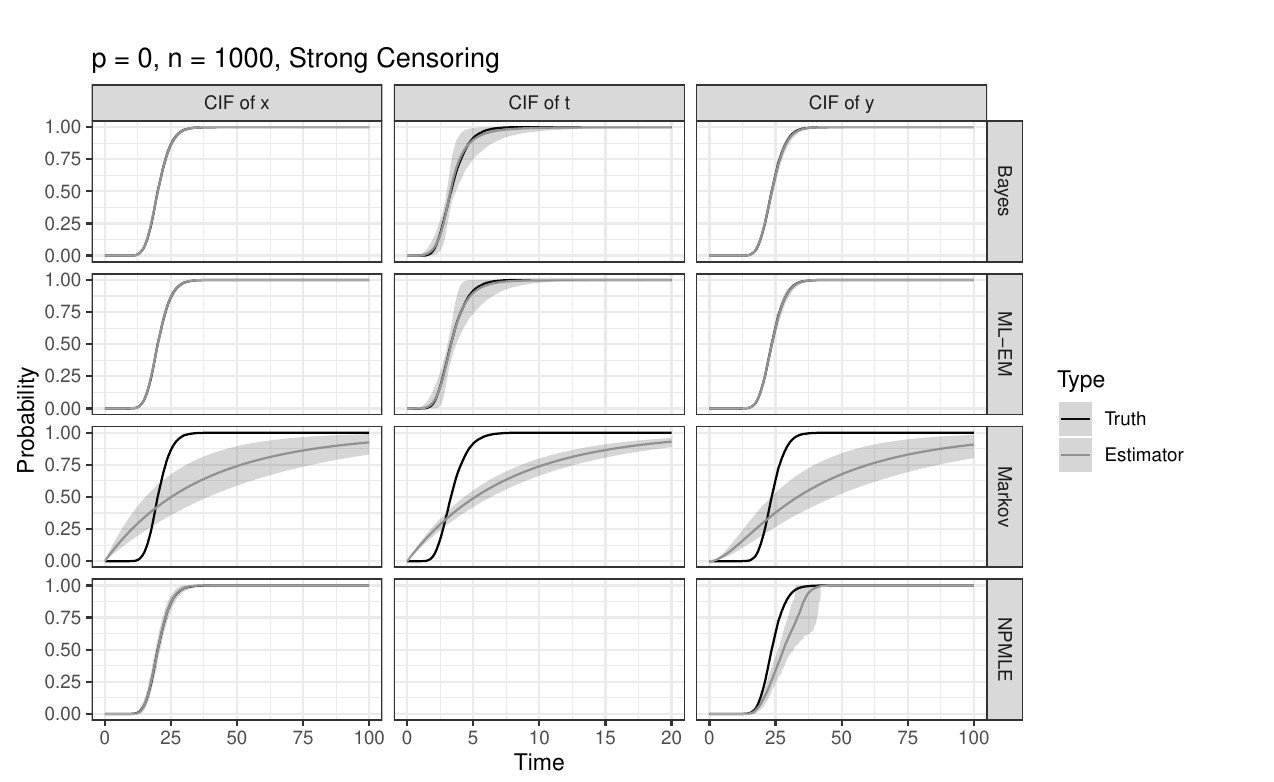}
	\caption{Marginal cumulative incidence functions (CIFs) compared for Bayesian estimation, EM, Markov model, and NPMLE. CIFs are point-wise averaged over 500 replications. For simulation setting, see information on top of the plot. }
\end{figure}
\begin{figure}[h]
		\centering
	\includegraphics[scale=.60, trim=130 120 100 100, clip, page =3]{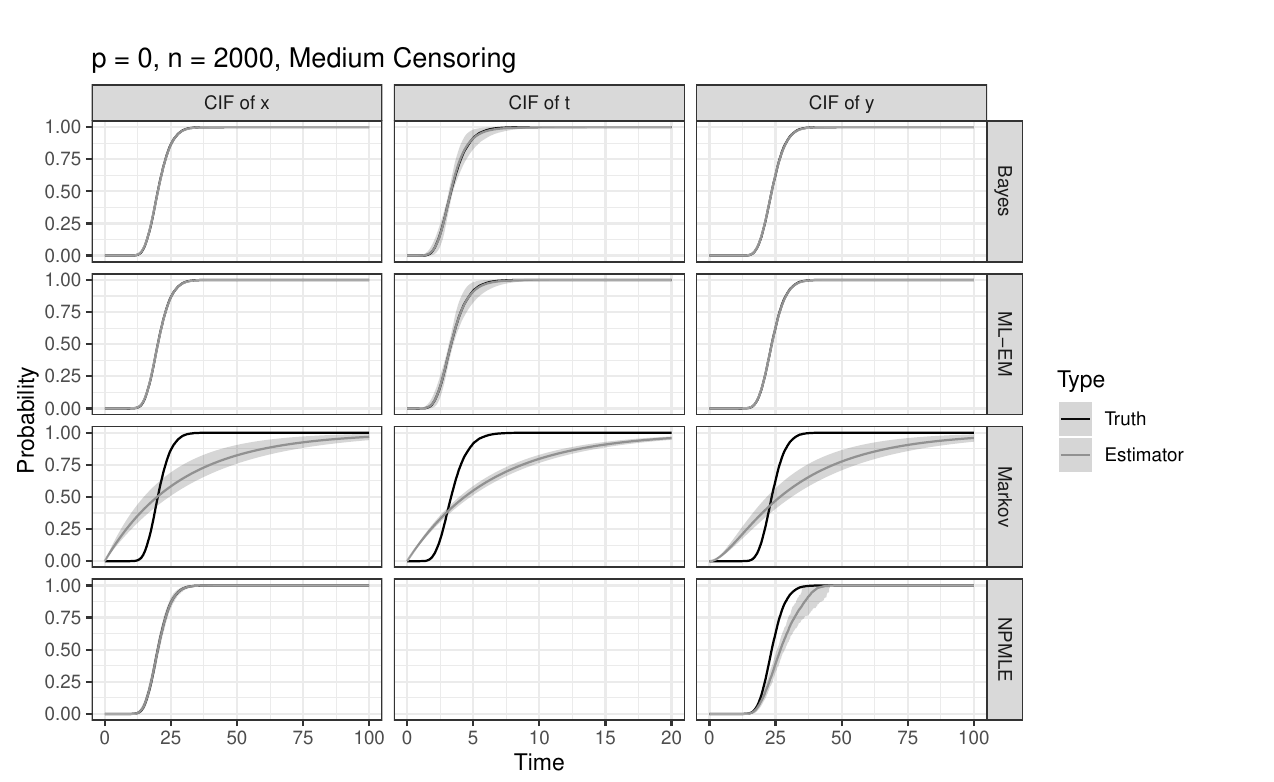}
	\caption{Marginal cumulative incidence functions (CIFs) compared for Bayesian estimation, EM, Markov model, and NPMLE. CIFs are point-wise averaged over 500 replications. For simulation setting, see information on top of the plot. }
\end{figure}
\begin{figure}[h]
		\centering
	\includegraphics[scale=.60, trim=130 120 100 100, clip, page =4]{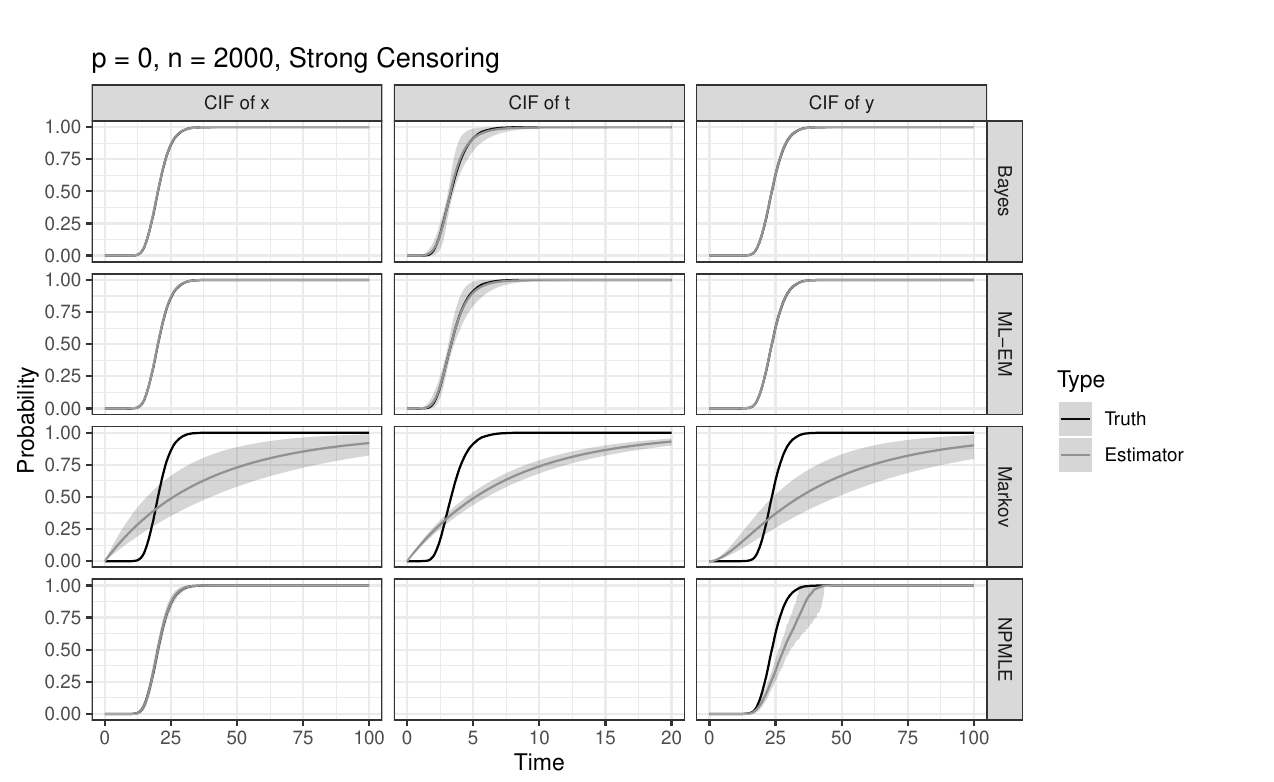}
	\caption{Marginal cumulative incidence functions (CIFs) compared for Bayesian estimation, EM, Markov model, and NPMLE. CIFs are point-wise averaged over 500 replications. For simulation setting, see information on top of the plot. }
\end{figure}
\begin{figure}[h]
		\centering
	\includegraphics[scale=.60, trim=130 120 100 100, clip, page =5]{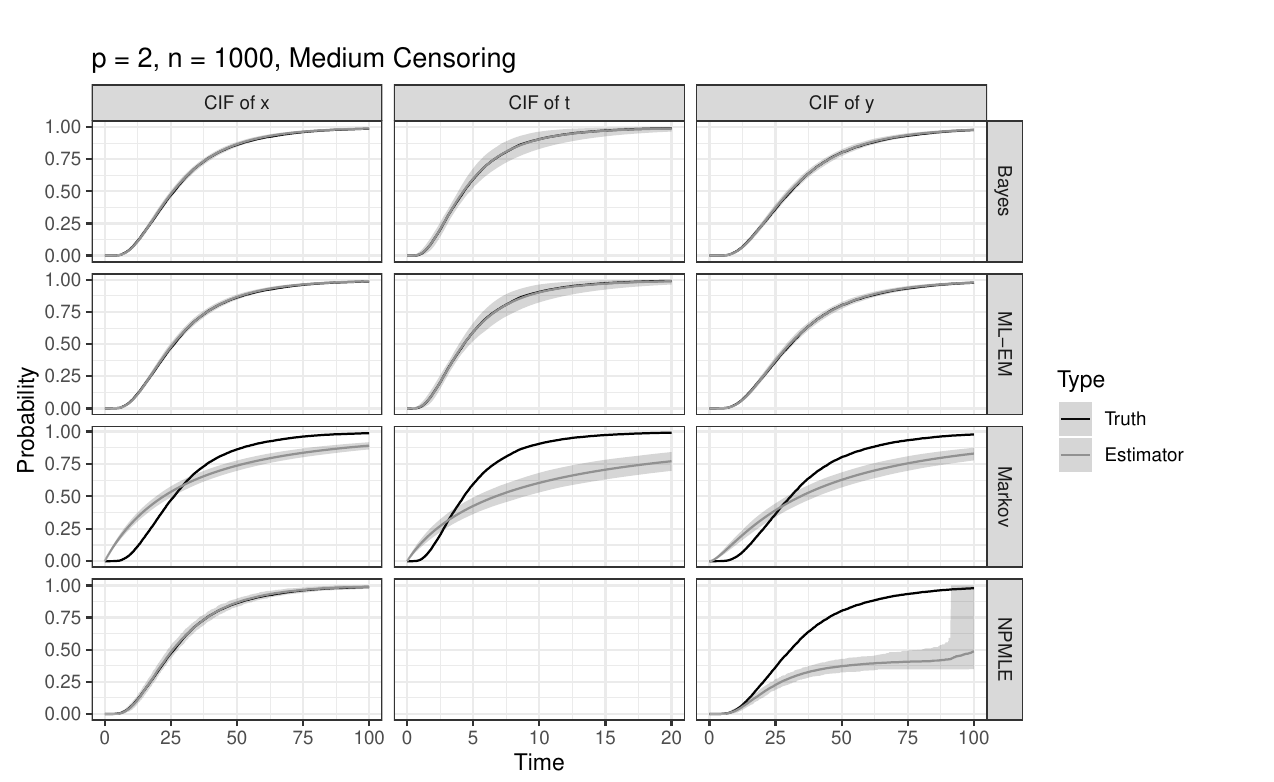}
	\caption{Marginal cumulative incidence functions (CIFs) compared for Bayesian estimation, EM, Markov model, and NPMLE. CIFs are point-wise averaged over 500 replications. For simulation setting, see information on top of the plot. }
\end{figure}
\begin{figure}[h]
		\centering
	\includegraphics[scale=.60, trim=130 120 100 100, clip, page =6]{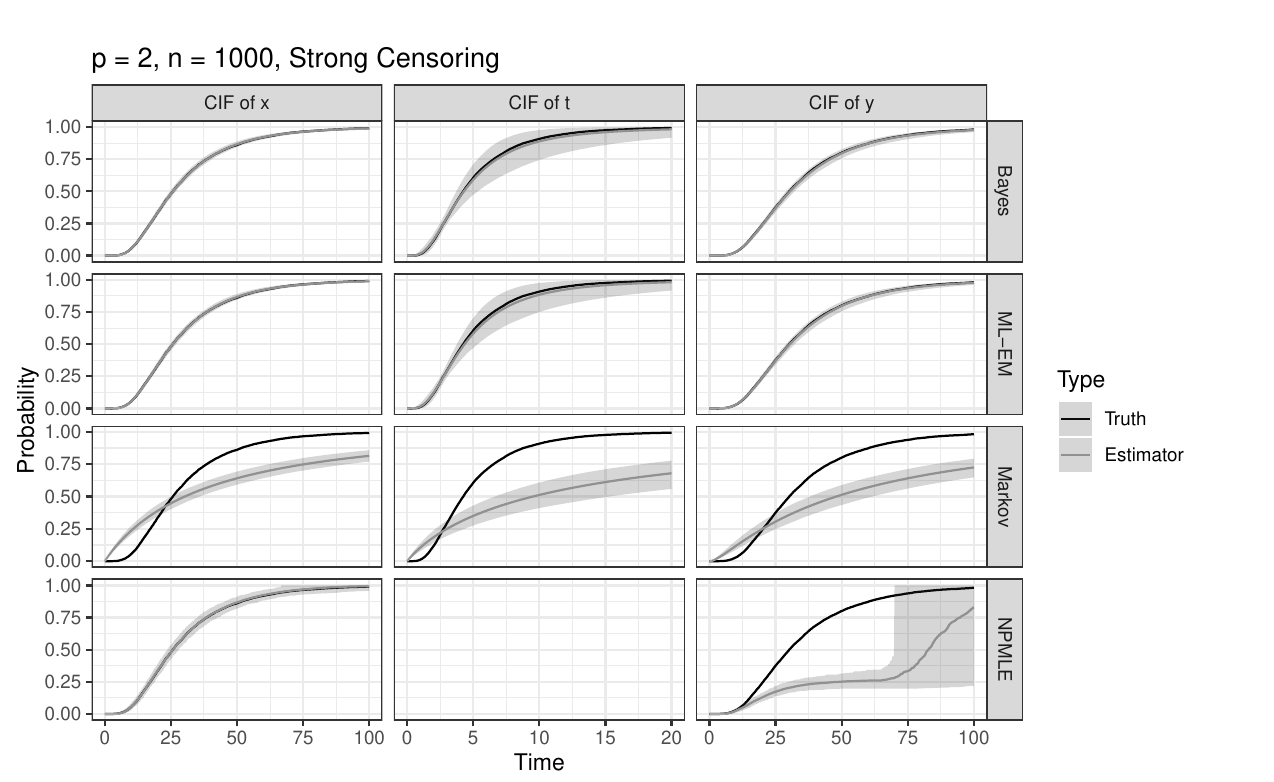}
	\caption{Marginal cumulative incidence functions (CIFs) compared for Bayesian estimation, EM, Markov model, and NPMLE. CIFs are point-wise averaged over 500 replications. For simulation setting, see information on top of the plot. }
\end{figure}
\begin{figure}[h]
		\centering
	\includegraphics[scale=.60, trim=130 120 100 100, clip, page =7]{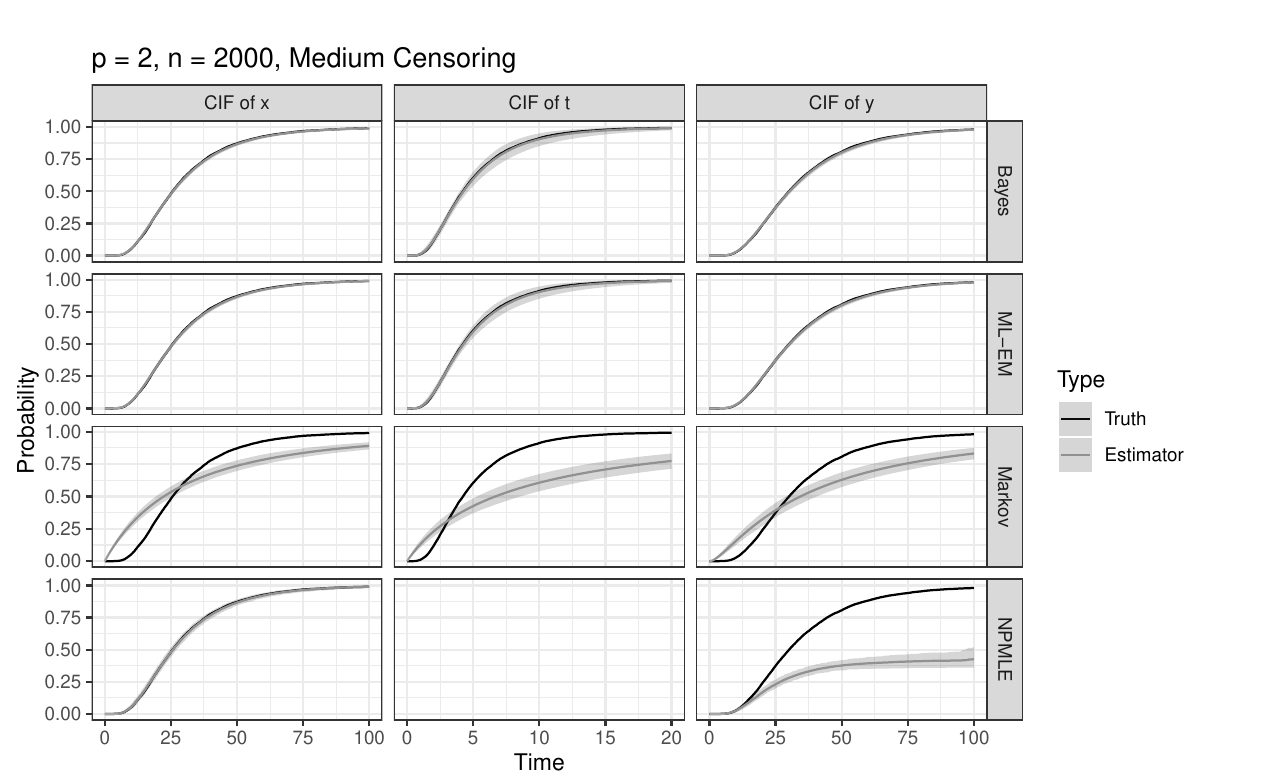}
	\caption{Marginal cumulative incidence functions (CIFs) compared for Bayesian estimation, EM, Markov model, and NPMLE. CIFs are point-wise averaged over 500 replications. For simulation setting, see information on top of the plot. }
\end{figure}
\begin{figure}[h]
		\centering
	\includegraphics[scale=.60, trim=130 120 100 100, clip, page =8]{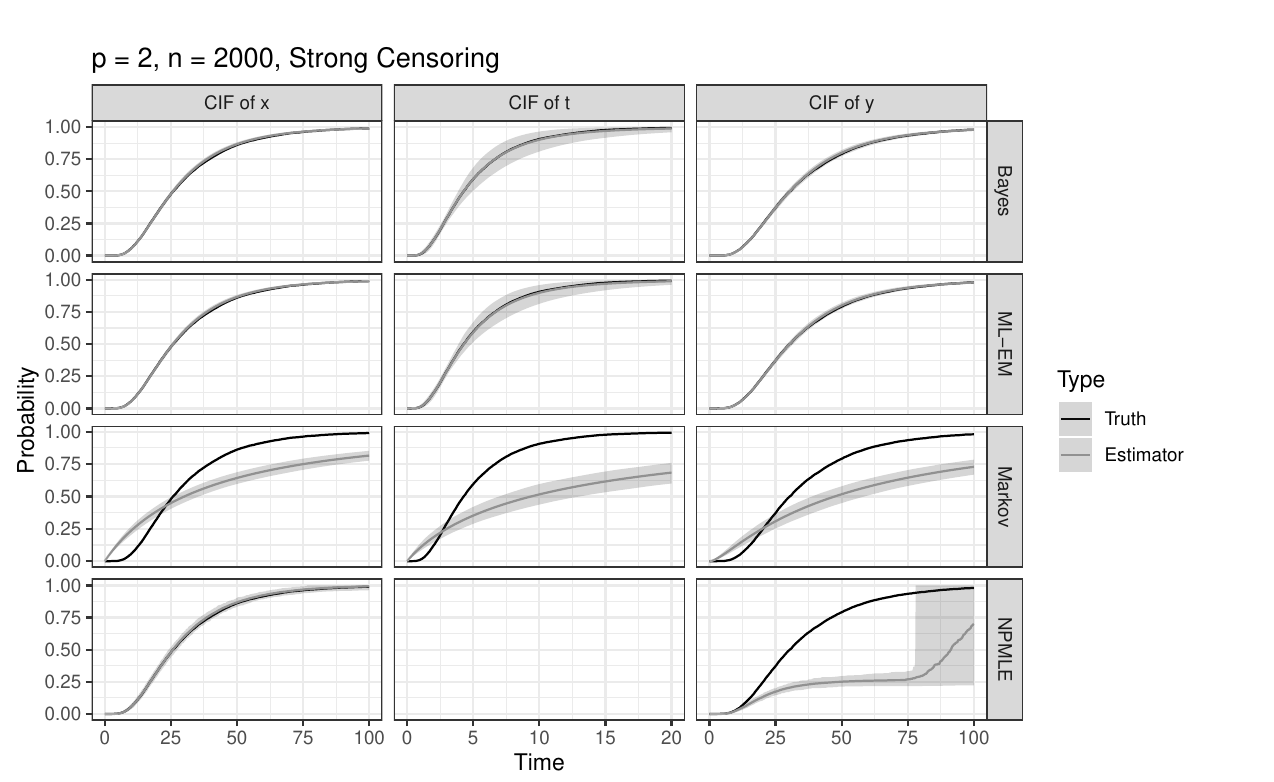}
	\caption{Marginal cumulative incidence functions (CIFs) compared for Bayesian estimation, EM, Markov model, and NPMLE. CIFs are point-wise averaged over 500 replications. For simulation setting, see information on top of the plot. }
\end{figure}

\clearpage
\newpage

\subsection{Estimation performance under misspecification}

In this section, we study the performance of the information criteria WAIC-1, WAIC-2, and DIC in selecting the correct distributions from a number of candidate models, where only one model is correctly specified. Subsequently, we study the bias in the model parameters and CIFs arising from distributional misspecification. For this we also used the simulation setting described in the main manuscript with lognormally distributed $x_i$ and lognormally distributed $t_i$ (in the following called the 'lognormal-lognormal' model). We specified and estimated the following models:
\begin{enumerate}
	\item lognormal - lognormal
	\item lognormal - Weibull
	\item lognormal - exponential
	\item Weibull - lognormal
	\item Weibull - Weibull
\end{enumerate}
We fit these models to the true lognormal-lognormal data from the simulation study. Models 2 to 5 misspecified the true model either with respect to $t_i$ (models 2 and 3),  $x_i$ (model 4), or both transition times (model 5). Subsequently, we ran the model until convergence and selected the model yielding smallest WAIC-1, WAIC-2, or DIC . We replicated this experiment 100 times on randomly simulated data sets for the setting with $p = 2$ covariates (Table \ref{tab::ICcomp}). Our results indicated that the true distribution of $x_i$ was very reliably identified, regardless of whether $t_i$ was correctly (model 4) or incorrectly specified (model 5). The results applied to all information criteria. The true distribution of $t_i$ was correctly selected in 66\% to 69\% of replicated runs correctly (WAIC-1), unless the sample was small and censoring was strong (50\%). WAIC-2 and DIC performed similarly. For $t_i$ we considered two alternative misspecified models, Weibull (model 2) and exponential (model 3). Among the two, the information criteria reliably selected the more complex misspecified model in 31\% to 34\% of all runs (Weibull / model 2) over the simpler misspecified model in none of the runs (exponential / model 3). 

\begin{table*} [h]
	\centering
	\caption{Performance of the information criteria WAIC-1, WAIC-2, and DIC in identifying the correct model among various misspecified models (1. lognormal - lognormal, 2. lognormal - Weibull, 3. lognormal - exponential, 4. Weibull - lognormal, 5. Weibull - Weibull). Setting with $p=2$ covariates and either $n = 1000$ or 2000 and medium or strong censoring. Numbers give the proportion out of all runs that a model is selected (in percent \% of 100 replicated runs). Model 1 is the correctly specified model (lognormal-lognormal). \\}
		\label{tab::ICcomp}
		\begin{tabular}{@{}l c c c c c l c c c c c l c c c c c @{}}
			& \multicolumn{5}{c}{WAIC-1} && \multicolumn{5}{c}{WAIC-2} && \multicolumn{5}{c}{DIC} \\
			\cline{2-6}
			\cline{8-12}
			\cline{14-18}
			Model & 1 & 2 & 3 & 4 & 5 && 1 & 2 & 3 & 4 & 5 && 1 & 2 & 3 & 4 & 5 \\ 
			\hline
			n = 1000, medium & 66 & 34 & 0 & 0 & 0 && 65 & 35 & 0 & 0 & 0 && 64 & 36 & 0 & 0 & 0 \\
			n = 1000, strong & 50 & 50 & 0 & 0 & 0 && 49 & 51 & 0 & 0 & 0 && 51 & 49 & 0 & 0 & 0 \\
			n = 2000, medium & 69 & 31 & 0 & 0 & 0 && 71 & 29 & 0 & 0 & 0 && 69 & 31 & 0 & 0 & 0 \\
			n = 2000, strong & 69 & 31 & 0 & 0 & 0 && 68 & 32 & 0 & 0 & 0 && 68 & 32 & 0 & 0 & 0 \\
			\hline
		\end{tabular}
\end{table*}

\begin{figure}[h]
	\centering
	\includegraphics[scale=.60, trim=130 120 140 120, clip, page =1]{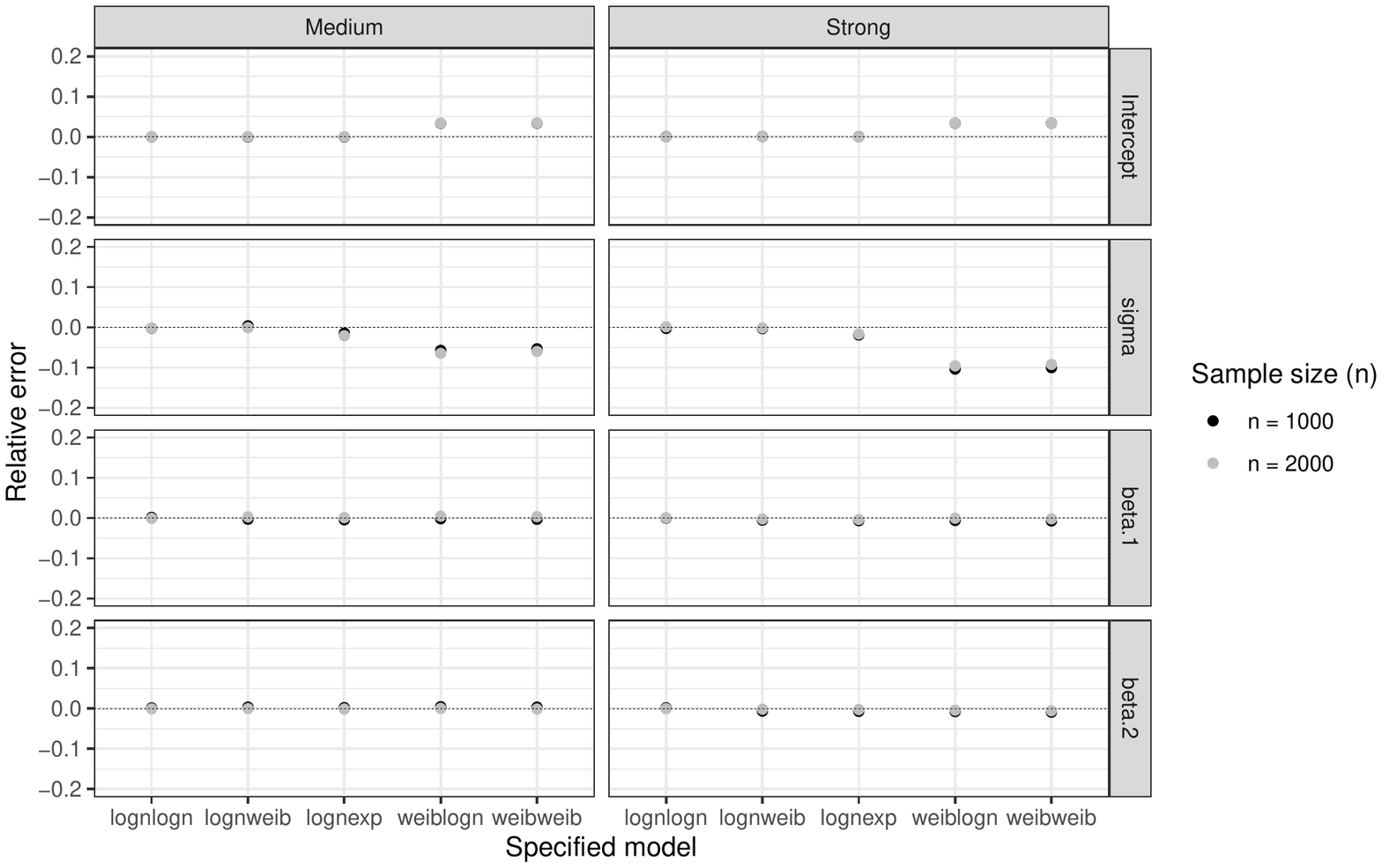}
	\caption{Mean relative error by misspecified models over 100 replicated data sets (500 sets for the lognormal-lognormal model). Parameters of the $x$-model shown. }
	\label{fig::modcomp_par1}
\end{figure}

\begin{figure}[h]
	\centering
	\includegraphics[scale=.60, trim=130 120 140 120, clip, page =2]{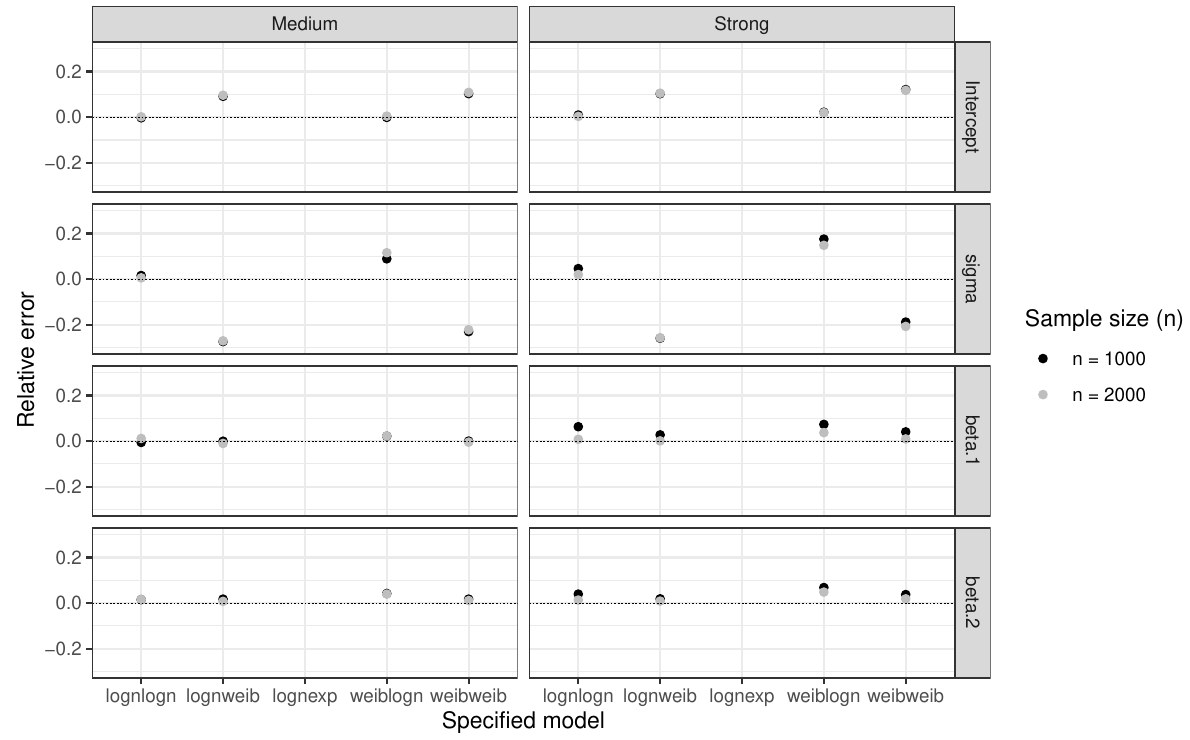}
	\caption{Mean relative error by misspecified models over 100 replicated data sets (500 sets for the lognormal-lognormal model). Parameters of the $t$-model shown. Mean relative error of the lognormal-exponential model off the scale.}
	\label{fig::modcomp_par2}
\end{figure}

\begin{figure}[h]
	\centering
	\includegraphics[scale=.80, trim=130 120 140 120, clip, page =1]{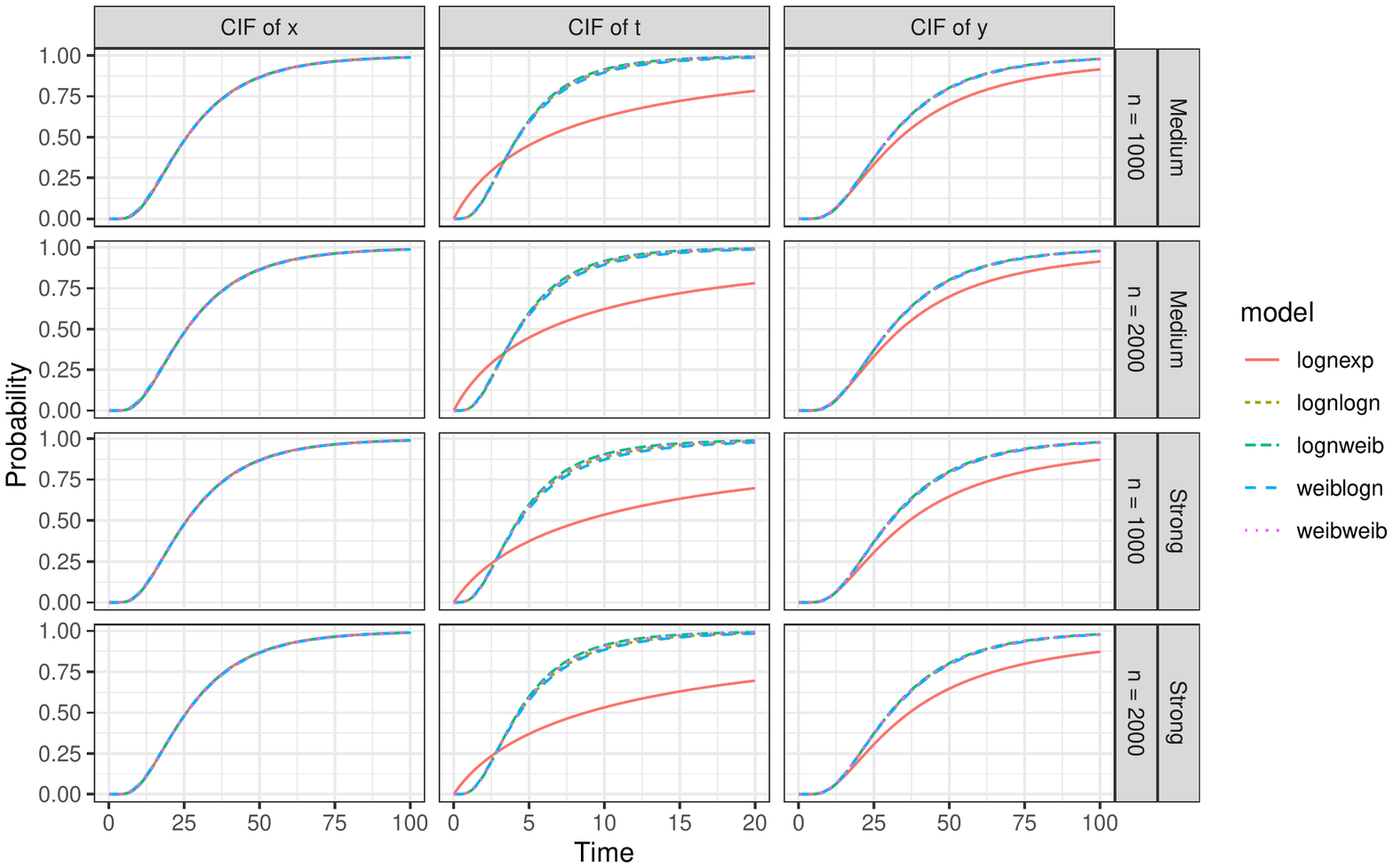}
	\caption{Cumulative Incidence functions (CIFs) averaged point-wise over 100 replicated data sets. }
	\label{fig::modcomp_cif}
\end{figure}

As the information criteria occasionally selected the wrong distribution for $t_i$, we also assessed parameter bias and accuracy of the estimated CIFs under misspecification. We found that the coefficients $\beta_1$ and $\beta_2$ were estimated with negligible bias (Figures \ref{fig::modcomp_par1}-\ref{fig::modcomp_par2}), unless the $t_i$ model was misspecified as exponential. The intercept $\beta_0$ and scale parameter $\sigma$ had larger error. However, it is important to note that these parameters are hard to compare across distributions. It is more important to assess, whether the cumulative incidence functions were comparable across misspecified models (Figure \ref{fig::modcomp_cif}). We found that the bias emerging from misspecification in the CIFs was, indeed, negligible, unless the transition was misspecified as exponentially distributed.


\clearpage
\newpage

\section{Supplement D: Additional information on the POBASCAM case study}
\subsection{MCMC chains of lognormal-loglogistic model}
\begin{figure}[h!]
		\centering
	\includegraphics[scale=.45, trim=20 40 20 80, clip]{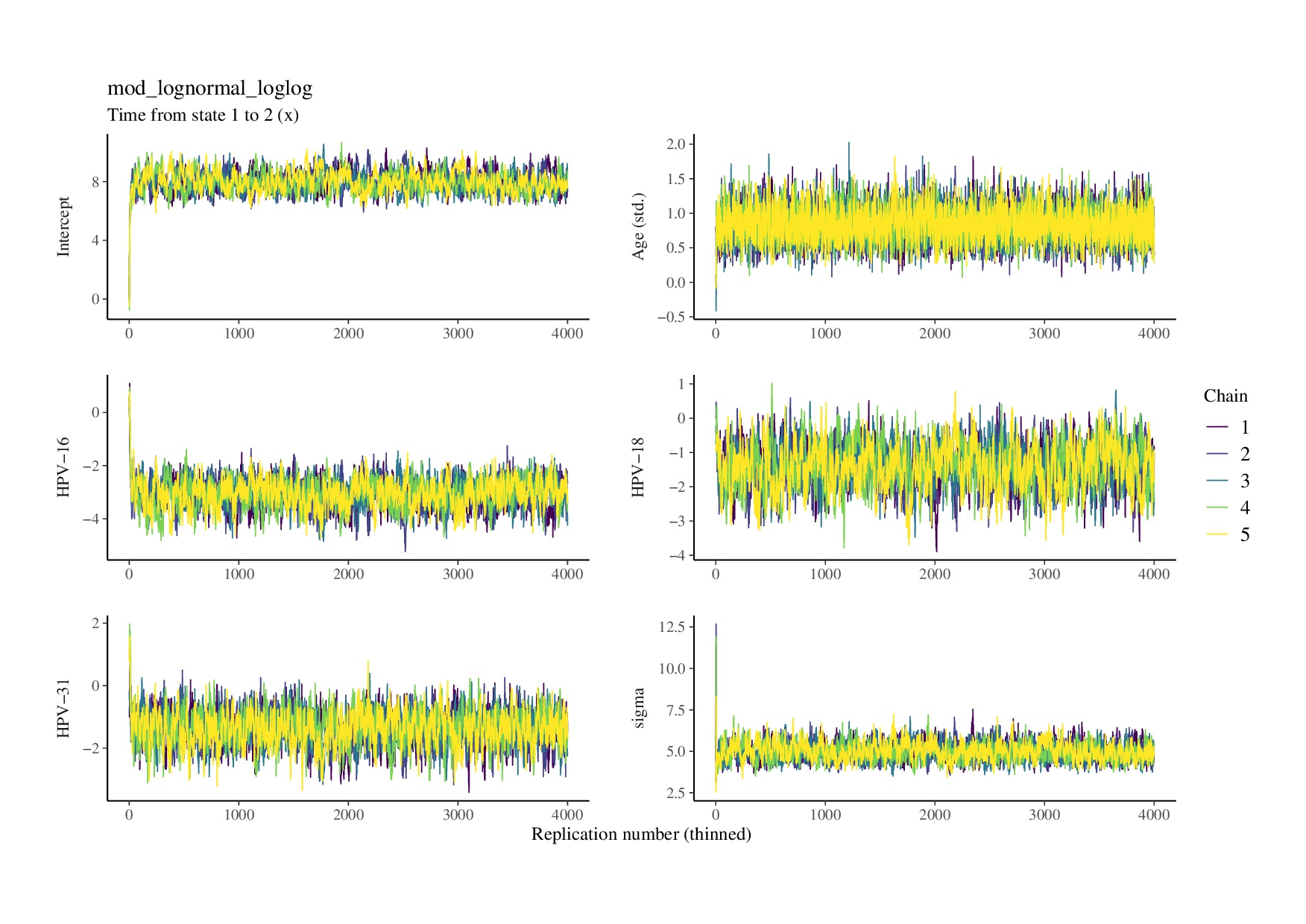}
	\caption{MCMC chains 1 through 5 (out of 15) of the lognormal-loglogistic model for the parameters of $\log x_i$. The chains were thinned using a step length of 200 replications. }
	\label{fig::mcmc_chain.x}
\end{figure}

\begin{figure}[h!]
		\centering
	\includegraphics[scale=.45, trim=20 40 20 80, clip]{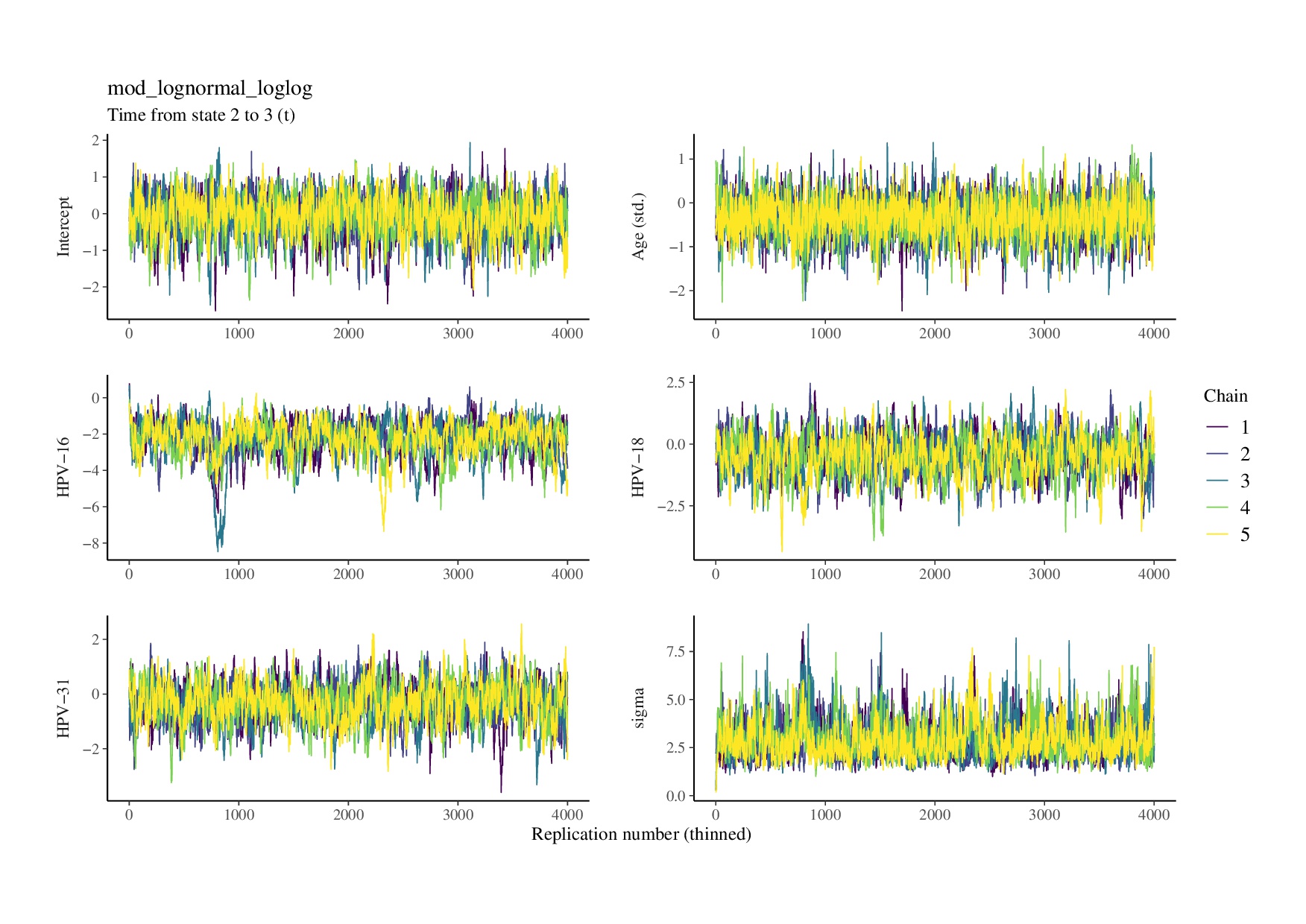}
	\caption{MCMC chains 1 through 5 (out of 15) of the lognormal-loglogistic model for the parameters of $\log t_i$. The chains were thinned using a step length of 200 replications.}
	\label{fig::mcmc_chain.t}
\end{figure}

To assess convergence of each models, we inspected the MCMC chains of the parameters of the models of $\log x_i$ and $\log t_i$. Figures \ref{fig::mcmc_chain.x} and \ref{fig::mcmc_chain.t} depict the MCMC chain of the lognormal-loglogistic model including the warm-up period. The figures depict the first five of fifteen MCMC chains, for clarity. The other chains looked similar.

 Convergence of all chains looked appropriate, as judged by a clear trend towards a single region of the posterior parameter space and good mixing of the MCMC chains. The thinned chains exhibited auto-correlation which is a well-known characteristic of MCMC chains when using data augmentation; see discussion Section of the main manuscript.

\subsection{Implementation notes for NPMLE and the Markov model}
The NPMLE for the CIF of $x_i$ and $y_i$ was implemented for the POBASCAM data using \texttt{R} package \texttt{interval} \citep{fay_exact_2010}, as in the simulation study (Section \ref{sec::simimpl}). The interval bounds were recoded before estimating $y_i$, as explained in Section \ref{sec::simimpl}. \\

A three-state Markov model was also implemented for the POBASCAM data using \texttt{R} package \texttt{msm} \citep{jackson_multi-state_2011}. The implementation was equivalent to that of the simulation study, Section  \ref{sec::simimpl}. 

\subsection{Lognormal-lognormal model fit by ML-EM }

We estimated the lognormal-lognormal model by the maximum likelihood expectation-maximization (ML-EM) algorithm described in Section B of the Supplemental Material.

\subsubsection{Implementation}
 We ran the ML-EM algorithm 100 times with random starting values. Each run first executed an approximate estimation by running the ML-EM algorithm with a low number of importance samples in the numerical integration ($m=100$) and a convergence limit of $10^{-4}$. By convergence limit we mean that the ML-EM algorithm was considered as converged when the change in the observed data likelihood came below the limit. The approximate solution can typically be obtained fast but may be inaccurate. Therefore, we refined the approximate solution by more precise ML-EM runs with higher integration accuracy, increasing $m$ to $2 \times 10^4$ importance samples and decreasing the convergence limit to $10^{-5}$. This run used the estimates from the approximate solution as starting values.

\subsubsection{Results}
\begin{figure}[h!]
	\centering
	\includegraphics[scale=1, trim=120 170 120 170, clip]{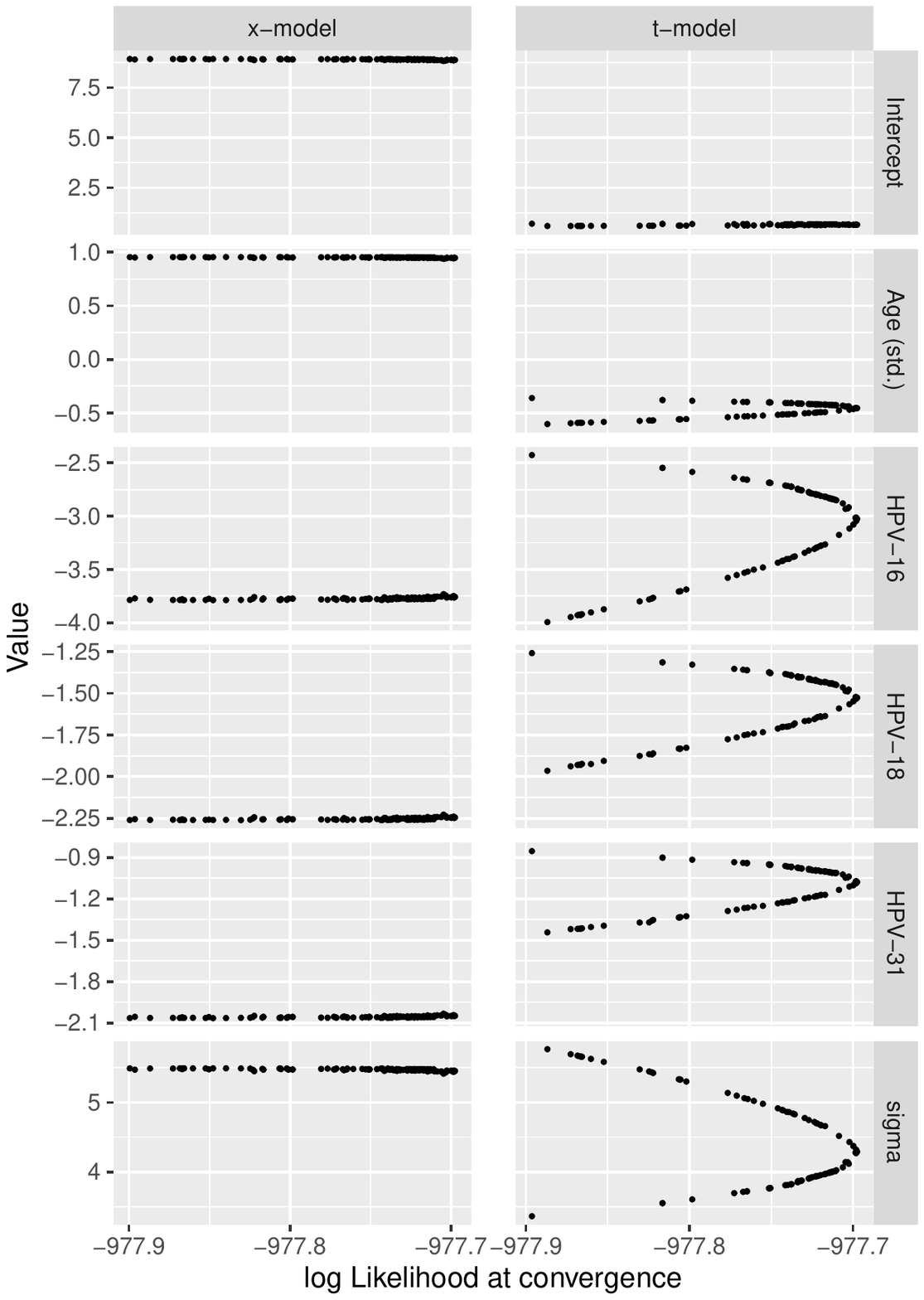}
	\caption{Parameter estimates by the EM algorithm (convergence limit $10^{-5}$) across 100 randomly initialized runs. Parameters $\betaf_t$ and $\sigma_t$ vary strongly across the iterations indicating a flat and/or multi-modal likelihood function.}
	\label{fig::MLEs}
\end{figure}

The repeated runs of the ML-EM algorithm converged at observed data log-likelihood values between -977.874 and -977.698. This result indicated a flat and/or multi-modal likelihood function, because we used a convergence limit of $10^{-5}$ and the differences in the obtained log-likelihood values after convergence had a range of values of 0.3. By this we mean that the algorithm either converged too early in a flat region before it reached the optimum or the algorithm found a local maximum. We, therefore, inspected the parameter estimates of the converged EM runs (Figure \ref{fig::MLEs}). The estimates of the parameters of the model of $x_i$ were stable across runs, but parameter estimates of the model of $t_i$ showed strong variation. In particular, $\sigma_t$ (minimum: 3.36, maximum: 5.76) and the regression coefficient of HPV-16 (-3.99, -2.43) varied strongly. However, visual inspection of the likelihood values against the parameter values indicated that the maximum log-likelihood value was approximately -977.698. We, therefore, used the estimates from this run as the maximum likelihood estimates (MLE). \\
\begin{table*}
	\centering
	\caption{Maximum likelihood estimates obtained by EM for the lognormal-lognormal model (log-likelihood = -977.698) compared to posterior median estimates from the Bayesian lognormal-lognormal model ($x_i$: time to CIN-2; $t_i$: time from CIN-2 to CIN-3+).}
	\label{tab::MLEs}
	\begin{tabular}{@{} l r r r r r @{}}
		& \multicolumn{2}{c}{MLE} && \multicolumn{2}{c}{Bayes} \\
		\cline{2-3}
		\cline{5-6}
		Parameter & $x_i$ & $t_i$ && $x_i$ & $t_i$ \\
		\hline		
		Intercept          & 8.89   &  0.64  &&  7.94 &  0.06  \\
		Age (standardized) & 0.95   & -0.45  &&  0.84 & -0.34  \\
		HPV-16*            & -3.76  & -3.03  && -3.04 & -2.01  \\
		HPV-18*            & -2.25  & -1.53  && -1.38 & -0.53  \\
		HPV-31*            & -2.05  & -1.08  && -1.35 & -0.33  \\
		sigma              & 5.45   &  4.29  &&  4.96 &  4.07  \\
		\hline
		\multicolumn{6}{l}{ * Reference: other HPV sub-type} \\
	\end{tabular}
\end{table*}

\begin{figure} 
	\includegraphics[scale=.60, trim=100 160 100 150, clip, page = 2]{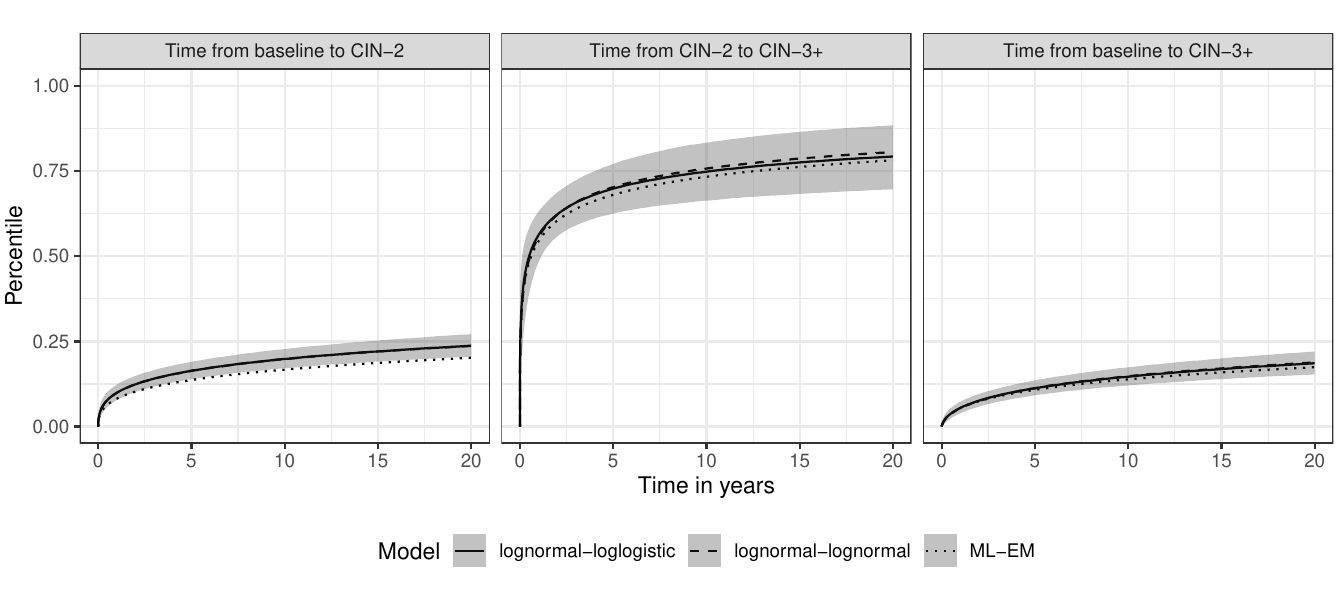}
	\caption{Posterior median of predictive cumulative incidence functions (CIF) for the Bayesian lognormal-loglogistic model (best performance acc. to WAIC-1), Bayesian lognormal-lognormal model, and lognormal-lognormal model fit by ML-EM. 95\% credible intervals of the lognormal-loglogistic model shown as shaded region.  }
	\label{fig::res_ppdplots_modcomp}
\end{figure}

Table \ref{tab::MLEs} compares the MLE with the posterior median estimates from the Bayesian lognormal-lognormal model. It is important to stress that the Bayesian estimates in Table \ref{tab::MLEs} are for the lognormal-lognormal model and they differ slightly from those in the main manuscript, because the model presented there is the lognormal-loglogistic model which fitted the data best. We found that the MLE and Bayesian estimates were comparable in magnitude and signs. In general, the  Bayesian estimates were slightly smaller, in absolute value than the MLE, which could be a result of the applied regularization. Figure \ref{fig::res_ppdplots_modcomp} compares the marginal cumulative incidence functions (CIF) of the Bayesian lognormal-lognormal model and the ML-EM fit to the Bayesian lognormal-logistic model (shown in the main manuscript). We find that the CIFs of the three models were almost identical on $t_i$ and $y_i$. The ML-EM CIF of $x_i$ had slightly stronger curvature than those of the Bayesian models.

%

\clearpage

\subsection{Conditional independence sensitivity analysis} \label{sec::application_sensitivity}

\begin{figure} [h]
	\centering
	\includegraphics[scale=.65,trim=100 170 100 150, clip, page = 1]{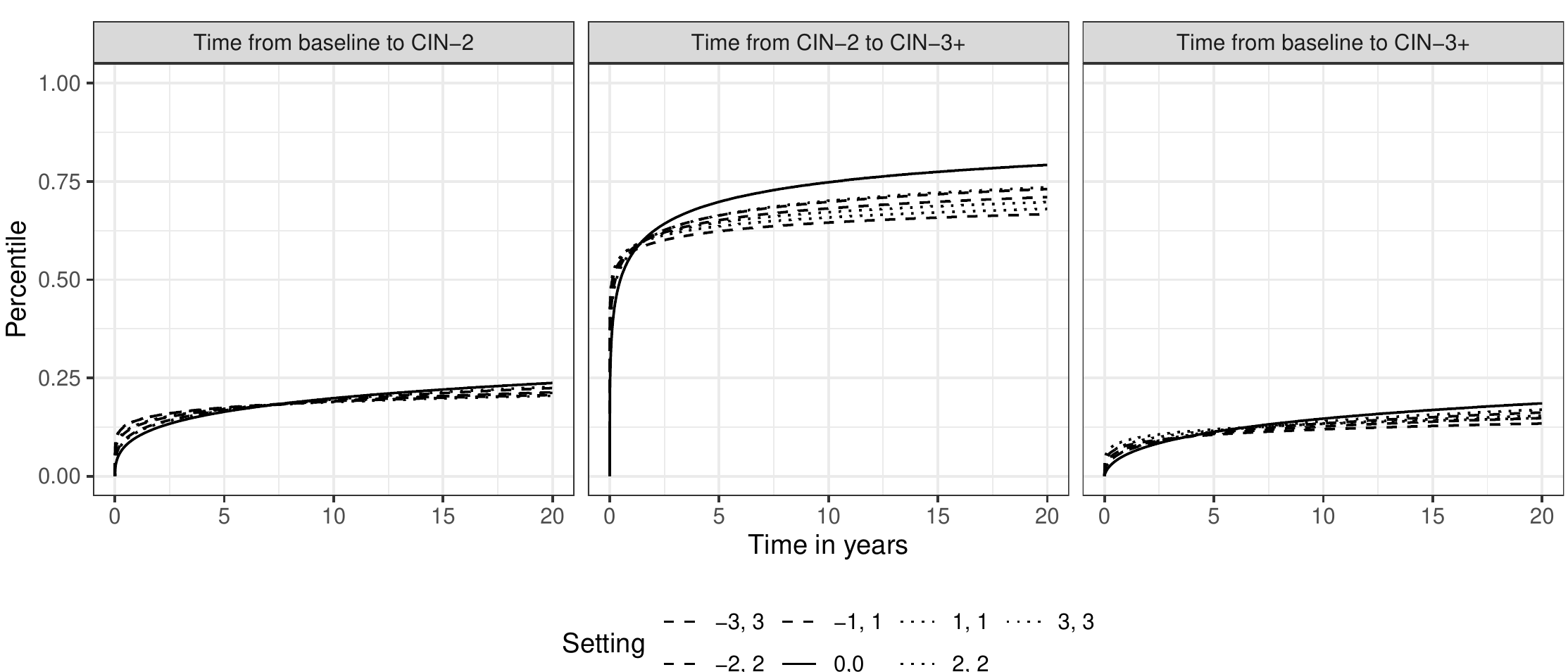}
	\caption{Posterior median of predictive cumulative incidence functions for different assumptions on $(\beta_{x,w},\beta_{t,w})$ with overlaid 95\% credible interval bands (lognormal-loglogistic model). The higher the absolute values of $(\beta_{x,w},\beta_{t,w})$ the stronger the curvature of the CIF as compared to the main model (black solid line).}
	\label{fig::sens_ppd}
\end{figure}

We assessed the sensitivity of our findings to the conditional independence assumption for the lognormal-loglogistic model. In particular, we specified a grid of parameters 
\begin{align*}
	(\beta_{x,w},\beta_{t,w}) = \{ (-3,3),(-2,2),(-1,1),(0,0),(1,1),(2,2),(3,3) \}
\end{align*}
in model 
\begin{align} 
	\log x_i = \z_{x,i}' \betaf_x + \beta_{x,w} w_i + \sigma_x \epsilon_{i} \label{eq::model_eq1_sens1} \\
	\log t_i = \z_{t,i}' \betaf_t + \beta_{t,w} w_i + \sigma_t \xi_{i} \label{eq::model_eq2_sens1},
\end{align}
where $(0,0)$ denotes the conditional independence model. We included both positive and negative $\beta_{x,w}$ to allow for an induced positive or negative partial association of $x_i$ and $t_i$. Condition $(1,1)$ generates, for example, a moderately strong unobserved confounding variable $w_i \sim N(0,1)$ for all $i$: individuals with $w_i$ one and two standard deviations above the mean have $\exp(1) = 2.71$ and $\exp(2) = 7.38$ times greater transitions times. Condition $(3,3)$ denotes a very strong effect, similar to that of HPV-16 in the conditional independence model for $\log x_i$.  All considered deviations from conditional independence caused a stronger curvature in the predictive CIF, especially that of $t_i$ (Figure \ref{fig::sens_ppd}). However, the conclusion of a rapid transition from CIN-2 to CIN-3+ by over half of the population was not changed, and the overall findings on the transition from baseline to CIN-2 and CIN-3+ remained very similar. The overlaid 95\% credible bands of all sensitivity conditions in Figure \ref{fig::sens_ppd} quantify the uncertainty added by relaxing conditional independence. \\

\begin{figure} [h]
	\centering
	\includegraphics[scale = .6, trim=100 160 50 185, clip, page = 3]{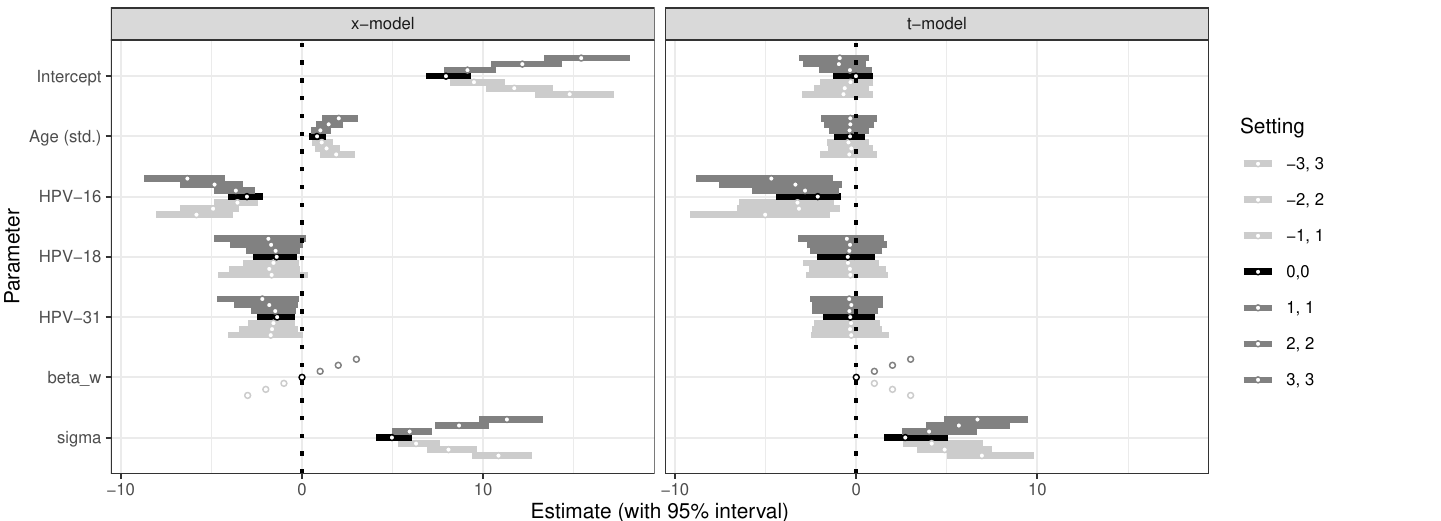}	
	\caption{Posterior parameter sensitivity to the conditional independence assumption. Black lines equivalent to main analysis model (lognormal-loglogistic).}
	\label{fig::sens_cind_par}
\end{figure}

The sign and significance of the regression coefficients did not change as compared to conditional independence (Figure \ref{fig::sens_cind_par}). The posterior distribution of the strongest effect (HPV-16) was sensitive to $(\beta_{x,w},\beta_{t,w})$ but it moved away from zero (i.e., the estimated effect was stronger) for larger absolute values of  $(\beta_{x,w},\beta_{t,w})$, while uncertainty increased. This held for both positive and negative $\beta_{x,w}$. In addition, scale parameters $\sigmaf$ and the intercept were more extreme for larger absolute values of  $(\beta_{x,w},\beta_{t,w})$, resulting in the stronger curvature observed in Figure \ref{fig::sens_ppd}. 

\clearpage
\newpage
\bibliographystyle{plainnat} 
\bibliography{bibliography_bayes3S}